\documentclass[twocolumn]{aastex63}

\usepackage{epsf}
\usepackage{amsmath}                
\usepackage{amsfonts}               
\usepackage{amssymb}                
\usepackage{epsfig}                 
\usepackage{float}
\usepackage{color}
\usepackage{tabularx}
\usepackage{comment}
\usepackage{makecell}
\usepackage{lineno}
\usepackage{longtable}
\usepackage{makecell}
\usepackage{enumitem}

\def \s{~\rm{s}}
\def \km{~\rm{km}}

\def \K{~\rm{K}}

\def \g{~\rm{g}}
\def \G{~\rm{G}}

\def \yr{~\rm{yr}}
\def \Myr{~\rm{Myr}}
\def \Gyr{~\rm{Gyr}}
\def \pc{~\rm{pc}}
\def \kpc{~\rm{kpc}}

\begin{document}



\title{Double Neutron Star Mergers and Fast Blue Optical Transients: A Circumbinary Disk Connection}

\author[0000-0002-2215-1841]{Aldana Grichener}
\affiliation{Steward Observatory and Department of Astronomy, University of Arizona, 933 North Cherry Avenue, Tucson, AZ 85721, USA}
\correspondingauthor{Aldana Grichener}
\email{agrichener@arizona.edu}

\author[0000-0003-3456-3349]{Ruggero Valli}
\affiliation{Max Planck Institute for Astrophysics, Karl-Schwarzschild-Str. 1, 85748 Garching, Germany}
\affiliation{Institute of Science and Technology Austria (ISTA), Am Campus 1, 3400 Klosterneuburg, Austria}

\author[0000-0002-6718-9472]{Mathieu Renzo}
\affiliation{Steward Observatory and Department of Astronomy, University of Arizona, 933 North Cherry Avenue, Tucson, AZ 85721, USA}

\author[0000-0002-6134-8946]{Ilya Mandel}
\affiliation{School of Physics and Astronomy, Monash University, Clayton, Victoria 3800, Australia}
\affiliation{The ARC Centre of Excellence for Gravitational Wave Discovery—OzGrav, Australia}

\author[0000-0001-5253-1338]{Kaitlin M. Kratter}
\affiliation{Steward Observatory and Department of Astronomy, University of Arizona, 933 North Cherry Avenue, Tucson, AZ 85721, USA}

\author[0000-0002-1530-9778]{Magdalena Siwek}
\affiliation{Center for Cosmology and Particle Physics, New York University, 726 Broadway, New York, NY 10003, USA}

\author[0000-0002-8134-4854]{Koushik Sen}
\affiliation{Steward Observatory and Department of Astronomy, University of Arizona, 933 North Cherry Avenue, Tucson, AZ 85721, USA}

\author[0000-0002-8134-4854]{Neev Shah}
\affiliation{Steward Observatory and Department of Astronomy, University of Arizona, 933 North Cherry Avenue, Tucson, AZ 85721, USA}


\begin{abstract}
Post-common-envelope evolution (CEE) circumbinary disks (CBDs) form when part of the ejected envelope remains bound and circularizes around the stripped core and its companion. Leveraging results of scale-free hydrodynamical simulations, we develop a framework to model the effect of such disks on subsequent binary evolution, and implement it in the population synthesis code \textsc{compas}. Applying our framework to double neutron star (DNS) merger progenitors, we find that CBDs generally induce substantial orbital shrinkage, changing the merging DNS population and lowering the overall merger rate. We explore an extensive grid of binary-evolution and disk models, and find that CBD interactions reduce the DNS merger rate by up to $28\%$, primarily through `premature' mergers between the neutron star (NS) and the giant’s core. These mergers may power luminous fast blue optical transients (LFBOTs), occurring at a fiducial rate of $\simeq11 \:\rm Gpc^{-3} \yr^{-1}$. Our results indicate that natal kicks contribute little to LFBOT offsets, and that they should be associated with young stellar populations, though not necessarily active star-forming regions. Since LFBOT progenitors would otherwise produce short-delay DNS mergers, CBDs further challenge DNS mergers as the source of early r-process enrichment. For surviving DNS mergers, accretion from the disk leads to higher chirp masses, potentially explaining the puzzling mass of GW190425. Our framework provides a general tool for studying CBD–binary interactions in post-CEE systems.
\end{abstract}


\keywords{binaries: general -- stars: neutron stars  -- stars: massive   -- stars: disks}

\section{INTRODUCTION}
\label{sec:intro}

Binaries are common across astrophysical scales, ranging from galaxy pairs to stellar and sub-stellar systems. In gas-rich environments, binarity often leads to the formation of circumbinary disks (CBDs), in which material settles into orbit around the binary as a whole rather than around either component individually (e.g., \citealt{Dutrey1994}; \citealt{Akeson2007}; \citealt{Hoard2009}; \citealt{Ragusa2021}; \citealt{Liao2021}; \citealt{GallardoCava2026}). The exchange of angular momentum and mass between the CBD and the binary through gravitational and accretion forces affects the binary's evolution (e.g., \citealt{ArtymowiczLubow1994}; \citealt{ArtymowiczLubow1996}), driving either orbital decay or expansion.  

The formation of CBDs around stellar binaries and their effect on the subsequent evolution of orbital separation and eccentricity has been extensively studied in recent years, with growing interest in CBDs around post-common-envelope evolution (CEE) binary systems (e.g., \citealt{KashiSoker2011}; \citealt{Dermineetal2013}; \citealt{ChenPodsiadlowski2017};  \citealt{IzzardJermyn2023}; \citealt{GagnierPejcha2023};  \citealt{Vetteretal2025}). If a binary survives dynamically unstable mass transfer, a fraction of the envelope material may remain bound to the system and fall back towards the inner binary. Since this gas carries substantial angular momentum, it cannot be re-accreted directly onto either component. Instead, it circularizes at radii comparable to a few times the post-CEE orbital separation, forming a rotationally supported CBD that can persist while it viscously spreads and is slowly accreted (e.g., see \citealt{Pringle1981}; \citealt{LaiMunoz2023} for reviews).

\begin{figure}
    \centering
    \includegraphics[width=8.5cm]{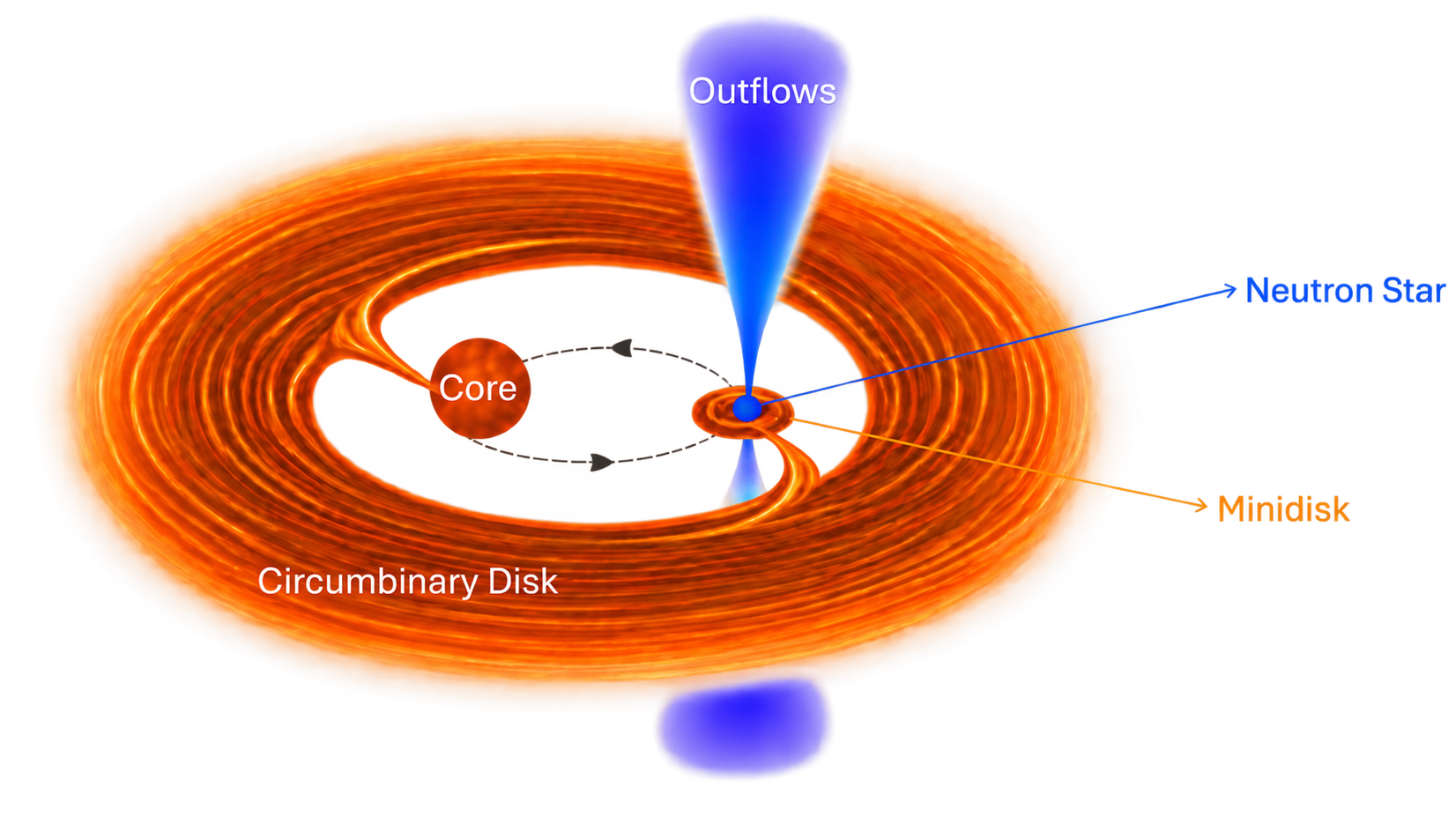} 
    \caption{Schematic illustration generated with ChatGPT of a CBD formed from leftover envelope material surrounding the exposed core and the NS at the end of the CEE phase. The CBD exchanges angular momentum with the inner binary, altering its subsequent orbital evolution. Material from the CBD may be re-accreted directly by the binary components or, in the case of the compact object, through a `minidisk' that can launch outflows. }
    \label{fig:CBD_3D}
\end{figure}

For binaries that involve a neutron star (NS) or a black hole (BH), the interaction with a CBD formed around the compact object and exposed core of its companion after CEE (see Fig.~\ref{fig:CBD_3D}) could affect the formation and merger rates of double compact objects (e.g., \citealt{Vetteretal2024}; \citealt{Weietal2024}; \citealt{Ungeretal2026}). If angular momentum is transferred from the binary to the disk, the orbit shrinks, potentially bringing its components into the gravitational-wave (GW)–driven merger regime, or shortening the merger timescale in systems that were already supposed to merge. If the compact object and core of the giant get even closer due to their interaction with the CBD, they may ultimately merge, producing energetic transients such as luminous fast blue optical transients (LFBOTs; e.g., \citealt{Soker2022}; \citealt{Metzger2022}; \citealt{TunaMetzger2023}; \citealt{CohenSoker2023}) and Ibn/Icn supernovae (SNe; e.g., \citealt{Weietal2024}); see \citealt{Grichener2025} for a recent review. Conversely, if the binary system extracts angular momentum from the disk, the orbit widens, driving the two components further apart. This orbital widening can delay or even prevent the merger of double compact objects that would otherwise coalesce within the age of the Universe through GW emission.

The effect of CBDs formed at the end of the CEE phase on binary compact object formation and mergers is particularly important for double neutron stars (DNSs). Most DNSs form through isolated binary evolution following a CEE phase between a giant star and an NS (e.g., \citealt{TutukovYungelson1993}; \citealt{DewivandenHeuvel2004}), after the exposed core of the giant explodes in a second SN event (e.g., \citealt{Taurisetal2015}; \citealt{Andrewsetal2015}; \citealt{Taurisetal2017}; \citealt{VignaGomezetal2018}; \citealt{Fragos2019}; \citealt{VignaGomezetal2020}; \citealt{Laplace2021}; \citealt{Vartanyan2021}; \citealt{Grichener2023}). Altering their merger timescales and rates due to CBD interactions could have important implications for future GW observations \citep{LIGO2025}, observations of Galactic DNSs through pulsar timing \citep{Manchester2005}, and the expected distribution of short gamma-ray bursts \citep{Goldsteinetal2017,Mandel:2025grb} and related transients, as well as for galactic chemical evolution \citep[e.g.,][]{Kobayashietal2023}. 

In this work, we incorporate the formation of CBDs right after the CEE phase, and their influence on the orbital evolution of the inner binaries, into the population synthesis code \textsc{compas} (\citealt{Stevensonetal2017}; \citealt{VignaGomezetal2018}; \citealt{TeamCOMPAS2022,Mandeletal2025}). Then, we study the effect of these CBDs on the evolution and mergers of DNS progenitors. In Section~\ref{sec:NumericalScheme}, we describe our CBD framework and population models. We then explore the implications of CBD interactions on DNS mergers (Section~\ref{sec:DNS_evolution_results}) and the occurrence of LFBOTs through `premature' mergers (Section~\ref{sec:FBOTs}). In Section~\ref{sec:OtherAssumptions}, we examine the sensitivity of our findings to different population synthesis assumptions. We find the resulting DNS chirp-mass distributions which may break degeneracies between model assumptions in Section~\ref{sec:ChirpMasses}. We discuss the implications of the reduction in prompt DNS mergers for r-process enrichment and the caveats of our work in Section~\ref{sec:Discussion} and summarize in Section~\ref{sec:Summary}.

\section{Methods}
\label{sec:NumericalScheme}

\subsection{CBD Framework in Population Synthesis}
\label{subsec:CBDformalsim}

\begin{figure}
    \centering
    \includegraphics[width=\linewidth]{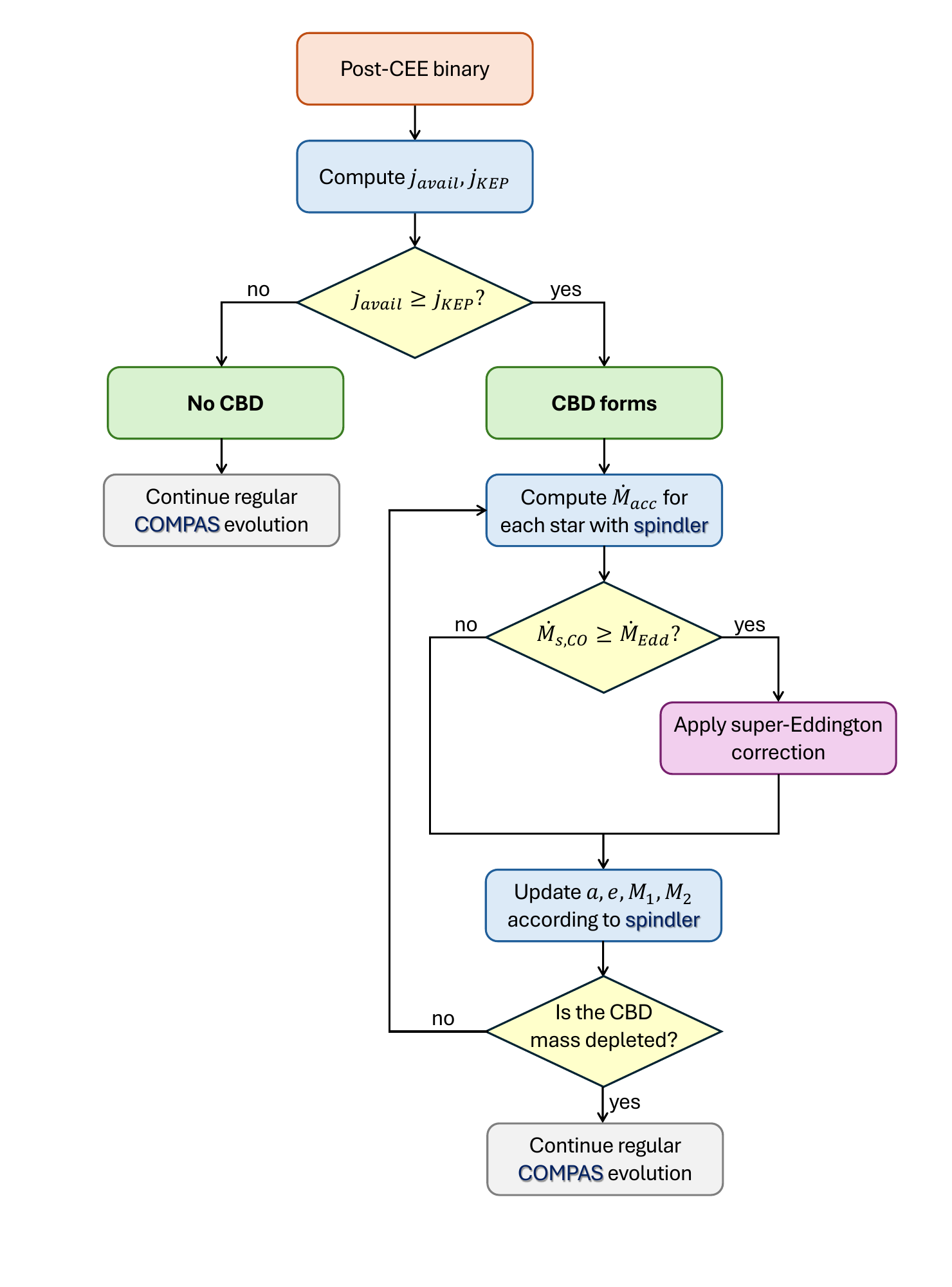} 
    \caption{Decision tree for implementing CBDs formed after CEE in the population synthesis code \textsc{compas}. $\dot{M}_{\rm s,CO}$ is short for $\dot{M}_{\rm supplied,CO}$, defined in Equation~\ref{eq:Mdot_acc_GE23}. $a,e,M_{\rm 1},M_{\rm 2}$ are the instantaneous orbital separation, eccentricity, and stellar masses at each iteration.}
    \label{fig:FlowTree}
\end{figure}

We build a formalism to study the effect of CBDs formed around the central binary from leftover envelope that remains bound after the CEE phase on a population level, and integrate it into the rapid population synthesis code \textsc{compas} (\citealt{Stevensonetal2017}; \citealt{VignaGomezetal2018}; \citealt{TeamCOMPAS2022}; \citealt{Mandeletal2025}). To model the evolution of single stars, \textsc{compas} uses the analytic fits from \citealt{Hurley2000} (see \citealt{Tout1996}), based on the stellar models of \cite{Pols1998}. Binary interactions are modeled semi-analytically following the general approach of \citealt{Hurleyetal2002} (see also \citealt{Tout1997}), but with significant changes in the treatment of mass transfer, CEE, winds, SNe, etc., as summarized by \cite{TeamCOMPAS2022}; \cite{Mandeletal2025}. We incorporate the effect of CBDs by combining an angular momentum based criterion for CBD formation \citep{Ungeretal2026} and the results of hydrodynamical simulations that study the interaction of the disks with the binaries they surround (\citealt{Siweketal2023,Siwek2023b}; as interpolated by \citealt{Vallietal2024}). Our framework is illustrated in Fig.~\ref{fig:FlowTree}. 

To determine whether a system can form a CBD, we evaluate $j_{\rm avail}$, the post-CEE specific angular momentum retained by the bound envelope material and available for disk formation. It depends on the fraction $f_{\rm M}$ of the pre-Roche-lobe-overflow (RLOF) giant’s envelope mass $M_{\rm env}$\footnote{We note that in \textsc{compas} the mass of the envelope remains unchanged from the onset of RLOF to the onset of CEE, while in reality the pre-CEE phase might occur on a thermal timescale (e.g., \citealt{Blagorodnova2021}), or mass could be removed through grazing envelope evolution (e.g., \citealt{Soker2015}).} that remains bound and may form a CBD of mass 
\begin{equation}
\begin{split}
M_{\rm CBD} =  f_{\rm M}M_{\rm env}, 
\label{eq:M_CBD}
\end{split}
\end{equation} 
and on the fraction $f_{\rm J}$ of the total angular momentum of the binary $J_{\rm total}$ at the onset of RLOF that this mass retains  
\begin{equation}
\begin{split}
J_{\rm avail} =  f_{\rm J}J_{\rm total}.  
\label{eq:J_CBD}
\end{split}
\end{equation}
Under this parametrization, $j_{\rm avail}$ is given by  
\begin{equation}
\begin{split}
j_{\rm avail} =  \frac{J_{\rm avail}}{M_{\rm CBD}}  = \beta \frac{J_{\rm total}}{M_{\rm env}}, 
\label{eq:CBD_AM}
\end{split}
\end{equation}
where $\beta \equiv f_{\rm J}/f_{\rm M}$ is a parameter that quantifies the specific angular momentum retention of the bound envelope material. 

We compare $j_{\rm avail}$ to the Keplerian specific angular momentum required to support a CBD 
\begin{equation}
j_{\rm KEP} = \zeta \sqrt{G(M_{\rm 1,post-CEE}+M_{\rm 2,post-CEE}) R_{\rm CBD,in}},
\label{eq:J_kep}
\end{equation}
where $M_{\rm 1,post-CEE}$ and $M_{\rm 2,post-CEE}$ are the masses of the binary components that survived the CEE phase, $R_{\rm CBD,in}$ is the inner disk radius of the potential CBD, and $\zeta$ is a dimensionless parameter that accounts for the mass distribution of the disk. We take $R_{\rm CBD,in}=2.5a_{\rm post,CEE}$, where $a_{\rm post,CEE}$ is the post-CEE orbital separation (e.g., \citealt{KashiSoker2011}; \citealt{TunaMetzger2023}; inspired by \citealt{ArtymowiczLubow1994}), and adopt $\zeta = 4$, which is roughly consistent with the density profile obtained in the CBD simulations of \cite{Vetteretal2024}. If
\begin{equation}
\begin{split}
j_{\rm avail} \ge j_{\rm KEP}, 
\label{eq:CBDcondition}
\end{split}
\end{equation}
we assume that the leftover envelope material has enough specific angular momentum to form a centrifugally supported CBD around the post-CEE binary. 

For systems where the condition in Equation~\ref{eq:CBDcondition} is satisfied, we incorporate the effect of CBDs on the evolution using \textsc{spindler} \citep{Vallietal2024}, an interpolation across the hydrodynamical simulations of \cite{Siweketal2023,Siwek2023b}, and additionally apply corrections that account for super-Eddington accretion. \cite{Siweketal2023,Siwek2023b} perform a large suite of two-dimensional scale-free hydrodynamical simulations of binaries immersed in CBDs using the moving-mesh code \textsc{arepo} (\citealt{Springel2010}; \citealt{Munoz2013}; \citealt{Pakmor2016}). The disk is treated using an $\alpha_{\rm visc}$-viscosity prescription \citep{ShakuraSunyaev1973} with $\alpha_{\rm visc}=0.1$ and fixed aspect ratio $h=0.1$. The disk mass is assumed to be small compared to the binary mass, so gravitational instability is not expected (though see discussion in Section~\ref{subsubsec:Disk_Properties_Caveats}). The binary components are represented by sink particles and kept on fixed Keplerian orbits, allowing the simulations to measure not only the gravitational and accretion torques exerted by the gas flow, but also the accretion onto each component individually. The secular evolution of the binary is then reconstructed from these quantities. \cite{Vallietal2024} interpolate across the results of these hydrodynamical simulations, and develop a model that predicts the orbital separation, eccentricity and mass ratio of a given binary following its interaction with a CBD. 

The parameter $\beta$ affects the interaction only through the CBD formation criterion: large values of $\rm \beta$ correspond to greater retained specific angular momentum and therefore allow CBDs to form in a larger fraction of systems. By contrast $f_{\rm M}$ sets the mass of the CBD, and therefore the amount of material available to accrete onto and exchange angular momentum with the binary. The inner binary in our simulations can accrete a fraction $f_{\rm acc} \le  f_{\rm M}$ from the pre-CEE envelope mass
\begin{equation}
\begin{split}
M_{\rm acc} =  f_{\rm acc}M_{\rm env}.
\label{eq:Macc}
\end{split}
\end{equation}
We use \textsc{spindler} to determine how the total accreted mass $M_{\rm acc}$ is partitioned between the two binary components, and subsequently correct the accreted amounts according to each stellar type. For the exposed core, accretion is governed mainly by disk viscosity, and its relatively long thermal timescale allows most of the inflowing material to be accreted. If one of the stars in the inner binary is a compact object, however, super-Eddington accretion is expected to drive winds from an accretion disk around the compact object fed by the CBD (also known as `minidisk', see Fig.~\ref{fig:CBD_3D}), expelling some fraction of the inflowing material. To check whether the \textsc{spindler} supplied mass corresponds to a super-Eddington accretion rate, we assume a constant mass accretion rate over the CBD lifetime, $\tau_{\rm CBD}$. We take a characteristic disk lifetime of \mbox{$\tau_{\rm CBD} = 10^{5} \yr$}, which corresponds to a timescale in which the outer edge of the disk has time to viscously spread to radii where photoevaporation becomes significant, and thereby drives wind mass loss that dissipates the disk  (e.g., \citealt{TunaMetzger2023}; \citealt{IzzardJermyn2023}). The CBD interaction proceeds in parallel with stellar and binary evolution, including wind mass loss and mass transfer, until the disk is depleted or the core undergoes a SN, which is assumed to eject the remaining CBD.  
The assumed timescales imply, however, that in most cases the CBD would be depleted well before the stripped core explodes in a stripped-envelope SN, forming the second NS in the DNS binary. 

We compare the constant accretion rate inferred from the \textsc{spindler} supplied mass with the compact object’s Eddington accretion rate. In case of super-Eddington accretion, we consider two prescriptions for the resulting compact-object accretion rate. First, we follow the results of \cite{GhodlaEldridge2023}, which allow for a strong super-Eddington accretion of the form
\begin{equation}
\begin{split}
\dot{M}_{\rm acc,CO} = f_{\rm GE} \dot{M}_{\rm supplied,CO},
\label{eq:Mdot_acc_GE23}
\end{split}
\end{equation}
where $\dot{M}_{\rm acc,CO}$ is the actual accretion rate by the compact object after accounting for mass loss through super-Eddington winds, $\dot{M}_{\rm supplied,CO}$ is the \textsc{spindler}-inferred accretion rate, and $f_{\rm GE}$ is the suppression factor, which is equal to $2/3$ in the case of NSs. To explore a more pessimistic scenario, we take the prescription of \cite{TunaMetzger2023} which results in
\begin{equation}
\begin{split}
\dot{M}_{\rm acc,CO} = 0.736\dot{M}_{\rm supplied,CO}^{2/5} \dot{M}_{\rm Edd}^{3/5},
\label{eq:Mdot_acc_TM23}
\end{split}
\end{equation}
where $\dot{M}_{\rm Edd}$ is the Eddington mass accretion rate. In both cases, since the mass is lost from the `minidisk' surrounding the accreting compact object, we assume that it carries away the accretor’s specific orbital angular momentum, and we remove the corresponding angular momentum from the binary together with the lost mass. Throughout the mass accretion and loss process, we update the post-CEE orbital separation, eccentricity, mass ratio, and component masses using the values calculated with \textsc{spindler} in each timestep.


\subsection{Population Synthesis Models}
\label{subsec:PopSynthModel}

After implementing the CBD formalism presented in Section~\ref{subsec:CBDformalsim}, we use \textsc{compas} to follow the evolution of massive binaries and study the effect of CBD interaction on the evolution and merger of DNS progenitors. Our `fiducial no CBD' model (model 1) follows the  \textsc{compas} code defaults \citep{Mandeletal2025}, as described in more detail below. We run a grid of populations varying different evolutionary assumptions (models 2-9), and build a `fiducial CBD' model by adding CBD interactions according to the formalism described in Section~\ref{subsec:CBDformalsim} to the `fiducial no CBD' case (model 10). Given the uncertainties in the retained specific angular momentum and the CBD mass, we evolve binaries across a range of CBD parameter choices (models 11-47). Then, we incorporate CBD interactions into the variations explored in models 2–9, allowing us to assess CBD effects across the different stellar and binary evolution assumptions (models 48-55). Each population contains $10^{7}$ massive binaries. The full list of simulations is described in Table~\ref{table:ModelVariation}.  

\begin{table*}
\begin{center}
\hspace{-1.5cm}
\setlength\tabcolsep{3 pt}
\scriptsize
\begin{tabular}{|| l l l ||}
 \hline
 \makecell{Number} & \makecell{Physics} & \makecell{Variation} \\ 
 \hline
  \makecell{1} & \makecell{`Fiducial no CBD'}  & \makecell{ } \\ 
 \hline
  \makecell{2} & \makecell{Metallicity distribution}  & \makecell{$f_{\rm Z}(Z) = \rm log-uniform$ } \\  
 \hline
 \makecell{3} & \makecell{SN recipe}  & \makecell{\cite{Fryeretal2012} remnant masses, bimodal kick} \\
 \hline
  \makecell{4} & \makecell{Mass transfer stability}  & \makecell{$q_{\rm crit}$ from \cite{Claeysetal2014}} \\
 \hline
  \makecell{5} & \makecell{Mass transfer stability}  & \makecell{`adiabatic' $q_{\rm crit}$ from \cite{Geetal2020}} \\
 \hline
  \makecell{6} & \makecell{Mass transfer stability}  & \makecell{`isentropic' $q_{\rm crit}$ from \cite{Geetal2020} } \\
 \hline
  \makecell{7} & \makecell{CEE}  & \makecell{ $\alpha = 0.5$} \\
 \hline
  \makecell{8} & \makecell{CEE}  & \makecell{ $\alpha = 2$ } \\
 \hline
   \makecell{9} & \makecell{CEE}  & \makecell{ $e_{\rm post,CEE}=0$ } \\
 \hline
   \makecell{10} & \makecell{`Fiducial CBD'}  & \makecell{$\beta=3$, $f_{\rm M}=0.05$} \\ 
 \hline
    \makecell{11-22} & \makecell{CBD disk mass fraction}  & \makecell{$f_{\rm M}=\{0.005, 0.01, 0.02, 0.03, 0.04, 0.05, 0.06, 0.07, 0.08,$\\ $0.09, 0.095 ,0.10$\}, $\beta=3$} \\
 \hline
   \makecell{23-46} & \makecell{CBD specific angular momentum}  & \makecell{$\beta=\{0.5, 1, 1.25, 1.4, 1.5, 1.55, 1.6, 1.65 ,1.7, 1.75, 1.9, 2,$ \\ $2.15, 2.25, 2.4, 2.5, 2.75, 3, 3.5, 4, 4.5, 5, 5.5, 6\}$, $f_{\rm M}=0.05$} \\
 \hline
    \makecell{47} & \makecell{Super-Eddington accretion mode}  & \makecell{\cite{TunaMetzger2023} super-Eddington accretion, $\beta=3$, $f_{\rm M}=0.05$} \\ 
 \hline
   \makecell{48} & \makecell{Metallicity distribution, CBD}  & \makecell{$f_{\rm Z}(Z) = \rm log-uniform$, $\beta=3$, $f_{\rm M}=0.05$ } \\ 
 \hline
 \makecell{49} & \makecell{SN recipe, CBD}  & \makecell{\cite{Fryeretal2012} remnant masses, bimodal kick, $\beta=3$, $f_{\rm M}=0.05$} \\
 \hline
  \makecell{50} & \makecell{Mass transfer stability, CBD}  & \makecell{ $q_{\rm crit}$ from \cite{Claeysetal2014}, $\beta=3$, $f_{\rm M}=0.05$} \\
 \hline
  \makecell{51} & \makecell{Mass transfer stability, CBD}  & \makecell{`adiabatic' $q_{\rm crit}$ from \cite{Geetal2020}, $\beta=3$, $f_{\rm M}=0.05$ } \\
 \hline
  \makecell{52} & \makecell{Mass transfer stability, CBD}  & \makecell{`isentropic' $q_{\rm crit}$ from \cite{Geetal2020}, $\beta=3$, $f_{\rm M}=0.05$ } \\
 \hline
  \makecell{53} & \makecell{CEE, CBD}  & \makecell{ $\alpha = 0.5$, $\beta=3$, $f_{\rm M}=0.05$ } \\
 \hline
  \makecell{54} & \makecell{CEE, CBD}  & \makecell{ $\alpha = 2$, $\beta=3$, $f_{\rm M}=0.05$ } \\
 \hline
    \makecell{55} & \makecell{CEE, CBD}  & \makecell{ $e_{\rm post,CEE}=0$, $\beta=3$, $f_{\rm M}=0.05$} \\
 \hline
 \end{tabular} 
\end{center}
\caption{Variations in \textsc{compas} simulations compared to the `fiducial no CBD' case.}
\centering
\label{table:ModelVariation}
\end{table*}

To focus on massive stars that could explode as SNe forming an NS remnant, we sample the zero-age main sequence (ZAMS) mass of the primary (initially heavier) star from the Kroupa initial mass function (IMF) $dN/dM_{\rm 1,ZAMS} \propto M_{\rm 1,ZAMS}^{-2.3}$ \citep{Kroupa2001} with masses between $5 \le M_{\rm 1,ZAMS}/ M_{\rm \odot} \le 100$. We draw the mass of the secondary (initially lighter star) from a flat distribution in mass ratio with $ 0.1 \le q_{\rm ZAMS} \equiv M_{\rm 2,ZAMS}/M_{\rm 1,ZAMS} \le 1$ \citep{Sana2012} and the initial orbital separation from a flat-in-the-log distribution with $ 0.1 \le a_{\rm ZAMS} / \rm au \le 1000$  (\citealt{Opik1924}). For simplicity, we assume circular orbits at the beginning of the evolution (see \citealt{deMinkBelczynski2015} for a study showing the negligible impact of initial eccentricities on the merger rates of double compact objects). 

All of the stars in our fiducial model have solar metallicity of $Z_{\rm \odot}=0.0142$ (\citealt{Asplund2009}, see discussion in Section~\ref{subsubsec:Stellat_Evolution_Caveats}). We relax this assumption by running models with a log-uniform metallicity distribution with a minimum value of $Z_{\rm min} =10^{-4}$ and a maximum value of $Z_{\rm max} =0.03$ (e.g., \citealt{Neijssel2019}; models 2, 48). This distribution is dominated by low-Z systems, as it heavily weights subsolar metallicities. 

NSs are formed in SN explosions that eject mass and may impart a natal kick to the compact remnant. In our fiducial models, we use the \cite{MandelMuller2020} prescription in which the natal kick amplitude depends on the pre-SN carbon-oxygen and helium shell mass. We run an additional model using the \cite{Fryeretal2012} delayed SN prescription and adopting a bimodal kick velocity distribution. We take a higher mode of $\sigma_{\rm high}=217 \km \s^{-1}$ for standard core-collapse SNe (one of the distributions from \citealt{DisbergMandel:2025}; correction to \citealt{Hobbs2005}) and a lower mode $\sigma_{\rm low}=30 \km \s^{-1}$ for electron capture SNe and for ultra stripped SNe where $\lesssim 0.2 M_{\rm \odot}$ of the envelope mass is bound to the progenitor at the moment of explosion (e.g., \citealt{VignaGomezetal2018}; models 3, 49). 

The vast majority of binaries that evolve towards DNS systems go through a CEE phase between the giant secondary and the NS that resulted from the SN explosion of the primary. The CEE phase is initiated when mass transfer from the giant star to its NS companion becomes dynamically unstable. We determine the stability of the mass transfer episode by comparing the donor's adiabatic response to mass loss with that of the Roche-lobe radius $R_{\rm RL}$. If the donor's response, quantified by $d \ln R_{\rm donor,MT}/ d \ln M_{\rm donor, MT}$, where $R_{\rm donor,MT}$ and $M_{\rm donor,MT}$ are the radius and mass of the donor at the onset of mass transfer, is smaller than that of the Roche lobe, $d \ln R_{\rm RL}/ d \ln M_{\rm donor, MT}$, the mass transfer episode is unstable and a CEE phase begins. We perform additional simulations varying the criteria of mass transfer stability, using critical mass ratios that separate between stable and unstable RLOF from \citealt{Claeysetal2014} (models 4, 50) and \citealt{Geetal2020} (models 5-6, 51-52). 

To determine the evolution of the binary during the CEE phase, we use the energy formalism (e.g., \citealt{Webbink1984}; \citealt{LivioSoker1988}), which compares the binding energy of the envelope prior to the CEE phase with the difference in orbital energies at the onset and at the termination of CEE, $E_{\rm bind} = \alpha\Delta E_{\rm orb}$. In our fiducial model, the efficiency parameter $\alpha$, which represents the fraction of the orbital energy budget available to unbind the envelope, is equal to one. We run two additional simulations with $\alpha=0.5$ (models 7, 53) and $\alpha=2$ (models 8, 54). The latter represents an additional energy source needed to unbind the envelope, and was favored by several double compact object studies (e.g., \citealt{Fragos2019}; \citealt{Garciaetal2021}; \citealt{Zevinetal2021}; \citealt{BroekgaardenBerger2021}). We adopt the standard \textsc{compas} prescriptions for CEE outcomes, without accounting for the effect of CBD torques on mass transfer (see discussion in Section~\ref{subsubsec:Stellat_Evolution_Caveats}).   

CEE inspiral typically tends to circularize the orbit through strong dissipation and drag, leaving at most a small residual eccentricity in most surviving post-CE binaries (e.g., \citealt{RickerTaam2012}; \citealt{Sand2020}; \citealt{Gonzales2022}; see discussion in Section~\ref{subsubsec:Stellat_Evolution_Caveats}). Due to the uncertainty in the exact eccentricity and its potential importance to the interaction of the CBD with the binary, in the fiducial case we adopt a simple phenomenological prescription in which the post-CEE eccentricity is set to $10\%$ of its pre-RLOF value. We relax this assumption and run additional models with circular post-CEE binaries (models 9, 55), following standard practice in population synthesis studies. 

We use the condition in Equation~\ref{eq:CBDcondition} to find which systems formed a CBD after the CEE phase. Motivated by the hydrodynamical simulations of \cite{Lau2022} and the analysis of \cite{TunaMetzger2023}, we take the fraction of envelope mass that forms the CBD (equation~\ref{eq:M_CBD}) to be $f_{\rm M}=0.05$, and adopt the specific angular momentum retention parameter (Equation~\ref{eq:CBD_AM}) to be $\beta = 3$ in our `fiducial CBD model' (model 10). This also coincides with the observations of AGB disks that point towards high values of $\beta$ \citep{IzzardJermyn2023}. Due to the large uncertainty in the amount of mass that remains bound after CEE and is available for CBD formation, we perform a parameter space exploration of $f_{\rm M}$, varying its value from 0.005 to 0.1 (models 11-22). We set its upper limit to keep the CBD gravitationally stable (e.g., \citealt{Teasdale2026}; see discussion in Section~\ref{subsubsec:Disk_Properties_Caveats}). We also run additional models varying the value of $\beta$ from half to six (models 23-46), keeping $f_{\rm J} = \beta f_{\rm M}$ below one, as required by angular-momentum conservation. In our `fiducial CBD model', we account for super-Eddington accretion using the prescription of \citealt{GhodlaEldridge2023} (Equation~\ref{eq:Mdot_acc_GE23}), taking $f_{\rm GE}=2/3$, as inferred for NS accretors. To explore the sensitivity of our results to this treatment, we perform an additional simulation using the prescription of \citealt{TunaMetzger2023} (Equation~\ref{eq:Mdot_acc_TM23}; model 47).

\section{CBD Interactions in DNS Progenitors}
\label{sec:DNS_evolution_results}

\subsection{Impact of CBD Mass and Angular Momentum}
\label{subsec:Mass_and_separation}
 
\begin{figure*}
    \centering
    \includegraphics[width=\linewidth]{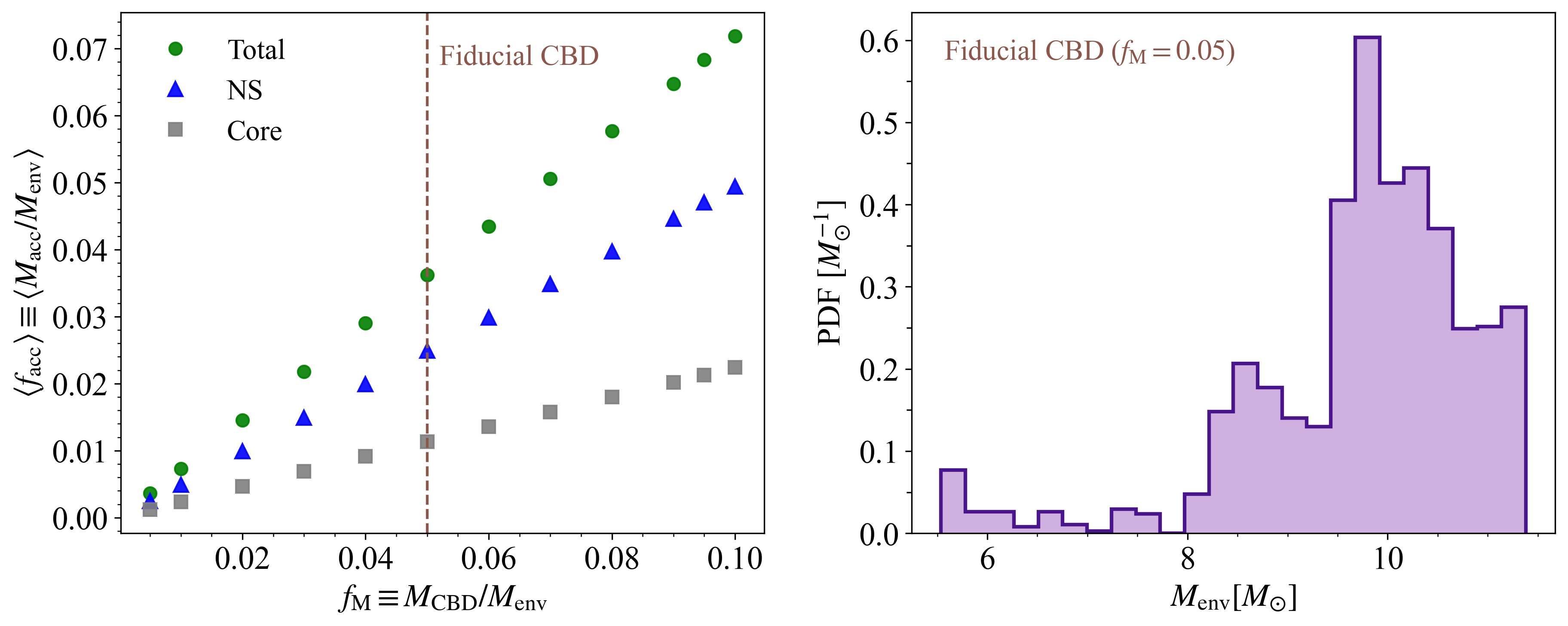}
    \caption{Mass accretion from a CBD onto the post-CEE binary in DNS merger progenitors: CBDs lead to significant mass growth. Left panel: average fraction of the envelope accreted by the binary $f_{\rm acc} \equiv M_{\rm acc}/M_{\rm env}$ shown as a function of the envelope fraction that remains bound after CEE and forms a CBD $f_{\rm M} \equiv M_{\rm CBD}/M_{\rm env}$. The blue triangles and gray squares show the fraction of envelope mass accreted by the NS and the core, respectively, while the green dots represent the fraction of envelope mass accreted by the central binary as a whole (sum of both components). The brown dashed vertical line marks our `fiducial CBD' model ($f_{\rm M}=0.05$; $\beta=3$). Right panel: envelope mass distribution for our `fiducial CBD' case prior to the CEE stage, normalized such that the area under the curve is unity.}
    \label{fig:BinaryAcrretion}
\end{figure*}

We study the interaction between CBDs formed after the CEE phase and the inner binaries in DNS progenitors. In the left panel of Fig.~\ref{fig:BinaryAcrretion}, we show how the average fraction of the envelope accreted by the binary and its components varies with the fraction of the envelope that is redistributed into a CBD. We note that even when accounting for the suppression in the accretion rate due to super-Eddington outflows, the NS, which is the less massive component of the binary, tends to accrete a larger fraction of the CBD mass. 

Let us explore our `fiducial CBD' model (model 10), where $5\%$ of the envelope remains bound and forms a CBD around the NS and the core, as an example. In this case, the central binary accretes on average $4\%$ of the envelope, which is about $76\%$ of the CBD mass. Taking an envelope mass of $\simeq 10M_{\rm \odot}$, motivated by the peak of its mass distribution (right panel of Fig.~\ref{fig:BinaryAcrretion}), implies about $0.38 M_{\rm \odot}$ is accreted by the binary system. Roughly $0.26 M_{\rm \odot}$ of this mass is accreted by the NS at a rate of $\simeq 80$ times the Eddington rate, and $0.12 M_{\rm \odot}$ is re-accreted by the giant’s core that would later explode and form a DNS. This mass gain leads to heavier NSs, and, consequently, higher binary chirp masses (see Section~\ref{sec:ChirpMasses}). The super-Eddington accretion rates onto the NS also raise the possibility that some of these post-CEE binaries could appear as ultraluminous X-ray sources during their interaction with the CBD (e.g., \citealt{Kaaretetal2017} for a review). However, their X-ray luminosities and observability would depend on the accretion flow geometry, beaming, and obscuration by surrounding material. Adopting the \cite{TunaMetzger2023} super-Eddington accretion allows for an average accretion rate by the NS of only five times the Eddington limit, reducing the amount of mass accreted by the NS to $\simeq 0.02 M_{\rm \odot}$, while keeping the mass accreted by the core mostly unchanged. 

\begin{figure*}
    \centering
    \includegraphics[width=\linewidth]{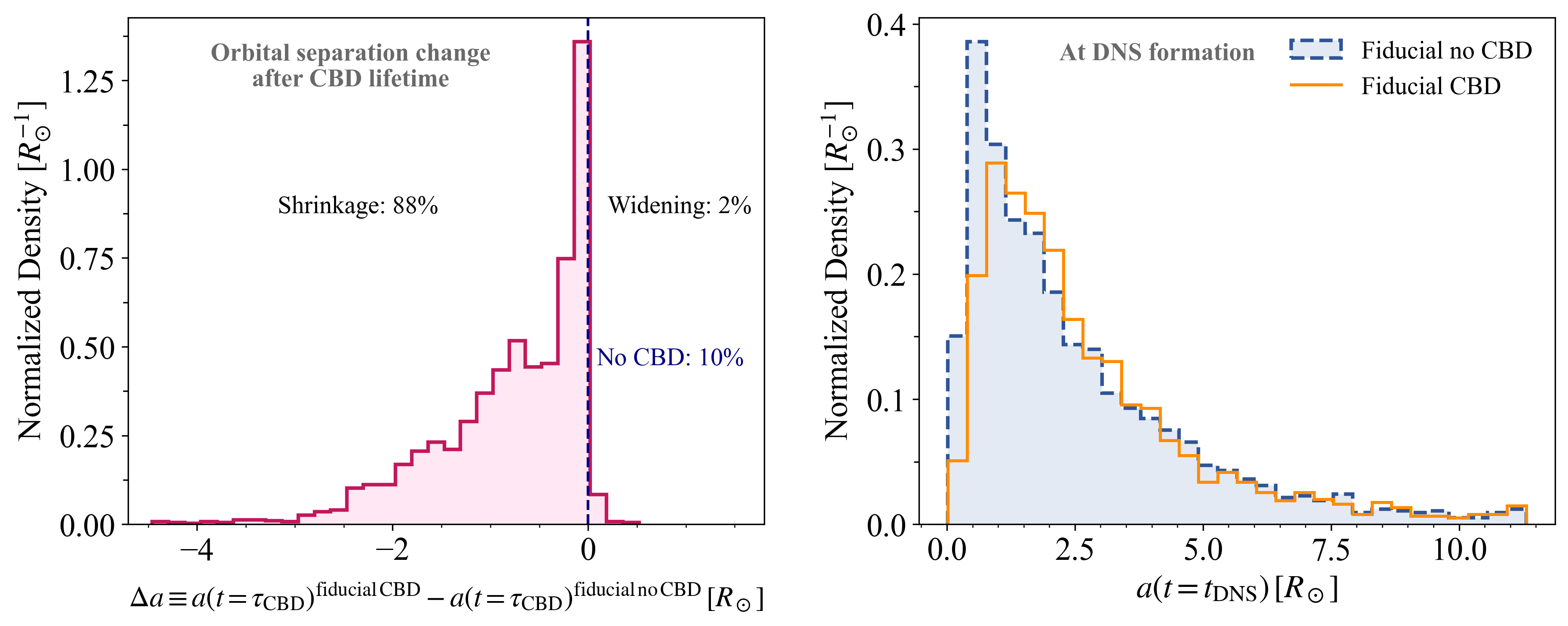}
    \caption{Effect of post-CEE CBD interaction on the subsequent semimajor axis evolution of DNS merger progenitors in the fiducial models: CBD interactions mostly shrink tight post-CEE binaries, reducing the number of DNS mergers. Left panel: CBD-induced change in orbital separation, defined as $\Delta a\equiv a(t=\tau_{\rm CBD})^{\rm fiducial \:CBD}-a(t=\tau_{\rm CBD})^{\rm fiducial \: no \:CBD}$, where $a(t=\tau_{\rm CBD})^{\rm fiducial \:CBD}$, $a(t=\tau_{\rm CBD})^{\rm fiducial \: no \:CBD}$ are the separation after the CBD lifetime in the `fiducial CBD' model and in the `fiducial no CBD' model, respectively. Negative values therefore indicate CBD-driven orbital shrinkage, while positive values indicate orbital widening. The blue dashed line marks $\Delta a=0$. Right panel: semimajor axis distribution at the moment of DNS formation $t_{\rm DNS}$ after the explosion of the core in a second SN event. Blue bins represent our `fiducial no CBD' case, while the orange outline shows the same model including CBD interactions. The no-CBD distribution integrates to unity, while the CBD distribution is shown using the same normalization, preserving the relative number of systems between the two models.}
    \label{fig:SemiMajorAxes}
\end{figure*}

Besides driving gas accretion in which the binary components gain mass, the CBD influences the evolution of the binary through gravitational torques that exchange angular momentum with the orbit. In Fig.~\ref{fig:SemiMajorAxes}, we show the CBD-induced change in the semimajor axis of DNS merger progenitors,  combining all binaries that evolve into DNSs that merge within 14 Gyr either with or without CBD interactions in our fiducial models. As shown in the left panel, $90\%$ of the DNS merger progenitors form CBDs after CEE, and most of them contract significantly as angular momentum is transferred from the binary to the disk. This reduces the number of systems that ultimately form DNSs, especially for tighter binaries, as can be seen in the right panel. Although CBD interactions also significantly alter the post-CEE eccentricity, producing a mean increase of $0.16$, this change is largely erased in our models by circularization after a case BB mass transfer episode that occurs between the core and the NS once the CBD is depleted.

\subsection{Populations of Merging DNSs}
\label{subsec:DNS_merger_populations}

\begin{figure*}
    \centering
    \includegraphics[width=14cm]{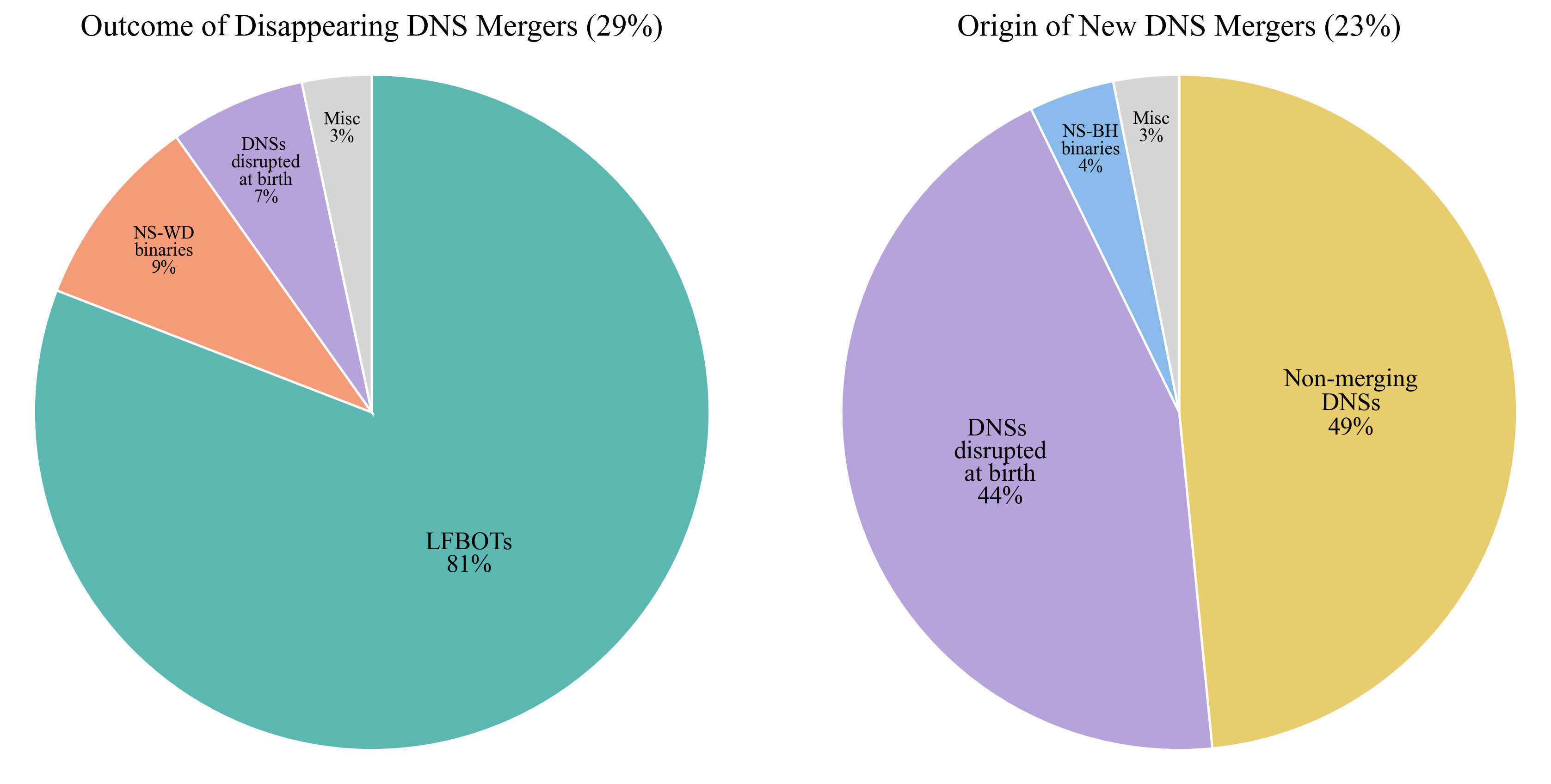}
    \caption{Effect of CBD interaction on which binary systems become merging DNSs in the fiducial model: CBD interactions substantially change the merging-DNS population by both preventing some systems from merging as DNSs, and enabling other systems to become merging DNSs. Left panel: Fate of systems that produce merging DNSs in the `fiducial no CBD' model but evolve into different outcomes once CBD interactions are included. Right panel: `fiducial no CBD' outcomes of systems that do not form merging DNSs without CBD interactions, but evolve into merging DNSs once CBD interactions are included. }
    \label{fig:DNS_evolution_Channels}
\end{figure*}

We explore how the CBD-driven changes in the mass and orbital separation of DNS merger progenitors shown in Figs.~\ref{fig:BinaryAcrretion} and \ref{fig:SemiMajorAxes} influence their subsequent evolution and final fate. As shown in Fig.~\ref{fig:DNS_evolution_Channels}, CBD interactions prevent $\simeq 29\%$ of the binaries that would otherwise have evolved to merging DNSs from doing so (left panel), while enabling the formation of new merging DNSs in systems that would otherwise result in different evolutionary outcomes, accounting for $\simeq 23\%$ of the new DNS merger population (right panel). 

In the presence of a CBD, many binaries that would have formed merging DNSs follow different evolutionary pathways instead (left panel of Fig.~\ref{fig:DNS_evolution_Channels}). First and most importantly, the orbital shrinkage in some DNS progenitors leads to `premature' mergers between the NSs and stellar cores mostly during the interaction with the CBD, preventing the binary systems from evolving to DNSs. These `premature' mergers account for $\simeq 81\%$ of the DNS mergers that no longer occur when CBD interactions are included, and can potentially power LFBOTs (see Section~\ref{sec:FBOTs}). Another $\simeq 9\%$ of the missing DNS mergers are instead lost to carbon--oxygen white dwarf (WD) -- NS binaries. In these systems, CBD-induced orbital shrinkage initiates or enhances subsequent mass transfer from the stripped core to the first-born NS, removing sufficient mass from the core such that it forms a WD rather than undergoing the stripped-envelope SN that would otherwise produce the second NS. A quarter of these systems pass through an ultracompact helium-star-donor X-ray phase (e.g., \citealt{Wang2021}), before forming detached NS–carbon–oxygen WD binaries. None of the resulting WDs subsequently transfers mass onto the NS in our simulations. If these systems were later driven into contact, their relatively high WD masses would make stable, long-lived NS–WD ultracompact X-ray binary evolution unlikely \citep{Bobrick2017}. A remaining $\simeq 6\%$ of binaries that would have evolved to DNS mergers without a CBD get disrupted during the second SN. In these systems, accretion from the CBD leads to secondaries with larger CO core masses (left panel of Fig.~\ref{fig:BinaryAcrretion}) that receive larger natal kicks when exploding as stripped envelope SNe in our model, and are therefore more likely to disrupt binaries. 

On the other hand, the loss of these DNS mergers is largely compensated by systems that form merging DNSs only when CBD interactions are included (right panel of Fig.~\ref{fig:DNS_evolution_Channels}). In $\simeq 44\%$ of these systems, CBD-driven orbital contraction leaves the binaries sufficiently tightly bound to survive the second SN and form DNSs in short-enough periods for future merger. Another $\simeq 45\%$ originate from binaries that would otherwise have evolved to non-merging DNSs; their reduced separations shorten the GW inspiral time sufficiently for them to coalesce within the age of the Universe. In a $\simeq 4\%$ minority, the CBD-induced orbital shrinkage leads to earlier mass transfer episodes between the stellar core and the NS in NS--BH progenitors, reducing the mass of the core and leading to the formation of an NS rather than a BH. 

\subsection{DNS Merger Rates and Timescales}
\label{subsec:DNS_merger_rate}

To estimate volumetric event rates from our population synthesis data, we normalize the fraction of simulated binaries producing each event to the local star-formation-rate density, giving
\begin{equation}
\begin{split}
R_{\rm event} =  f_{\rm event} \frac{n_{\rm SFR}}{M_{\rm SF, sim}}, 
\label{eq:R_DNS}
\end{split}
\end{equation}
where $f_{\rm event} = N_{\rm event}/N_{\rm total}$ is the number of events of interest divided by the total number of simulated binaries, $n_{\rm SFR} \simeq  10^{7} M_{\rm \odot} \rm Gpc^{-3} yr^{-1}$ is the local star formation rate normalized to the Kroupa IMF \citep{MadauDickinson2014}, and $M_{\rm SF, sim}\simeq 88 M_{\rm \odot}$ is the total stellar mass formed per simulated massive binary. Substituting these values in Equation~\ref{eq:R_DNS} results in 
\begin{equation}
\begin{split}
R_{\rm event} =  1.1 \times 10^{5} f_{\rm event} \rm Gpc^{-3} yr^{-1}. 
\label{eq:R_DNS_callibrated}
\end{split}
\end{equation}

\begin{figure*}
    \centering
    \includegraphics[width=\linewidth]{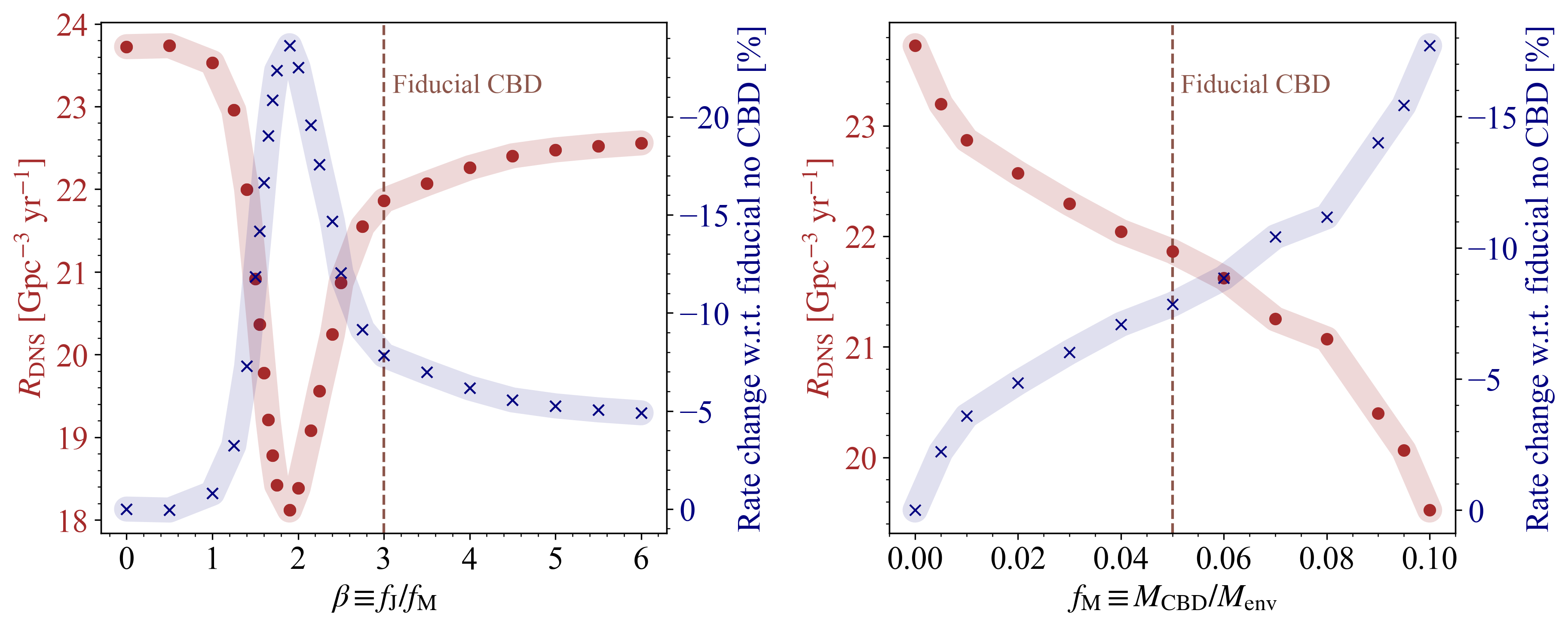}
    \caption{Rate of DNS mergers as a function of CBD model parameters: in most cases, the CBD-binary interactions reduce the overall rate of merging DNSs. Red dots (left y-axis) show the local merger rate density $R_{\rm DNS}$, while the blue X markers (right y-axis) present the percent change relative to a case with no CBD, computed as $100 \times \left(R_{\rm DNS} - R_{\rm DNS}^{\rm fiducial \: no \: CBD}\right) / R_{\rm DNS}^{\rm fiducial \: no \: CBD}$, where $R_{\rm DNS}^{\rm fiducial \: no \: CBD}$ is the DNS merger rate of the `fiducial no CBD' case. The brown dashed vertical line marks our `fiducial CBD' model ($f_{\rm M}=0.05$, $\beta=3$). Left panel: merger rates as a function of the specific angular momentum retention parameter $\beta$ assuming $f_{\rm M}=0.05$ of the envelope mass forms a CBD. Right panel: merger rates vs the fraction of envelope mass that forms the CBD $f_{\rm M}$ for a specific angular momentum parameter $\beta = 3$.}
    \label{fig:DNS_merger_rate_CBD_parameters}
\end{figure*}

We present both the volumetric DNS merger rate $R_{\rm DNS}$ (left/brown y axis) for the `fiducial CBD' model, computed by substituting the simulated DNS merger fraction into Equation~\ref{eq:R_DNS_callibrated}, and the relative reduction in the DNS mergers compared to the `fiducial no CBD' model (right/blue y axis) in Fig.~\ref{fig:DNS_merger_rate_CBD_parameters}. For our `fiducial CBD' case, where the specific angular momentum budget parameter $\beta$ equals $3$ and CBDs that contain $5\%$ of the envelope mass exchange angular momentum with the central binary, the interaction of the binary with the CBD leads to a $\simeq 8 \%$ reduction in the local DNS merger rate compared to a case where the formation of CBDs is not taken into account (from $R_{\rm DNS}=23$ to $21 \: \rm Gpc^{-3}\yr^{-1}$). The net rate change reflects the variety of competing evolutionary effects that reshape the population of merging DNSs, as illustrated in Fig.~\ref{fig:DNS_evolution_Channels}. Overall, we find that CBD interactions prevent the formation of more merging DNSs than they create, resulting in a lower merger rate.

Furthermore, we show the effect of varying different CBD parameters on the merger rate of DNSs. The left panel of Fig.~\ref{fig:DNS_merger_rate_CBD_parameters} presents its dependence on the CBD specific angular momentum budget parameter $\beta$. For most values of $\beta$, we find a decrease in the DNS merger rate relative to the no CBD case, driven by CBD-induced orbital shrinkage. This reduction is strongest for $\beta\lesssim 1.9$. The reason is that, in this regime, fewer DNS progenitors form CBDs, and those that do are generally in tighter systems, making a `premature' merger between the NS and the giant's core more likely. For larger values of $\beta$, wider binaries that would otherwise evolve into non-merging DNSs form CBDs as well (see \citealt{Ungeretal2026}), and the orbital shrinkage they induce can drive the DNS systems to merger. As a result, the number of mergers increases relative to the minimum, but still remains lower than in the no CBD case. When exploring how the disk mass affects the DNS merger rate (right panel of Fig.~\ref{fig:DNS_merger_rate_CBD_parameters}), we find that increasing it generally lowers the rate density, as more massive disks exert stronger gravitational torques, driving greater orbital shrinkage (Fig.~\ref{fig:SemiMajorAxes}) that makes the systems more prone to `premature' mergers. 

Under the \citet{TunaMetzger2023} super-Eddington prescription (model 47; Equation~\ref{eq:Mdot_acc_TM23}), the population of merging DNSs changes in a manner similar to that shown in Fig.~\ref{fig:DNS_evolution_Channels}. However, the overall DNS merger rate decreases only slightly, because the formation of new merging DNSs is approximately balanced by the loss of existing ones.

\begin{figure}
    \centering
    \includegraphics[width=\linewidth]{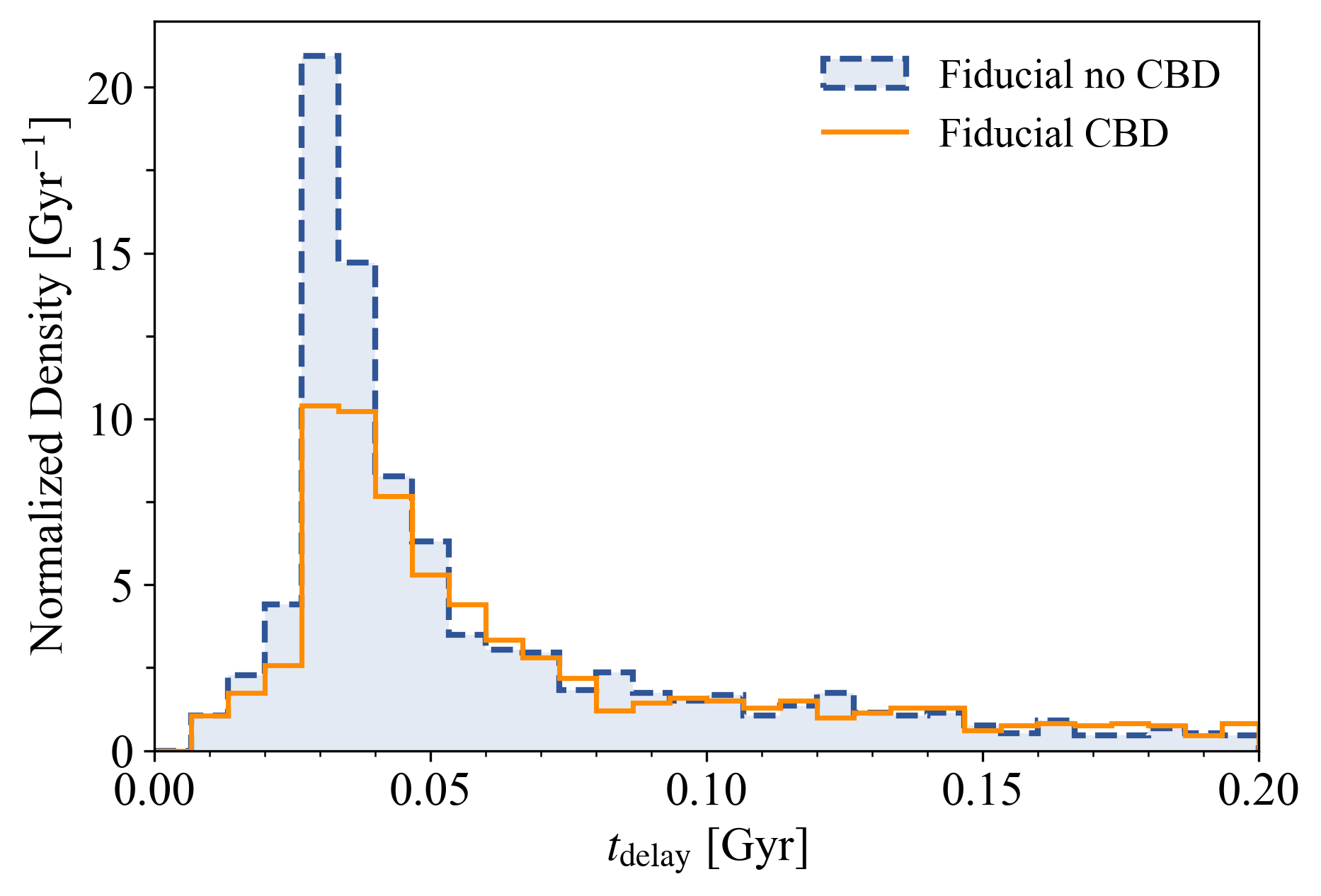}
    \caption{Delay time distribution of DNS mergers in our fiducial models capped at $0.2 \Gyr$: CBD interactions mostly suppress mergers at short delay times. The outline shows the population synthesis models that include the interaction of CBDs that form after the CEE phase with the binary system, while blue bins represent the same binary systems not including CBD interactions. The no-CBD distribution sums to one, while the CBD distribution is shown using the same normalization, as in the right panel of Fig.~\ref{fig:SemiMajorAxes}. The delay time distributions are shown only in the time region where the CBD interactions lead to noticeable differences. }
    \label{fig:Delay_Time_Distribution_with_inset}
\end{figure}

The interaction between the CBD and the inner binary affects not only the origin of the merging population and the total merger rate, but also the merger timescales of the systems that still manage to merge. Fig.~\ref{fig:Delay_Time_Distribution_with_inset} presents the delay time distribution of DNS mergers for the fiducial models. We find that CBDs significantly reduce the number of DNS mergers prior to $\simeq 0.2 \Gyr$, lowering their merger rates below $\simeq 40 \Myr$ by $40\%$. This deficit arises since the tightest post-CEE binaries, which would otherwise form DNSs with the shortest delay times, are the most likely to undergo `premature' mergers between the NS and the companion’s core. This could influence r-process nucleosynthesis at early times, as discussed in Section~\ref{subsec:R_process}.

Based on these results, we conclude that CBDs tend to decrease the numbers of merging DNSs, especially at early times. Across the CBD parameter variations shown in Fig.~\ref{fig:DNS_merger_rate_CBD_parameters}, the simulated rates we find span $17-23\:\rm Gpc^{-3} \yr^{-1}$, well within the range inferred by \cite{LIGO2026} of $5.1-154.7 \; \rm Gpc^{-3} \yr^{-1}$, so current observations do not yet allow to discriminate different CBD parameters. We note, however, that the DNS merger rate inferred from the Galactic DNS population is somewhat higher; converting the Galactic estimate of \cite{Grunthal2021} to volumetric units using a Milky Way star-formation rate of $1.65 M_{\rm \odot} \rm \yr^{-1}$ \citep{Licquia2015} and the same local star-formation-rate density adopted here gives $\simeq140-310 \: \rm Gpc^{-3} \yr^{-1}$. Therefore, our absolute merger rates in both CBD and no-CBD models are lower than those favored by Galactic DNS observations, although these Galactic rates remain in some tension with the current GW-inferred rate. 

\section{CBD-induced LFBOTs}
\label{sec:FBOTs}

As discussed in Section~\ref{sec:DNS_evolution_results}, the decrease in the DNS merger rate is primarily driven by `premature' mergers between the NS and the giant’s core, caused by CBD-induced orbital hardening. Such mergers have been proposed as a possible channel for producing AT2018cow-like LFBOTs (\citealt{Sokeretal2019}; \citealt{Metzger2022}; \citealt{Soker2022}; \citealt{TunaMetzger2023}; \citealt{CohenSoker2023}). LFBOTs are characterized by relatively short optical rise times of a few days and luminosities comparable to SN explosions (e.g., \citealt{Margutti2019}; \citealt{Coppejans2020}; \citealt{Ho2020}), with evidence for dense circumstellar material that can include hydrogen-depleted gas \citep{FoxSmith2019}. In CBD-induced NS–core mergers, rapid accretion of the disrupted core onto the NS can power luminous emission through accretion-driven outflows, and the circumstellar material can consist of hydrogen-depleted gas lost by the stripped core at the onset of the merger, as well as hydrogen-rich material from the CBD and its photoevaporative winds (\citealt{Metzger2022}; \citealt{TunaMetzger2023}). CBD-fed accretion onto the NS could also lead to the launching of jets that interact with the previously ejected envelope on a much shorter timescale, providing another route to powering LFBOTs \citep{Soker2022}.

\begin{figure*}
    \centering
    \includegraphics[width=\linewidth]{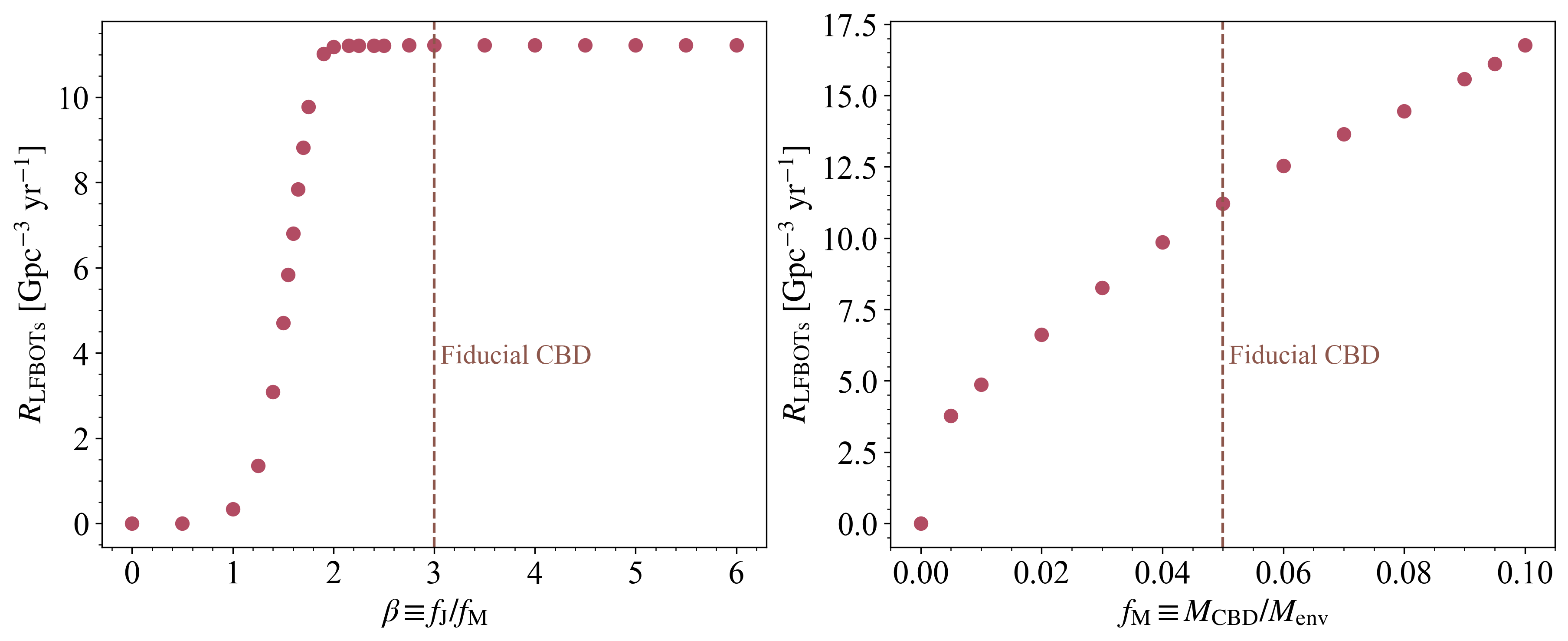}
    \caption{Local merger rate density of CBD-induced NS--core mergers that might power LFBOTs as a function of CBD model parameters: missing DNS mergers give rise to a substantial number of LFBOTs. The pink dots represent the LFBOT candidates in our population synthesis simulations, with the brown dashed vertical line presenting our `fiducial CBD' model ($f_{\rm M}=0.05$, $\beta=3$). Left panel: LFBOT rate $R_{\rm LFBOTs}$ as a function of the specific angular momentum parameter $\beta$ assuming $f_{\rm M}=0.05$ of the envelope mass forms a CBD.  Right panel: LFBOT rate vs the fraction of envelope mass that forms the CBD $f_{\rm M}$ for a specific angular momentum parameter $\beta = 3$. }
    \label{fig:LFBOTs}
\end{figure*}

In Fig.~\ref{fig:LFBOTs} we show the rate of CBD-induced NS--core mergers that can power LFBOTs as a function of our CBD parameters, computed by substituting their simulated number into Equation~\ref{eq:R_DNS_callibrated}. We find that the rate of these `premature' mergers initially rises with the specific angular momentum budget parameter $\beta$, before plateauing for $\beta \gtrsim 2$ (left panel). At this point, most of the DNS progenitors that are close enough to experience an early merger do so, and after that, the wider systems in which CBDs are formed shift mostly from non-merging to merging DNSs. In contrast, the rate rises consistently with increasing CBD mass (right panel), as more massive disks exert stronger forces on the binary, and therefore drive more efficient orbital shrinkage. Moreover, we find that in the fiducial case about $87\%$ of the `premature' mergers happen within the CBD lifetime, allowing for a denser circumstellar material around the transient. Performing a similar analysis for binary BH progenitors, we find they have a negligible contribution to the population of LFBOTs. 

\cite{Ho2023} compute the observed volumetric rate of  LFBOTs based on data from the Census of the Local Universe survey and find it to be $R_{\rm LFBOTs,obs}\simeq 70 \: \rm Gpc^{-3} \:yr^{-1}$, $\simeq15\%$ of which could be explained by CBD-induced NS--core mergers with fiducial model parameters. However, we note that the observed rate is highly uncertain, as combining estimates from Census of the Local Universe survey and the Bright Transient Survey give a conservative range of $R_{\rm LFBOTs,obs}\simeq 0.3-420 \: \rm Gpc^{-3} \:yr^{-1}$ (\citealt{Coppejans2020}; \citealt{Ho2023}). Therefore, a larger observational sample will be essential for assessing the contribution of CBD-induced NS--core mergers to the LFBOT population, which Rubin/LSST may provide by detecting tens to hundreds of LFBOTs per year \citep{Nugent2026}. Moreover, making a cleaner comparison between the simulated and observed rates would require integrating over the metallicity distribution representative of the local star-forming Universe. 

\begin{figure}
    \centering
    \includegraphics[width=\linewidth]{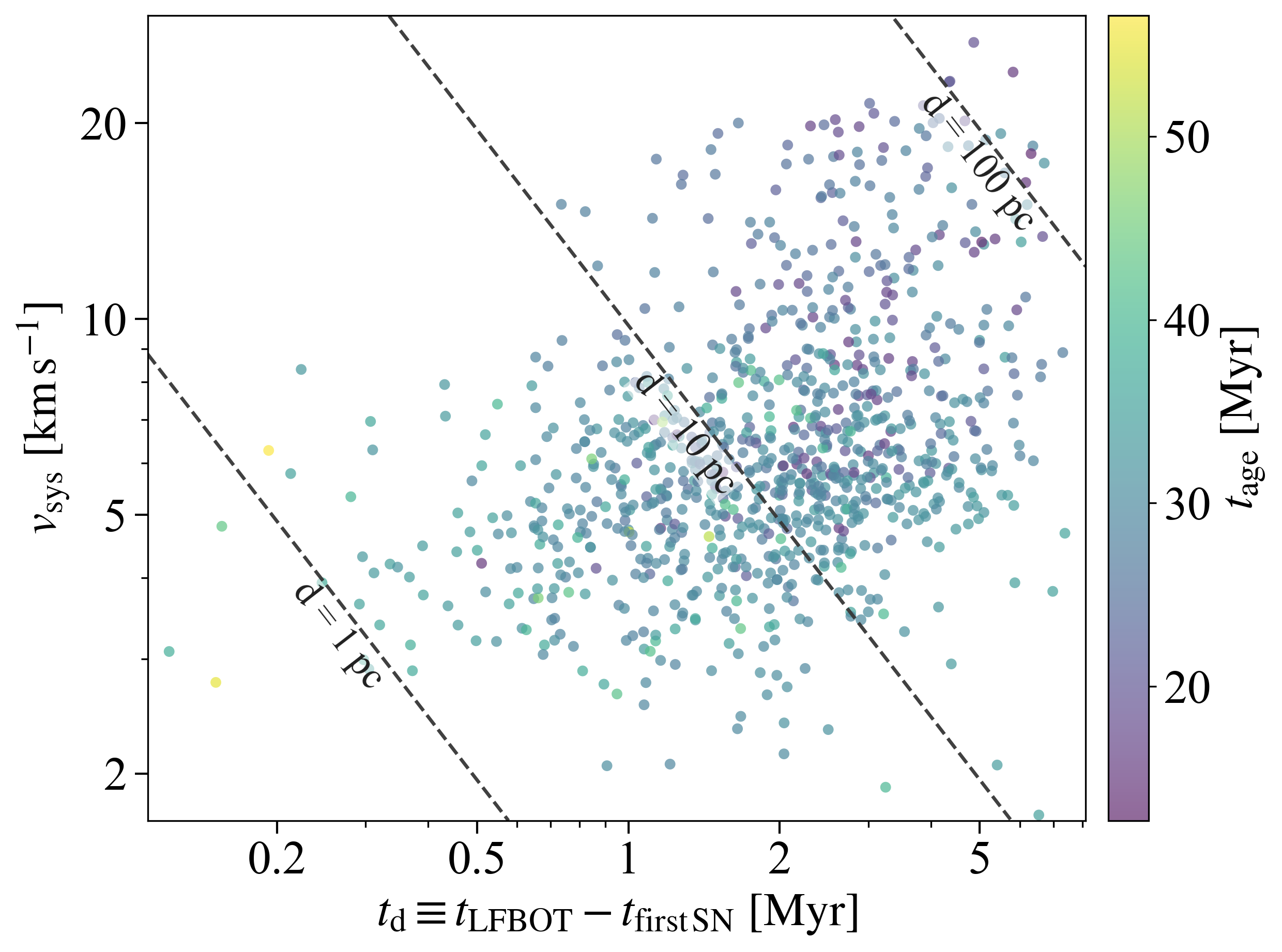}
    \caption{Systemic velocities of LFBOT progenitors as a function of the time elapsed between the first SN and the `premature' merger between the NS and the stellar core on a logarithmic scale: most binaries remain close to their local stellar population when the LFBOT occurs. The color bar represents the age of the binary at merger, and the dashed diagonal black lines mark constant displacements since the first SN due to the post-SN systemic velocity.}
    \label{fig:LFBOTs_displacement}
\end{figure}
 
In Fig.~\ref{fig:LFBOTs_displacement} we present the post-first-SN systemic velocities of the LFBOT progenitors, $v_{\rm sys}$, vs the time elapsed between the first SN and the `premature' merger of the NS with the stripped stellar core, $t_{\rm d} \equiv t_{\rm LFBOT} - t_{\rm first \ SN}$. Together, these quantities provide a first-order estimate of the typical offset of LFBOTs from their local stellar population following the first SN, due to the post-SN systemic velocity, $d\simeq v_{\rm sys}t_{\rm d}$, without taking the galactic potential into consideration. The combination of relatively small systemic velocities, $v_{\rm sys}\simeq 2-27 \km\s^{-1}$, and short post-SN delay times, $t_{\rm d}\simeq 0.01-7 \Myr$, results in typically small displacements before the LFBOT occurs, with $40\%$ having offsets smaller than $10\pc$, and $99\%$ smaller than $100\pc$ relative to the local stellar population. Therefore, we find that natal kicks are expected to make only a minor contribution to the observed locations of CBD-induced LFBOTs, and they are therefore expected to remain spatially associated with their local progenitor populations.

Furthermore, our results indicate that LFBOTs should occur near young or intermediate-age stellar populations, but not necessarily within regions of ongoing star formation, despite their small post-SN displacements. While massive stars are born in star forming regions, over the several tens of $\Myr$ preceding the LFBOT (color bar in Fig.~\ref{fig:LFBOTs_displacement}), the original H II regions are expected to have faded (e.g., \citealt{Haydon2020}) and their natal molecular cloud to have been dispersed by stellar feedback (e.g., \citealt{Chevance2020}; \citealt{Kim2022}). This is consistent with the results of \citet{Nugent2026}, who find that although LFBOTs occur in actively star-forming galaxies, their locations do not always coincide with the brightest regions of their hosts. Events located several kilo-parsecs from any young stellar population would, however, be difficult to explain through the small post-SN displacements estimated from our model, and would require substantial pre-SN motion of the progenitor binary, potentially following dynamical ejection from its birth cluster (e.g., \citealt{Oh2016}). 

We emphasize that the displacements estimated here represent the offset produced following the first SN by the post-SN systemic velocity relative to the local stellar population, and are distinct from the offsets from host galaxy centers commonly reported for LFBOTs. Therefore, any large observed galactic offsets, such as $\simeq1.7\kpc$ for AT2018cow \citep{Perley2019} and $\simeq16.5\kpc$ for AT2023fhn \citep{Chrimes2024}, primarily reflect the location of the progenitor population within its host galaxy. Moreover, pre-SN dynamical interactions, which are not included in our isolated binary evolution simulations, may also contribute to the observed displacements.

\section{Sensitivity to additional population synthesis assumptions}
\label{sec:OtherAssumptions}

\subsection{Stellar Variations with CBD Interactions}
\label{subsec:OtherAssumptions_CBD}

In Section~\ref{subsec:DNS_merger_rate}, we have shown that accounting for the presence of a CBD in DNS progenitors can decrease the expected merger rate. Here, we test this result under additional variations of uncertain population synthesis parameters. 

\begin{figure*}
    \centering
    \includegraphics[width=\linewidth]{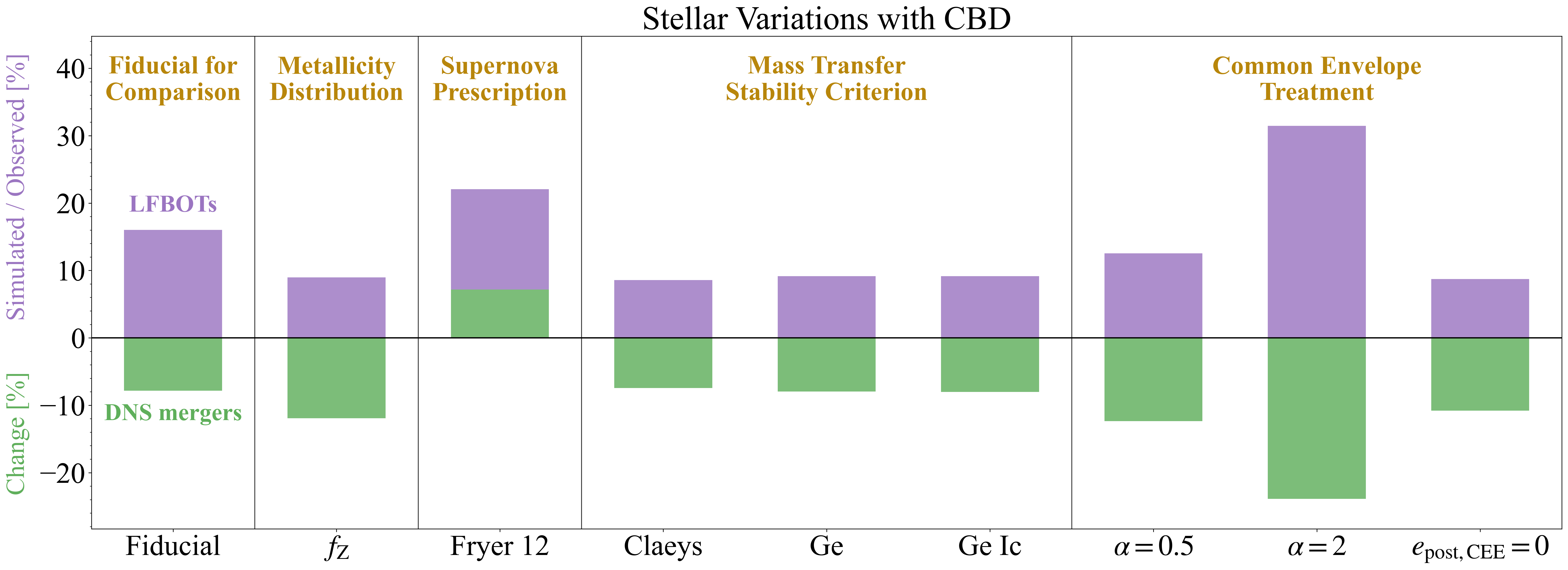}
    \caption{Effect of CBD interactions on DNS mergers and LFBOTs across variations in our population synthesis assumptions: regardless of the parameter variation considered, the inclusion of CBDs almost always results in a decrease in the DNS merger rate, which is usually dominated by CBD-induced `premature' NS--core mergers. Green bars show the percent change in the DNS merger rate comparing a model with CBDs to a model with no CBDs for different stellar evolution treatments, computed as $100 \times \left(R_{\rm DNS}^{\rm CBD} - R_{\rm DNS}^{\rm no \: CBD}\right) / R_{\rm DNS}^{\rm no \: CBD}$, where $R_{\rm DNS}^{\rm CBD}$ is the DNS merger rate of a model that includes CBD interactions, while $R_{\rm DNS}^{\rm no\:CBD}$ is the DNS merger rate of a model that is not taking CBDs into account. Purple bars show the predicted rate of CBD-induced LFBOT in each model, $R_{\rm LFBOTs}$, expressed as a percentage of the fiducial observed LFBOT rate,  $R_{\rm LFBOTs,obs}\simeq 70 \: \rm Gpc^{-3} \:yr^{-1}$ \citep{Ho2023}. Variations from left to right with model numbers we are comparing from Table~\ref{table:ModelVariation} - Fiducial (models 1, 10), $f_{\rm Z}$: metallicity following a log-uniform distribution (models 2, 48), Fryer: SN remnant recipes from \cite{Fryeretal2012} and bimodal Maxwellian distribution for natal kicks (models 3, 49), Claeys: critical mass ratio for stable mass transfer from \citealt{Claeysetal2014} (models 4, 50), Ge and Ge Ic: critical mass ratios for stable mass transfer from \citealt{Geetal2020} for the full adiabatic response and under the assumption of artificially isentropic envelopes, respectively (models 5-6, 51-52), $\alpha = 0.5$ and $\alpha = 2$: CEE efficiency parameter of 0.5 and 2, respectively (models 7-8, 53-54), $e_{\rm post,CEE}=0$: circular orbit after CEE (models 9, 55).}
    \label{fig:DNS_merger_rate_stellar_parameters_with_CBD}
\end{figure*}

We evaluate the sensitivity of our findings to the underlying assumptions by varying model parameters related to the initial metallicity distribution (model 48), SN explosion prescription (models 49), mass transfer stability criterion (models 50-52), and CEE treatment (models 53-55), relative to the `fiducial CBD' case, as described in Section~\ref{subsec:PopSynthModel}. We explore the effect of these variations on the DNS merger rate and the occurrence of LFBOTs, and find a nearly robust overall trend: CBDs lower the DNS merger rate (green bars in Fig.~\ref{fig:DNS_merger_rate_stellar_parameters_with_CBD}), often by an amount comparable to or larger than the difference between our fiducial models. This reduction is mainly driven by `premature', LFBOT-producing mergers (purple bars in Fig.~\ref{fig:DNS_merger_rate_stellar_parameters_with_CBD}).   

We note that in the \cite{Fryeretal2012} case (models 3, 49), however, CBD interactions increase the rate of DNS mergers. The reason is that in this case CBDs produce two large competing effects. While they drive tight NS+stripped core binaries to `premature' stellar merger before the second SN, as seen in the purple bars, they also create a larger amount of new DNS mergers. These mostly originate from previously non-merging and unbound DNS systems. 

On the other hand, when adopting a CEE efficiency parameter of $\alpha=2$ (model 8, 54), the decrease in the DNS merger rate due to CBD interactions is notably larger than in our `fiducial CBD model'. As in the fiducial models, in this case most of the missing DNSs are due to CBD-induced `premature' mergers that lead to LFBOTs. Since for $\alpha=2$ many more systems survive the CEE but end up as tight binaries (see Section~\ref{subsec:OtherAssumptions_no_CBD} for more details), the CBD-induced orbital shrinkage can drive the systems to an early merger more easily. As a result, the merger rate is reduced more strongly relative to the corresponding no-CBD model. 

\subsection{Stellar Variations without CBD Interactions}
\label{subsec:OtherAssumptions_no_CBD}

\begin{figure*}
    \centering
    \includegraphics[width=\linewidth]{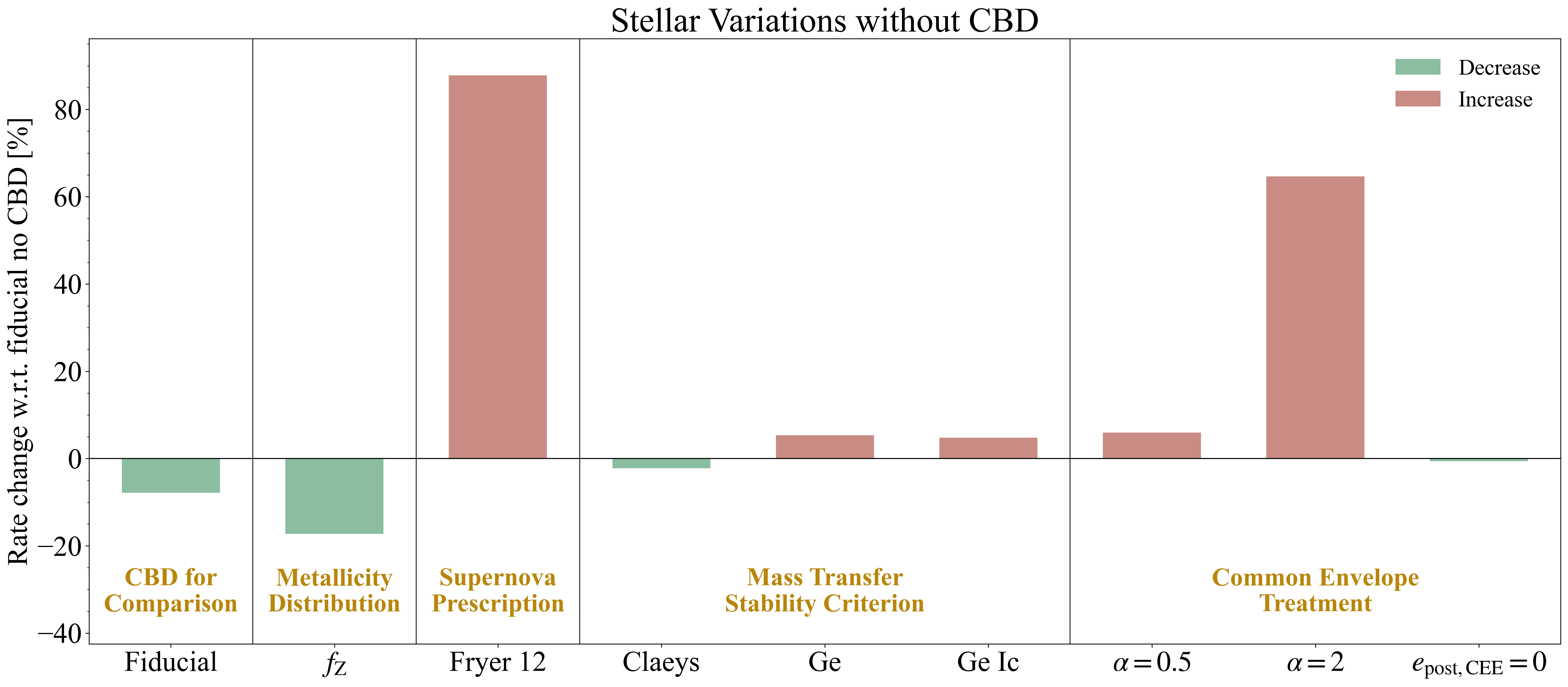}
    \caption{Percent change in the DNS merger rate relative to the fiducial model with no CBDs for different variations in our population synthesis assumptions, computed as in Fig. \ref{fig:DNS_merger_rate_CBD_parameters}.  Several variations result in comparable DNS merger rates to the `fiducial CBD' case. Brown indicates an increase compared to the `fiducial no CBD' case, while green indicates a decrease. Variations from left to right with model numbers we are comparing from Table~\ref{table:ModelVariation} - Fiducial: adding CBD interactions with our fiducial model parameters (models 1, 10), $f_{\rm Z}$: metallicity following a log-uniform distribution (models 1, 2), Fryer: SN remnant recipes from \cite{Fryeretal2012} and bimodal Maxwellian distribution for natal kicks (models 1, 3), Claeys: critical mass ratio for stable mass transfer from \citealt{Claeysetal2014} (models 1, 4), Ge and Ge Ic: critical mass ratios for stable mass transfer from \citealt{Geetal2020} for the full adiabatic response and under the assumption of artificially isentropic envelopes, respectively (models 1, 5-6), $\alpha = 0.5$ and $\alpha = 2$: CEE efficiency parameter of 0.5 and 2, respectively (models 1, 7-8), $e_{\rm post,CEE}=0$: circular orbit after CEE (models 1, 9). }
    \label{fig:DNS_merger_rate_stellar_parameters_no_CBD}
\end{figure*}

To assess the significance of our result given the current population synthesis accuracy, and to explore whether variations in the stellar-evolution prescriptions can affect the DNS merger rate in a similar way to CBDs, we perform simulations with the aforementioned variations relative to the `fiducial no CBD' case (models 2-9). We find that changes in the assumed metallicity distribution, critical mass ratio for unstable mass transfer and post-CEE eccentricity have little to moderate effect on the DNS merger rate, whereas the treatment of SN prescription and common envelope removal can significantly affect the DNS merger rate, as shown in Fig.~\ref{fig:DNS_merger_rate_stellar_parameters_no_CBD}. 

We find that assuming a log-uniform metallicity distribution lowers the DNS merger rate by $\simeq 17\%$ compared to the `fiducial no CBD case' in which all binaries are formed at solar metallicity, a result that is in line with recent studies finding a weaker metallicity dependence for DNS mergers than for BH-containing binaries (e.g., \citealt{GiacobboMapelli2018}; \citealt{Baibhav2019}; \citealt{Tang2020}; \citealt{vanSonetal2024}). Lower metallicity stars are typically more compact due to their lower opacity. The smaller radii at low metallicities tend to shift the onset of mass transfer to later evolutionary stages; as a result, stripping is less efficient, or can occur after the donor core has had more time to grow, implying larger core masses and a shorter Wolf--Rayet phase (e.g., \citealt{Renzo2019}; \citealt{Klencki2020}). This can reduce the amount of wind-driven mass loss, which is also weaker at lower metallicities (e.g., \citealt{Vink2001}; \citealt{Vink2005}; \citealt{Mokiem2007}), and drive the formation of heavier SN progenitors. These higher-mass progenitors typically receive larger natal kicks within the \cite{MandelMuller2020} formalism, leading more binaries to unbind during the first SN, and therefore to the formation of fewer DNS systems, and fewer mergers as a result. Heavier progenitors also form more BHs. Another effect of smaller stars is that their envelopes are more tightly bound, giving an increase in stellar mergers during CEE. We note, however, that the effect of metallicity on the DNS merger rate is non-monotonic, with the largest reduction in rate occurring around $0.1Z_{\rm \odot}$.

Adopting the \cite{Fryeretal2012} compact remnant mass function, together with a bimodal Maxwellian distribution for the imparted natal kick, increases the DNS merger rate by $\simeq 90\%$. This is a result of a large reshuffling of the DNS merger population. In this model, the second SN that forms the DNS receives substantially lower natal kicks, leading to fewer binary disruptions at this stage, and consequently more systems that evolve to DNS binaries and later merge. In addition, the \cite{MandelMuller2020} prescription allows for BH formation in lower mass progenitors, turning some NS--BHs into DNS mergers.

We also explore how varying the CEE efficiency parameter $\alpha$ affects the DNS merger rate, and find that relative to the fiducial model with $\alpha=1$, $\alpha=2$ leads to a substantial increase in the DNS merger rate by about $65\%$. In this case, more energy is available to unbind the envelope, making envelope ejection easier and leading to a higher CEE survival rate, and consequently more DNS formation and mergers. This effect also allows heavier progenitors to form double compact objects, instead of ending their evolution as compact object-core mergers (e.g., \citealt{FryerWoosley1998}; \citealt{SokerGilkisiPTF14hls}; \citealt{GrichenerSoker2019}; \citealt{Schroder2020}; \citealt{Grichener2023}; \citealt{Grichener2025}). 

\section{Shift in the Chirp-Mass Distribution of Merging DNSs}
\label{sec:ChirpMasses}

As in the case of binary BHs, the overall merger rate of DNSs is not sufficient to distinguish the underlying physics. In Section~\ref{subsec:OtherAssumptions_no_CBD} we found that varying other population synthesis assumptions, such as the SN prescription and the mass transfer stability criteria, could produce rate changes comparable to the inclusion of CBDs at the end of the CEE phase. Therefore, to provide an additional observational constraint, we turn now to the chirp masses of merging DNSs.

\begin{figure*}
    \centering
    \includegraphics[width=\linewidth]{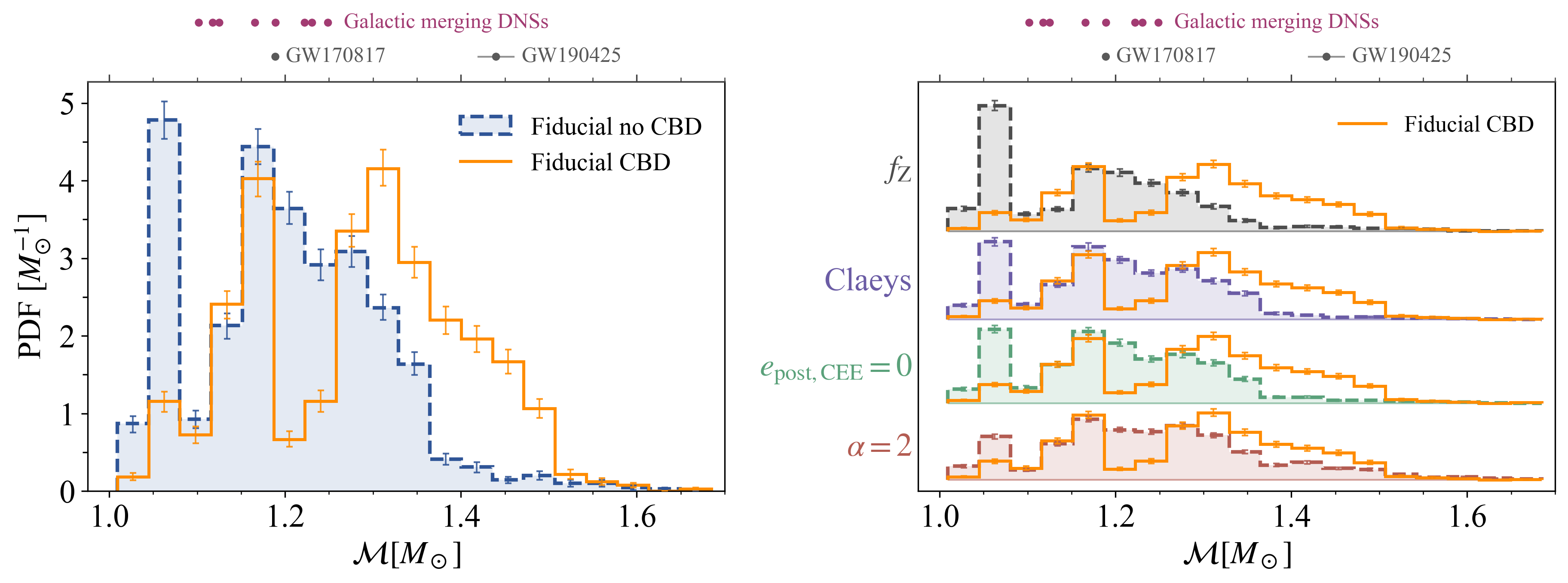}
    \caption{Chirp-mass distribution of the merging DNS population: CBD interactions shift the distribution to higher masses, and may explain GW190425. The orange outline represents the `fiducial CBD' model. The gray points above the upper $x$ axis are the observed chirp masses of the two detected DNS mergers: GW170817 and GW190425, and the magenta points mark the chirp masses of observed Galactic DNSs that are expected to merge within Hubble time, computed from their component masses. Left panel: comparison of the `fiducial no CBD' (model 1; blue bins) and the `fiducial CBD' (model 10) cases. Error bars in the left panel show the $16^{\rm th}-84^{\rm th}$ percentile range from bootstrap resampling of each simulated chirp-mass distribution at fixed sample size. Right panel: comparison between the `fiducial CBD' case and non-CBD model variations that reduce the DNS merger rates: $f_{\rm Z}$: metallicity following a log-uniform distribution (model 2; grey bins), Claeys: critical mass ratio for stable mass transfer from \citealt{Claeysetal2014} (model 4; purple bins), $e_{\rm post,CEE}=0$: circular orbit after CEE (model 9; green bins). The `$\alpha =2$' model, which adopts a CEE efficiency parameter of $\alpha =2$ (model 8; red bins), is shown since it produces a significantly larger average chirp mass than the rest.}
    \label{fig:ChirpMass}
\end{figure*}

In the left panel of Fig.~\ref{fig:ChirpMass} we show the pre-merger chirp-mass distribution of DNSs in our fiducial models. As discussed in Section~\ref{subsec:Mass_and_separation}, the post-CEE core can re-accrete part of the envelope through a CBD. As a result, it can gain mass prior to exploding in the second SN event, usually leading to a heavier NS (see discussion on potential post-CBD mass loss in Section~\ref{subsubsec:Stellat_Evolution_Caveats}). The first born NS can accrete mass as well, retaining material that is not launched through super-Eddington outflows. Therefore, when taking CBD interactions into consideration, the chirp-mass distribution shifts to larger values, with the prominent peaks moving by $\simeq 0.05-0.1 M_{\rm \odot}$, and the mean value changing from $\simeq 1.2M_{\rm \odot}$ to $\simeq 1.28M_{\rm \odot}$. We find a deficit of mergers with $\mathcal{M} \lesssim 1.08 M_{\rm \odot}$ compared to the models with no CBD, and a substantial excess of mergers with $\mathcal{M} \gtrsim 1.35 M_{\rm \odot}$. 

We next examine whether this shift toward higher chirp masses is specific to CBD interactions. For comparison, we consider the three additional non-CBD models that yield DNS merger rates comparable to that of the `fiducial CBD' model (See Fig. \ref{fig:DNS_merger_rate_stellar_parameters_no_CBD}): log-uniform metallicity distribution (model 2), critical mass ratio for stable mass transfer from \cite{Claeysetal2014} (model 4), and circular post-CEE eccentricity (model 9). The right panel of Fig.~\ref{fig:ChirpMass} shows the resulting chirp-mass distribution of these models, and that of a CEE efficiency parameter $\alpha=2$ (model 8). Although adopting the latter distribution yields a substantially higher DNS merger rate than the `fiducial CBD' model, it is also the only non-CBD variation in our grid that results in a higher mean chirp mass ($\simeq 1.24M_{\rm \odot}$), which might be harder to distinguish from the effect of CBD-driven accretion. This arises since the higher CEE efficiency allows more massive progenitors to survive CEE and form DNSs, instead of merging within the envelope. Therefore, we examine the chirp-mass of this case as well. We find substantial differences between the `fiducial CBD' case and all presented models, which exhibit lower chirp-mass peaks, and significantly fewer systems with $\mathcal{M} \gtrsim 1.35 M_{\rm \odot}$ than in the `fiducial CBD' model. 

To assess whether the structure in the predicted chirp-mass distributions is affected by the finite number of merging DNSs in our \textsc{compas} populations, we bootstrap each model by resampling its merging-DNS population with replacement at fixed sample size and recomputing the chirp-mass PDF using the same bins as in Fig.~\ref{fig:ChirpMass}. The resulting $16^{\rm th}-84^{\rm th}$ percentile intervals quantify the finite-sample uncertainty in each bin and are shown as error bars for the fiducial models in the left panel. We find that the broad features of the distributions and the shift toward higher chirp masses in models including CBD interactions are robust to these sampling uncertainties, while some of the smaller-scale variations between adjacent bins may arise from finite-sample fluctuations.

Unfortunately, only two DNS mergers have been detected so far, GW170817 with $\mathcal{M} = 1.188^{+0.004}_{-0.002}\,M_{\odot}$ \citep{Abbotetal2017}  and GW190425 with  $\mathcal{M} = 1.44^{+0.02}_{-0.02}\,M_{\odot}$ \citep{Abbottetal2020}, so it is not yet possible to construct an observational chirp-mass distribution for a meaningful comparison with our models. We note, however, that the high chirp mass of GW190425 is much more compatible with models that include CBD interactions. As in previous population-synthesis studies (e.g., \citealt{Oslowski2011}; \citealt{VignaGomezetal2018}; \citealt{Mandel2021}; \citealt{Chattaraj2026}), this mass remains challenging to explain within most of our no CBD models (model 1-7, 9), for which fewer than $1\%$ of merging DNSs have chirp masses in the range $ 1.42 M_{\rm \odot}\le \mathcal{M} \le 1.46 M_{\odot}$. However, when CBD-driven accretion is included, the fraction of DNS mergers with GW190425-like chirp masses increases to $\simeq 10\%$, making such systems considerably more common in the simulated population. By contrast, the $\alpha=2$ model (model 8), despite having the second-highest mean chirp mass, produces GW190425-like DNS mergers at a fraction lower by a factor of two relative to the `fiducial CBD model'. We note, however, that allowing super-Eddington accretion during the CEE phase or the subsequent case BB mass transfer could result in higher chirp masses as well.  

For additional observational context, we also show in Fig.~\ref{fig:ChirpMass} the eight Galactic DNS systems with measured individual masses that are expected to merge within a Hubble time (see Appendix~\ref{app:galactic_merging_dns} for more details). The chirp masses estimated from the component mass measurements are in the range $\mathcal{M} \simeq 1.10-1.25 M_{\rm \odot}$. While four Galactic DNSs lie in regions populated comparably by the fiducial CBD and no-CBD models, for the other four the no-CBD model predicts a higher local probability density. To date, no merging Galactic DNSs have been observed near the high-mass peak of the `fiducial CBD' chirp-mass distribution. We note, however, that more massive Galactic DNSs tend to have shorter GW inspiral times \citep{Grichener2026}, and are therefore less likely to be represented in the present-day Galactic DNS population. 

The substantial mass accretion onto NSs in our `fiducial CBD' model (Section~\ref{subsec:Mass_and_separation}), which might explain the high mass of GW190425, is difficult to reconcile with the measured component masses of Galactic DNSs, suggesting that CBDs may form in a smaller fraction of systems and/or that less mass is accreted from them than found in our `fiducial CBD' model. However, the `fiducial no CBD' model also fails to reproduce the observed component masses, indicating that neither model fully captures their distribution. Interestingly, examining the individual masses reveals that in most systems, the recycled pulsar, generally identified as the first-born NS, is the more massive component. This could potentially be explained by additional mass gained through accretion from a CBD.

\section{Discussion}
\label{sec:Discussion}

\subsection{R-process Enrichment}
\label{subsec:R_process}

It remains unclear whether DNS mergers alone can account for all r-process nucleosynthesis (e.g., \citealt{Ji2019}; \citealt{HolmbeckAndrews2024}; \citealt{Soker2026}). Recent studies have suggested that due to their long GW inspiral times, DNS mergers might not occur early enough to explain the production of heavy elements in the early Galaxy (e.g., \citealt{Coteetal2019}; \citealt{Wehmeyer2019}; \citealt{Kobayashietal2020}; \citealt{Kobayashietal2023}; \citealt{ChenHsinYu2025}; \citealt{Saleem2025}). The strong decrease in the DNS merger rates at timescales of $\lesssim 40 \Myr$ due to CBD-induced `premature' mergers (Fig.~\ref{fig:Delay_Time_Distribution_with_inset}) makes this tension even more severe. Such a substantial suppression of short-delay mergers strengthens the need for an additional, more prompt r-process source, and may extend the time over which such a source is needed.

On this timescale, which is comparable to the lifetime of massive stars, r-process nucleosynthesis could be dominated by the merger of an NS with the core of a giant star during CEE (e.g., \citealt{Papishetal2015}; \citealt{GrichenerSoker2019}; \citealt{Gricheneretal2022}; \citealt{Grichener2023}; \citealt{JinSoker2024}; \citealt{Grichener2025}; \citealt{Soker2025}; \citealt{Soker2026}; \citealt{ThielemannCowan2026}). In this case, the r-process nucleosynthesis occurs in the self-neutronizing accretion disk that is formed around the NS from the destroyed core material. These mergers are significantly faster than DNS mergers since they are dominated by the dynamical friction of the envelope rather than caused by GW emission. As CBD interactions drive a large fraction of DNS progenitors to undergo a `premature' merger between the NS and the stellar core, thereby introducing an additional NS–core merger channel, their relative contribution to r-process nucleosynthesis may become even more significant, as will be explored in a future study. 



\subsection{Limitations of the Physical Models}
\label{subsec:Caveats}


\subsubsection{Disk Properties Uncertainties}
\label{subsubsec:Disk_Properties_Caveats}

\textit{Mass:}
The amount of envelope mass that remains bound after CEE and can subsequently be redistributed into a CBD varies significantly across hydrodynamical simulations, with some studies suggesting that the majority of the envelope is ejected (e.g., \citealt{Lau2022}), and others indicating most of it may remain bound (e.g., \citealt{Ohlmann2016}). Since the effects of the disk self-gravity are not included in the hydrodynamical simulations on which our model is based, when varying the CBD mass in our population-synthesis study, we take $10\%$ of the envelope mass as an upper limit. For this value, the Toomre parameter (e.g., \citealt{Toomre1964}; \citealt{KratterLodato2016}) satisfies $Q\gtrsim 1$ when assuming a constant aspect ratio of $h=0.1$, and the disks are expected to be marginally stable against local gravitational instability. Higher CBD masses would reduce $Q$, increasing the importance of self-gravity, and potentially leading to clumping or fragmentation if the disk cools sufficiently efficiently (e.g., \citealt{Gammie2001}; \citealt{Kratter2008}; \citealt{Rice2005}). However, due to uncertain disk properties, including its temperature, surface-density profile, and radial extent, effects of self-gravity cannot be ruled out for all systems within our explored parameter space.

\textit{Lifetime:}
The CBD simulations performed by \cite{Siweketal2023,Siwek2023b} provide the mass accreted towards the central binary, but are agnostic to the accretion rate and the disk lifetime, as the effect of photoevaporation and disk wind are not included. We take a characteristic lifetime of $\tau_{\rm CBD} = 10^{5} \yr$, comparable to the thermal timescale of a stripped helium core and, in some cases, even to the nuclear timescale of a more evolved core. For a fixed CBD mass, a shorter disk lifetime implies a higher initial accretion rate. In most of the models explored here, which adopt the super-Eddington correction of \cite{GhodlaEldridge2023}, the accretion timescale does not affect the total mass accreted once the flow becomes super-Eddington (Equation~\ref{eq:Mdot_acc_GE23}). By contrast, under the prescription of \cite{TunaMetzger2023}, higher accretion rates drive stronger super-Eddington winds and may therefore result in less mass being accreted onto the NS (Equation~\ref{eq:Mdot_acc_TM23}). The resulting NS mass growth can therefore depend appreciably on the assumed disk lifetime. However, because the total NS mass gain is already small under this prescription, this additional variation has little impact on the population-level outcomes considered here. Reducing the disk lifetime by an order of magnitude, from $\tau_{\rm CBD}=10^{5} \yr$ to $\tau_{\rm CBD}=10^{4} \yr$, changes the merger rates explored here by less than $3\%$. 
 
\textit{Thickness and Viscosity:}
There are additional disk properties that might affect the interaction of a CBD with a binary system. The aspect ratio (set to $h=0.1$ for the CBDs explored in \citealt{Siweketal2023,Siwek2023b}) and viscosity (treated through the disk prescription of \citealt{ShakuraSunyaev1973} with $\alpha_{\rm visc}=0.1$) of the disk play a crucial role in shaping its interaction with the inner binary and its subsequent orbital evolution, as they control the structure of the inner cavity and the accretion streams. Thicker disks have greater pressure support that smooths the cavity edge, weakening the tendency toward orbital contraction and, in many cases, instead driving expansion of the binary. Thinner disks preserve a sharper, more strongly non-axisymmetric cavity morphology, which enhances the gravitational pull from disk gas near the binary and can drive stronger orbital contraction (e.g., \citealt{Tiede2020}; \citealt{DittmannRyan2022}; \citealt{Dittmannetal2025}). The viscosity then modulates how efficiently gas is transported across the cavity edge: higher viscosity refills and mixes the inner region more effectively, lowering density contrasts and weakening the gas-driven force, whereas lower viscosity maintains a deeper, higher-contrast cavity with more coherent, lagging gas streams, strengthening the gravitational torque that removes angular momentum from the binary and driving stronger orbital decay. Moreover, the adopted viscosity prescription might influence the effect of CBDs on the eccentricity evolution of the binary in our estimated post-CEE eccentricity regime (see Section~\ref{subsubsec:Stellat_Evolution_Caveats}). \cite{DOrazioDuffell2021} find that equal-mass binaries with initial eccentricities $e\lesssim 0.1$ are circularized by the CBD, while the simulations of \cite{Siweketal2023} underlying our framework instead predict eccentricity growth for equal-mass binaries with $e\gtrsim 0.01$. This disagreement illustrates that the eccentricity evolution, particularly for low eccentricity binaries, remains sensitive to the adopted disk physics. The thickness and viscosity of post-CEE disks are not observationally constrained and might vary throughout the disk lifetime, with the disk being initially thicker just after the CEE and becoming thinner as it cools. While the values in the simulations of \cite{Siweketal2023,Siwek2023b} are in the range of commonly used values for studying CBDs formed after CEE (e.g., \citealt{TunaMetzger2023}; \citealt{Weietal2024}), a parameter-space exploration of these values would be valuable for testing the robustness of our results, particularly as preparation for future observations that may help constrain their relevant range. 

\subsubsection{Stellar Evolution Uncertainties}
\label{subsubsec:Stellat_Evolution_Caveats}

\textit{Effect of Metallicity:}
In our fiducial model we assume that binaries form at solar metallicity, and relax this assumption by adopting a log-uniform distribution in metallicity between $10^{-4}$ and $0.03$. In reality, DNS progenitors span a range of metallicities that depends on the star formation history and redshift of their host galaxy. Since the metallicity affects the stellar radii and wind mass loss rates (as discussed in Section~\ref{sec:OtherAssumptions}), it would influence the orbital evolution. Most cosmic star formation occurs at relatively high metallicities (e.g., \citealt{vanSon2023}).  Lower metallicities are found at higher redshifts, but unless there are large delays between formation and merger, mergers in these environments are generally harder to detect (e.g., \citealt{Abbotetal2017}). Therefore, we adopt solar metallicity as a representative choice for the distributions most likely to be observed in the near future. A more complete treatment, involving simulations across a metallicity grid and convolution with a cosmic star formation and metallicity evolution history, including selection effects, is beyond the scope of this work.

\textit{Post-CEE Circularization:}
Post-CEE systems typically exhibit small eccentricities (e.g., \citealt{Beuermann2012}; \citealt{Kruckow2021}), and population synthesis studies therefore commonly assume circular orbits after CEE. However, complete circularization during CEE is not guaranteed, and hydrodynamical simulations can leave binaries with non-zero residual eccentricities (e.g., \citealt{Passy2012}; \citealt{GlanzPerets2021}). Since CBD interactions and especially eccentricity pumping can be highly sensitive to the initial eccentricity of the binary, this uncertainty is important for our models. Moreover, the CBD-driven evolution of binaries at low but non-zero eccentricities is itself uncertain. Within \textsc{spindler}, binaries starting from exactly circular orbits do not develop eccentricity. We therefore adopt a simple phenomenological prescription in which after CEE the binaries retain $10\%$ of their pre-RLOF eccentricity in most models. We note that pre-RLOF tidal evolution was not included in our calculations (see e.g., \citealt{Kapil2026a} for a new prescription), so this assumption might overestimate the post-CEE eccentricity. Previous work, however, suggests that a significant fraction of DNS progenitors may retain non-negligible eccentricities at the onset of CEE, even when accounting for tidal circularization (e.g., \citealt{VignaGomezetal2020}). Nevertheless, within our adopted prescription, the binaries develop non-negligible eccentricities in the presence of a CBD, while the DNS merger rate remains similar to that obtained in simulations that assume CBD interactions and $e_{\rm post,CEE}=0$. For even more eccentric post-CEE orbits, the \citealt{Siweketal2023} prescription tends to enhance the eccentricity further, driving it towards the $e\sim0.5$ attractor they found (see also Fig. 1 in \citealt{Vallietal2024}), and thereby strengthening its impact on the merger rate and associated observables.

\textit{Effect of CBD Torques on Mass Transfer:} While our model includes CBD-driven orbital evolution, we rely on the standard \textsc{compas} prescriptions to determine mass-transfer stability, without explicitly accounting for the influence of ongoing disk torques. Continued angular momentum extraction from the binary by the disk could enhance the mass-transfer rate and destabilize mass transfer that would otherwise remain stable (e.g., \citealt{Metzger2022}; \citealt{TunaMetzger2023}). This may be particularly relevant for case BB mass transfer, which is assumed to be stable in our simulations, potentially increasing the rate of NS--core mergers and associated LFBOTs, while further reducing the formation of merging DNSs. Quantifying the impact of disk torques would require a more self-consistent treatment of mass transfer in the presence of a CBD.

\textit{Post-CBD Mass Transfer:} In our models, stripped stars re-accrete hydrogen-rich envelope material through a CBD. This may cause them to re-expand, as in the case of stars retaining residual envelopes following CEE (e.g., \citealt{Laplace2020}; \citealt{VignaGomez2022}). Additional expansion may be driven by gravitational energy release and thermal readjustment during accretion (\citealt{Lauetal2024}; \citealt{Cohen2026}). Consequently, these stars might fill their Roche lobes, limiting further accretion and potentially redirecting material towards the NS companion, therefore altering the component masses and orbital evolution. These effects are not included in our models, and are left for future work.

\section{Summary}
\label{sec:Summary}

In this study, we developed a framework to model CBD formation in post-CEE systems and its impact on the evolution of the inner binaries, and implemented this framework in the population synthesis code \textsc{compas} (Section~\ref{sec:NumericalScheme}). We used it to explore the effect of CBD interactions on the population of merging DNSs and LFBOTs (Sections~\ref{sec:DNS_evolution_results}-\ref{sec:ChirpMasses}), and discussed the implications for r-process enrichment and the main limitations and caveats of our approach (Section~\ref{sec:Discussion}). 

To determine which systems form a CBD from leftover envelope gas surrounding the first-born NS and the exposed core, we followed \cite{Ungeretal2026} and compared the specific angular momentum available for CBD formation (equation~\ref{eq:CBD_AM}) to the Keplerian specific angular momentum required to support the disk (equation~\ref{eq:J_kep}). If the former is larger than the latter, we assumed the bound material has enough angular momentum to form a centrifugally supported CBD around the binary (equation~\ref{eq:CBDcondition}). In this case, we used \textsc{spindler} \citep{Vallietal2024}, which interpolates the hydrodynamical CBD simulations of   \cite{Siweketal2023,Siwek2023b}, to model the CBD-driven evolution of the semimajor axis, eccentricity, and component masses of the binary shaped by gravitational and accretion forces. For compact objects, we reduced the amount of mass that is ultimately accreted and accounted for the angular momentum carried away by super-Eddington winds (equations~\ref{eq:Mdot_acc_GE23}-\ref{eq:Mdot_acc_TM23}). 

We used \textsc{compas} to generate populations of massive binaries and to quantify how CBD interactions affect DNS mergers. We ran a suite of models, varying metallicities, SN and mass transfer treatments, and CBD parameters (Table~\ref{table:ModelVariation}). Our main findings are as follows. 
\begin{enumerate}[label=(\roman*), leftmargin=0pt, labelindent=0pt, itemindent=!, labelsep=0.5em]   
    \item \textbf{\textit{CBDs change the merging DNS population and lead to a decrease in the merger rate}}: The interactions between the CBD and the central binary typically shrink the orbit (Fig.~\ref{fig:SemiMajorAxes}), changing the binary population that results in DNS mergers (Fig.~\ref{fig:DNS_evolution_Channels}). This occurs mostly through `premature' mergers between the NS and the giant’s core in systems that would otherwise evolve into merging DNSs. The reduction in the number of DNS mergers is partly compensated by systems that would be disrupted during the second SN or form DNSs too wide to merge within the age of the Universe, but instead become merging DNSs due to CBD-driven orbital contraction. On balance, however, we found that CBD interactions reduce the predicted merger rate relative to analogous models without CBDs in most cases (Figs.~\ref{fig:DNS_merger_rate_CBD_parameters}, \ref{fig:DNS_merger_rate_stellar_parameters_with_CBD}). In our fiducial model, the DNS merger rate decreases by $8\%$, while in the most strongly affected case it can be reduced by as much as $28\%$.
    \item \textbf{\textit{The missing DNSs can power LFBOTs}}: The `premature' NS--core mergers, driven by orbital shrinkage during the binary-CBD interaction, generally occur before the disk material is depleted and can therefore give rise to LFBOTs (Section~\ref{sec:FBOTs}). This links DNS mergers and LFBOTs through a common progenitor pathway. We found that CBD-induced LFBOTs occur at a fiducial rate of $\simeq11 \:\rm Gpc^{-3} \yr^{-1}$, and are robust across our population study (Figs.~\ref{fig:LFBOTs}, \ref{fig:DNS_merger_rate_stellar_parameters_with_CBD}). They are expected to remain spatially associated with their local progenitor environment, with natal kicks causing only a minor displacement relative to the local stellar population (Fig.~\ref{fig:LFBOTs_displacement}). This implies the transients typically arise near relatively young stellar populations, without necessarily coinciding with sites of ongoing star formation. 
    \item \textbf{\textit{CBDs preferentially suppress DNS mergers at short delay times}}:  Since most DNS progenitors that undergo CBD-induced NS–-core mergers originate from initially tight binaries, CBD interactions mostly suppress DNS mergers at short delay times, lowering their rate by $40\%$ for delay times below $\simeq40 \Myr$ (Fig.~\ref{fig:Delay_Time_Distribution_with_inset}). This may further challenge the ability of DNS mergers to account for early r-process enrichment (Section~\ref{subsec:R_process}). However, these `premature' mergers provide a new avenue for forming NS--core mergers, which might contribute to r-process nucleosynthesis on timescales comparable to those of massive-star evolution.
    \item \textbf{\textit{CBD interactions produce more massive merging DNSs}}: CBDs influence DNS progenitors by mass accretion, which shifts the chirp-mass distribution of merging DNSs to higher values (Fig.~\ref{fig:ChirpMass}), with an average fiducial growth of $\simeq0.08M_{\rm \odot}$. This may help explain the high mass of GW190425, which is not accounted for by standard population models to date (Section~\ref{sec:ChirpMasses}). 
\end{enumerate}

While the parameter space exploration showed that our results are robust to most stellar evolution variations, the uncertainties in CBD parameters could still significantly affect our predictions. The thickness and viscosity of the disk, for instance, could alter the amount of mass that is accreted by the central binary, and lead to a stronger inspiral. A natural follow-up, beyond the scope of this paper, would be to relax our assumptions about disk thickness, either by interpolating over a broader suite of hydrodynamical simulations that spans a wider range of aspect ratios (e.g., \citealt{Tiede2020}; \citealt{DittmannRyan2022}), or by adopting an analytic CBD framework in which these parameters can be varied directly (e.g., \citealt{IzzardJermyn2023}; \citealt{TunaMetzger2023}; \citealt{Weietal2024}). Combined observations of DNS merger properties, LFBOTs, and r-process enrichment could provide an indication of the presence of CBDs in post-CEE massive binaries. 

In this work we focused on the impact of CBDs formed after the CEE phase on the population of merging DNSs, since the majority of DNS progenitors are expected to undergo at least one common envelope phase (e.g., \citealt{Taurisetal2017}; \citealt{VignaGomezetal2018}; \citealt{Grichener2023}; \citealt{Nathaniel2025}). More broadly, the same framework can be applied to any type of stellar binary in which a CBD forms following CEE, with potential implications for post-asymptotic giant branch systems, luminous red novae, interacting binaries that contain WDs, and binary BHs.

\section*{ACKNOWLEDGMENTS}
\label{sec:acknowledgments}

We thank Mike Lau, Michaela Mapelli, Smadar Naoz, Fabian Schneider, and Fritz R\"{o}pke for helpful discussions. A.G. acknowledges support from the Steward Observatory Fellowship in Theoretical and Computational Astrophysics, the IAU-Gruber Fellowship, and the CHE Fellowship. R.V. was supported by the NOMIS-Fellowship Program. M.R. acknowledges support from NASA (ATP: 95380NSSC24K0932). I.M. acknowledges support from the Australian Research Council (ARC) Centre of Excellence for Gravitational Wave Discovery (OzGrav), through project number CE230100016. 

\section*{Data availability}
\label{sec:DataAvailability}

The population synthesis data generated for this project is publicly available on Zenodo at \href{https://doi.org/10.5281/zenodo.23851460}{10.5281/zenodo.23851460}.

\section*{Software}
\label{sec:Software}

Simulations in this paper used a modified version of the open-source rapid binary population synthesis code \href{https://github.com/TeamCOMPAS/COMPAS}{\textsc{compas}}, incorporating the CBD framework developed in this work. It is publicly available on \href{https://github.com/AldanaGrichener/COMPAS/tree/CBD-paper}{GitHub}. The CBD framework is integrated in \textsc{compas}, starting from version 03.31.00.
This work also made use of \textsc{spindler} \citep{Vallietal2024}.

\bibliography{refs} 

@ARTICLE{Asplund2009,
       author = {{Asplund}, Martin and {Grevesse}, Nicolas and {Sauval}, A. Jacques and {Scott}, Pat},
        title = "{The Chemical Composition of the Sun}",
      journal = {\araa},
         year = 2009,
        month = sep,
       volume = {47},
       number = {1},
        pages = {481-522},
          doi = {10.1146/annurev.astro.46.060407.145222},
archivePrefix = {arXiv},
       eprint = {0909.0948},
 primaryClass = {astro-ph.SR},
       adsurl = {https://ui.adsabs.harvard.edu/abs/2009ARA&A..47..481A}
}

@ARTICLE{ChenPodsiadlowski2017,
       author = {{Chen}, Wen-Cong and {Podsiadlowski}, Philipp},
        title = "{Rapid Orbital Decay in Detached Binaries: Evidence for Circumbinary Disks}",
      journal = {\apjl},
         year = 2017,
        month = mar,
       volume = {837},
       number = {2},
          eid = {L19},
        pages = {L19},
          doi = {10.3847/2041-8213/aa624a},
archivePrefix = {arXiv},
       eprint = {1702.06311},
 primaryClass = {astro-ph.SR},
       adsurl = {https://ui.adsabs.harvard.edu/abs/2017ApJ...837L..19C}
}

@ARTICLE{DOrazioDuffell2021,
       author = {{D'Orazio}, Daniel J. and {Duffell}, Paul C.},
        title = "{Orbital Evolution of Equal-mass Eccentric Binaries due to a Gas Disk: Eccentric Inspirals and Circular Outspirals}",
      journal = {\apjl},
         year = 2021,
        month = jun,
       volume = {914},
       number = {1},
          eid = {L21},
        pages = {L21},
          doi = {10.3847/2041-8213/ac0621},
archivePrefix = {arXiv},
       eprint = {2103.09251},
 primaryClass = {astro-ph.HE},
       adsurl = {https://ui.adsabs.harvard.edu/abs/2021ApJ...914L..21D}
}

@ARTICLE{GagnierPejcha2023,
       author = {{Gagnier}, Damien and {Pejcha}, Ond{\v{r}}ej},
        title = "{Post-dynamical inspiral phase of common envelope evolution. Binary orbit evolution and angular momentum transport}",
      journal = {\aap},
         year = 2023,
        month = jun,
       volume = {674},
          eid = {A121},
        pages = {A121},
          doi = {10.1051/0004-6361/202346057},
archivePrefix = {arXiv},
       eprint = {2302.00691},
 primaryClass = {astro-ph.SR},
       adsurl = {https://ui.adsabs.harvard.edu/abs/2023A&A...674A.121G}
}

@ARTICLE{Grichener2023,
       author = {{Grichener}, Aldana},
        title = "{Mergers of neutron stars and black holes with cores of giant stars: a population synthesis study}",
      journal = {\mnras},
         year = 2023,
        month = jul,
       volume = {523},
       number = {1},
        pages = {221-232},
          doi = {10.1093/mnras/stad1449},
archivePrefix = {arXiv},
       eprint = {2302.06663},
 primaryClass = {astro-ph.HE},
       adsurl = {https://ui.adsabs.harvard.edu/abs/2023MNRAS.523..221G}
}

@ARTICLE{Hurley2000,
       author = {{Hurley}, Jarrod R. and {Pols}, Onno R. and {Tout}, Christopher A.},
        title = "{Comprehensive analytic formulae for stellar evolution as a function of mass and metallicity}",
      journal = {\mnras},
         year = 2000,
        month = jul,
       volume = {315},
       number = {3},
        pages = {543-569},
          doi = {10.1046/j.1365-8711.2000.03426.x},
archivePrefix = {arXiv},
       eprint = {astro-ph/0001295},
 primaryClass = {astro-ph},
       adsurl = {https://ui.adsabs.harvard.edu/abs/2000MNRAS.315..543H}
}

@ARTICLE{Hurleyetal2002,
       author = {{Hurley}, Jarrod R. and {Tout}, Christopher A. and {Pols}, Onno R.},
        title = "{Evolution of binary stars and the effect of tides on binary populations}",
      journal = {\mnras},
         year = 2002,
        month = feb,
       volume = {329},
       number = {4},
        pages = {897-928},
          doi = {10.1046/j.1365-8711.2002.05038.x},
archivePrefix = {arXiv},
       eprint = {astro-ph/0201220},
 primaryClass = {astro-ph},
       adsurl = {https://ui.adsabs.harvard.edu/abs/2002MNRAS.329..897H}
}

@ARTICLE{KashiSoker2011,
       author = {{Kashi}, Amit and {Soker}, Noam},
        title = "{A circumbinary disc in the final stages of common envelope and the core-degenerate scenario for Type Ia supernovae}",
      journal = {\mnras},
         year = 2011,
        month = oct,
       volume = {417},
       number = {2},
        pages = {1466-1479},
          doi = {10.1111/j.1365-2966.2011.19361.x},
archivePrefix = {arXiv},
       eprint = {1105.5698},
 primaryClass = {astro-ph.SR},
       adsurl = {https://ui.adsabs.harvard.edu/abs/2011MNRAS.417.1466K}
}

@ARTICLE{Kroupa2001,
       author = {{Kroupa}, Pavel},
        title = "{On the variation of the initial mass function}",
      journal = {\mnras},
         year = 2001,
        month = apr,
       volume = {322},
       number = {2},
        pages = {231-246},
          doi = {10.1046/j.1365-8711.2001.04022.x},
archivePrefix = {arXiv},
       eprint = {astro-ph/0009005},
 primaryClass = {astro-ph},
       adsurl = {https://ui.adsabs.harvard.edu/abs/2001MNRAS.322..231K}
}

@ARTICLE{LivioSoker1988,
       author = {{Livio}, Mario and {Soker}, Noam},
        title = "{The Common Envelope Phase in the Evolution of Binary Stars}",
      journal = {\apj},
         year = 1988,
        month = jun,
       volume = {329},
        pages = {764},
          doi = {10.1086/166419},
       adsurl = {https://ui.adsabs.harvard.edu/abs/1988ApJ...329..764L}
}

@ARTICLE{Pols1998,
       author = {{Pols}, Onno R. and {Schr{\"o}der}, Klaus-Peter and {Hurley}, Jarrod R. and {Tout}, Christopher A. and {Eggleton}, Peter P.},
        title = "{Stellar evolution models for Z = 0.0001 to 0.03}",
      journal = {\mnras},
         year = 1998,
        month = aug,
       volume = {298},
       number = {2},
        pages = {525-536},
          doi = {10.1046/j.1365-8711.1998.01658.x},
       adsurl = {https://ui.adsabs.harvard.edu/abs/1998MNRAS.298..525P}
}

@ARTICLE{TeamCOMPAS2022,
       author = {{Team COMPAS: J. Riley} and {Agrawal}, Poojan and {Barrett}, Jim W. and {Boyett}, Kristan N.~K. and {Broekgaarden}, Floor S. and {Chattopadhyay}, Debatri and {Gaebel}, Sebastian M. and {Gittins}, Fabian and {Hirai}, Ryosuke and {Howitt}, George and {Justham}, Stephen and {Khandelwal}, Lokesh and {Kummer}, Floris and {Lau}, Mike Y.~M. and {Mandel}, Ilya and {de Mink}, Selma E. and {Neijssel}, Coenraad and {Riley}, Tim and {van Son}, Lieke and {Stevenson}, Simon and {Vigna-G{\'o}mez}, Alejandro and {Vinciguerra}, Serena and {Wagg}, Tom and {Willcox}, Reinhold},
        title = "{Rapid Stellar and Binary Population Synthesis with COMPAS}",
      journal = {\apjs},
         year = 2022,
        month = feb,
       volume = {258},
       number = {2},
          eid = {34},
        pages = {34},
          doi = {10.3847/1538-4365/ac416c},
archivePrefix = {arXiv},
       eprint = {2109.10352},
 primaryClass = {astro-ph.IM},
       adsurl = {https://ui.adsabs.harvard.edu/abs/2022ApJS..258...34R}
}

@ARTICLE{Sana2012,
       author = {{Sana}, H. and {de Mink}, S.~E. and {de Koter}, A. and {Langer}, N. and {Evans}, C.~J. and {Gieles}, M. and {Gosset}, E. and {Izzard}, R.~G. and {Le Bouquin}, J. -B. and {Schneider}, F.~R.~N.},
        title = "{Binary Interaction Dominates the Evolution of Massive Stars}",
      journal = {Science},
         year = 2012,
        month = jul,
       volume = {337},
       number = {6093},
        pages = {444},
          doi = {10.1126/science.1223344},
archivePrefix = {arXiv},
       eprint = {1207.6397},
 primaryClass = {astro-ph.SR},
       adsurl = {https://ui.adsabs.harvard.edu/abs/2012Sci...337..444S}
}

@ARTICLE{Siweketal2023,
       author = {{Siwek}, Magdalena and {Weinberger}, Rainer and {Hernquist}, Lars},
        title = "{Orbital evolution of binaries in circumbinary discs}",
      journal = {\mnras},
         year = 2023,
        month = jun,
       volume = {522},
       number = {2},
        pages = {2707-2717},
          doi = {10.1093/mnras/stad1131},
archivePrefix = {arXiv},
       eprint = {2302.01785},
 primaryClass = {astro-ph.HE},
       adsurl = {https://ui.adsabs.harvard.edu/abs/2023MNRAS.522.2707S}
}

@ARTICLE{Soker2022,
       author = {{Soker}, Noam},
        title = "{A Common Envelope Jets Supernova (CEJSN) Impostor Scenario for Fast Blue Optical Transients}",
      journal = {Research in Astronomy and Astrophysics},
         year = 2022,
        month = may,
       volume = {22},
       number = {5},
          eid = {055010},
        pages = {055010},
          doi = {10.1088/1674-4527/ac5b40},
archivePrefix = {arXiv},
       eprint = {2201.07728},
 primaryClass = {astro-ph.HE},
       adsurl = {https://ui.adsabs.harvard.edu/abs/2022RAA....22e5010S}
}

@ARTICLE{Stevensonetal2017,
       author = {{Stevenson}, Simon and {Vigna-G{\'o}mez}, Alejandro and {Mandel}, Ilya and {Barrett}, Jim W. and {Neijssel}, Coenraad J. and {Perkins}, David and {de Mink}, Selma E.},
        title = "{Formation of the first three gravitational-wave observations through isolated binary evolution}",
      journal = {Nature Communications},
         year = 2017,
        month = apr,
       volume = {8},
          eid = {14906},
        pages = {14906},
          doi = {10.1038/ncomms14906},
archivePrefix = {arXiv},
       eprint = {1704.01352},
 primaryClass = {astro-ph.HE},
       adsurl = {https://ui.adsabs.harvard.edu/abs/2017NatCo...814906S}
}

@ARTICLE{TunaMetzger2023,
       author = {{Tuna}, Semih and {Metzger}, Brian D.},
        title = "{Long-term Evolution of Massive-star Post-common-envelope Circumbinary Disks and the Environments of Fast Luminous Transients}",
      journal = {\apj},
         year = 2023,
        month = oct,
       volume = {955},
       number = {2},
          eid = {125},
        pages = {125},
          doi = {10.3847/1538-4357/acef17},
archivePrefix = {arXiv},
       eprint = {2306.10111},
 primaryClass = {astro-ph.HE},
       adsurl = {https://ui.adsabs.harvard.edu/abs/2023ApJ...955..125T}
}

@ARTICLE{Vallietal2024,
       author = {{Valli}, Ruggero and {Tiede}, Christopher and {Vigna-G{\'o}mez}, Alejandro and {Cuadra}, Jorge and {Siwek}, Magdalena and {Ma}, Jing-Ze and {D'Orazio}, Daniel J. and {Zrake}, Jonathan and {de Mink}, Selma E.},
        title = "{Long-term evolution of binary orbits induced by circumbinary disks}",
      journal = {\aap},
         year = 2024,
        month = aug,
       volume = {688},
          eid = {A128},
        pages = {A128},
          doi = {10.1051/0004-6361/202449421},
archivePrefix = {arXiv},
       eprint = {2401.17355},
 primaryClass = {astro-ph.HE},
       adsurl = {https://ui.adsabs.harvard.edu/abs/2024A&A...688A.128V}
}

@ARTICLE{VignaGomezetal2018,
       author = {{Vigna-G{\'o}mez}, Alejandro and {Neijssel}, Coenraad J. and {Stevenson}, Simon and {Barrett}, Jim W. and {Belczynski}, Krzysztof and {Justham}, Stephen and {de Mink}, Selma E. and {M{\"u}ller}, Bernhard and {Podsiadlowski}, Philipp and {Renzo}, Mathieu and {Sz{\'e}csi}, Dorottya and {Mandel}, Ilya},
        title = "{On the formation history of Galactic double neutron stars}",
      journal = {\mnras},
         year = 2018,
        month = dec,
       volume = {481},
       number = {3},
        pages = {4009-4029},
          doi = {10.1093/mnras/sty2463},
archivePrefix = {arXiv},
       eprint = {1805.07974},
 primaryClass = {astro-ph.SR},
       adsurl = {https://ui.adsabs.harvard.edu/abs/2018MNRAS.481.4009V}
}

@ARTICLE{Weietal2024,
       author = {{Wei}, Dandan and {Schneider}, Fabian R.~N. and {Podsiadlowski}, Philipp and {Laplace}, Eva and {R{\"o}pke}, Friedrich K. and {Vetter}, Marco},
        title = "{Evolution and final fate of massive post-common-envelope binaries}",
      journal = {\aap},
         year = 2024,
        month = aug,
       volume = {688},
          eid = {A87},
        pages = {A87},
          doi = {10.1051/0004-6361/202348560},
archivePrefix = {arXiv},
       eprint = {2311.07278},
 primaryClass = {astro-ph.HE},
       adsurl = {https://ui.adsabs.harvard.edu/abs/2024A&A...688A..87W}
}

@ARTICLE{Vetteretal2024,
       author = {{Vetter}, Marco and {R{\"o}pke}, Friedrich K. and {Schneider}, Fabian R.~N. and {Pakmor}, R{\"u}diger and {Ohlmann}, Sebastian T. and {Lau}, Mike Y.~M. and {Andrassy}, Robert},
        title = "{From spherical stars to disk-like structures: 3D common-envelope evolution of massive binaries beyond inspiral}",
      journal = {\aap},
         year = 2024,
        month = nov,
       volume = {691},
          eid = {A244},
        pages = {A244},
          doi = {10.1051/0004-6361/202451579},
archivePrefix = {arXiv},
       eprint = {2410.07841},
 primaryClass = {astro-ph.SR},
       adsurl = {https://ui.adsabs.harvard.edu/abs/2024A&A...691A.244V}
}

@ARTICLE{vanSonetal2024,
       author = {{van Son}, L.~A.~C. and {Roy}, S.~K. and {Mandel}, I. and {Farr}, W.~M. and {Lam}, A. and {Merritt}, J. and {Broekgaarden}, F.~S. and {Sander}, A. and {Andrews}, J.~J.},
        title = "{Winds of change: why binary black hole formation is metallicity dependent, while binary neutron star formation is not}",
      journal = {arXiv e-prints},
         year = 2024,
        month = nov,
          eid = {arXiv:2411.02484},
        pages = {arXiv:2411.02484},
archivePrefix = {arXiv},
       eprint = {2411.02484},
 primaryClass = {astro-ph.HE},
       adsurl = {https://ui.adsabs.harvard.edu/abs/2024arXiv241102484V}
}

@ARTICLE{Sokeretal2019,
       author = {{Soker}, Noam and {Grichener}, Aldana and {Gilkis}, Avishai},
        title = "{Diversity of common envelope jets supernovae and the fast transient AT2018cow}",
      journal = {\mnras},
         year = 2019,
        month = apr,
       volume = {484},
       number = {4},
        pages = {4972-4979},
          doi = {10.1093/mnras/stz364},
archivePrefix = {arXiv},
       eprint = {1811.11106},
 primaryClass = {astro-ph.HE},
       adsurl = {https://ui.adsabs.harvard.edu/abs/2019MNRAS.484.4972S}
}

@ARTICLE{SokerGilkisiPTF14hls,
       author = {{Soker}, Noam and {Gilkis}, Avishai},
        title = "{Explaining iPTF14hls as a common-envelope jets supernova}",
      journal = {\mnras},
         year = 2018,
        month = mar,
       volume = {475},
       number = {1},
        pages = {1198-1202},
          doi = {10.1093/mnras/stx3287},
archivePrefix = {arXiv},
       eprint = {1711.05180},
 primaryClass = {astro-ph.HE},
       adsurl = {https://ui.adsabs.harvard.edu/abs/2018MNRAS.475.1198S}
}

@ARTICLE{VignaGomezetal2020,
       author = {{Vigna-G{\'o}mez}, Alejandro and {MacLeod}, Morgan and {Neijssel}, Coenraad J. and {Broekgaarden}, Floor S. and {Justham}, Stephen and {Howitt}, George and {de Mink}, Selma E. and {Vinciguerra}, Serena and {Mandel}, Ilya},
        title = "{Common envelope episodes that lead to double neutron star formation}",
      journal = {\pasa},
         year = 2020,
        month = sep,
       volume = {37},
          eid = {e038},
        pages = {e038},
          doi = {10.1017/pasa.2020.31},
archivePrefix = {arXiv},
       eprint = {2001.09829},
 primaryClass = {astro-ph.SR},
       adsurl = {https://ui.adsabs.harvard.edu/abs/2020PASA...37...38V}
}

@ARTICLE{IzzardJermyn2023,
       author = {{Izzard}, Robert G. and {Jermyn}, Adam S.},
        title = "{Circumbinary discs for stellar population models}",
      journal = {\mnras},
         year = 2023,
        month = may,
       volume = {521},
       number = {1},
        pages = {35-50},
          doi = {10.1093/mnras/stac2899},
archivePrefix = {arXiv},
       eprint = {2401.14315},
 primaryClass = {astro-ph.SR},
       adsurl = {https://ui.adsabs.harvard.edu/abs/2023MNRAS.521...35I}
}

@ARTICLE{Taurisetal2015,
       author = {{Tauris}, Thomas M. and {Langer}, Norbert and {Podsiadlowski}, Philipp},
        title = "{Ultra-stripped supernovae: progenitors and fate}",
      journal = {\mnras},
         year = 2015,
        month = aug,
       volume = {451},
       number = {2},
        pages = {2123-2144},
          doi = {10.1093/mnras/stv990},
archivePrefix = {arXiv},
       eprint = {1505.00270},
 primaryClass = {astro-ph.SR},
       adsurl = {https://ui.adsabs.harvard.edu/abs/2015MNRAS.451.2123T}
}

@ARTICLE{Taurisetal2017,
       author = {{Tauris}, T.~M. and {Kramer}, M. and {Freire}, P.~C.~C. and {Wex}, N. and {Janka}, H. -T. and {Langer}, N. and {Podsiadlowski}, Ph. and {Bozzo}, E. and {Chaty}, S. and {Kruckow}, M.~U. and {van den Heuvel}, E.~P.~J. and {Antoniadis}, J. and {Breton}, R.~P. and {Champion}, D.~J.},
        title = "{Formation of Double Neutron Star Systems}",
      journal = {\apj},
         year = 2017,
        month = sep,
       volume = {846},
       number = {2},
          eid = {170},
        pages = {170},
          doi = {10.3847/1538-4357/aa7e89},
archivePrefix = {arXiv},
       eprint = {1706.09438},
 primaryClass = {astro-ph.HE},
       adsurl = {https://ui.adsabs.harvard.edu/abs/2017ApJ...846..170T}
}

@ARTICLE{Mandeletal2025,
       author = {{Team COMPAS: I. Mandel} and {Riley}, Jeff and {Boesky}, Adam and {Brcek}, Adam and {Hirai}, Ryosuke and {Kapil}, Veome and {Lau}, Mike Y.~M. and {Merritt}, JD and {Rodr{\'\i}guez-Segovia}, Nicol{\'a}s and {Romero-Shaw}, Isobel and {Song}, Yuzhe and {Stevenson}, Simon and {Vajpeyi}, Avi and {van Son}, L.~A.~C. and {Vigna-G{\'o}mez}, Alejandro and {Willcox}, Reinhold},
        title = "{Rapid Stellar and Binary Population Synthesis with COMPAS: Methods Paper II}",
      journal = {\apjs},
         year = 2025,
        month = sep,
       volume = {280},
       number = {1},
          eid = {43},
        pages = {43},
          doi = {10.3847/1538-4365/adf8d0},
archivePrefix = {arXiv},
       eprint = {2506.02316},
 primaryClass = {astro-ph.SR},
       adsurl = {https://ui.adsabs.harvard.edu/abs/2025ApJS..280...43T}
}

@ARTICLE{Ungeretal2026,
       author = {{Unger}, Lotem and {Grichener}, Aldana and {Soker}, Noam},
        title = "{Circumbinary Disks in Post Common Envelope Binary Systems with Compact Objects}",
      journal = {\apj},
         year = 2026,
        month = feb,
       volume = {998},
       number = {1},
          eid = {79},
        pages = {79},
          doi = {10.3847/1538-4357/ae3241},
archivePrefix = {arXiv},
       eprint = {2411.15652},
 primaryClass = {astro-ph.SR},
       adsurl = {https://ui.adsabs.harvard.edu/abs/2026ApJ...998...79U}
}

@ARTICLE{MandelMuller2020,
       author = {{Mandel}, Ilya and {M{\"u}ller}, Bernhard},
        title = "{Simple recipes for compact remnant masses and natal kicks}",
      journal = {\mnras},
         year = 2020,
        month = dec,
       volume = {499},
       number = {3},
        pages = {3214-3221},
          doi = {10.1093/mnras/staa3043},
archivePrefix = {arXiv},
       eprint = {2006.08360},
 primaryClass = {astro-ph.HE},
       adsurl = {https://ui.adsabs.harvard.edu/abs/2020MNRAS.499.3214M}
}

@ARTICLE{Claeysetal2014,
       author = {{Claeys}, J.~S.~W. and {Pols}, O.~R. and {Izzard}, R.~G. and {Vink}, J. and {Verbunt}, F.~W.~M.},
        title = "{Theoretical uncertainties of the Type Ia supernova rate}",
      journal = {\aap},
         year = 2014,
        month = mar,
       volume = {563},
          eid = {A83},
        pages = {A83},
          doi = {10.1051/0004-6361/201322714},
archivePrefix = {arXiv},
       eprint = {1401.2895},
 primaryClass = {astro-ph.SR},
       adsurl = {https://ui.adsabs.harvard.edu/abs/2014A&A...563A..83C}
}

@ARTICLE{Geetal2020,
       author = {{Ge}, Hongwei and {Webbink}, Ronald F. and {Chen}, Xuefei and {Han}, Zhanwen},
        title = "{Adiabatic Mass Loss in Binary Stars. III. From the Base of the Red Giant Branch to the Tip of the Asymptotic Giant Branch}",
      journal = {\apj},
         year = 2020,
        month = aug,
       volume = {899},
       number = {2},
          eid = {132},
        pages = {132},
          doi = {10.3847/1538-4357/aba7b7},
archivePrefix = {arXiv},
       eprint = {2007.09848},
 primaryClass = {astro-ph.SR},
       adsurl = {https://ui.adsabs.harvard.edu/abs/2020ApJ...899..132G}
}

@ARTICLE{LIGO2025,
       author = {{The LIGO Scientific Collaboration} and {the Virgo Collaboration} and {the KAGRA Collaboration} and {Abac}, A.~G. and {Abouelfettouh}, I. and {Acernese}, F. and {Ackley}, K. and {Adamcewicz}, C. and {Adhicary}, S. and {Adhikari}, D. and {Adhikari}, N. and {Adhikari}, R.~X. and {Adkins}, V.~K. and {Afroz}, S. and {Agarwal}, D. and {Agathos}, M. and {Aghaei Abchouyeh}, M. and {Aguiar}, O.~D. and {Ahmadzadeh}, S. and {Aiello}, L. and {Ain}, A. and {Ajith}, P. and {Akutsu}, T. and {Albanesi}, S. and {Alfaidi}, R.~A. and {Al-Jodah}, A. and {All{\'e}n{\'e}}, C. and {Allocca}, A. and {Al-Shammari}, S. and {Altin}, P.~A. and {Alvarez-Lopez}, S. and {Amarasinghe}, O. and {Amato}, A. and {Amra}, C. and {Ananyeva}, A. and {Anderson}, S.~B. and {Anderson}, W.~G. and {Andia}, M. and {Ando}, M. and {Andrade}, T. and {Andr{\'e}s-Carcasona}, M. and {Andri{\'c}}, T. and {Anglin}, J. and {Ansoldi}, S. and {Antelis}, J.~M. and {Antier}, S. and {Aoumi}, M. and {Appavuravther}, E.~Z. and {Appert}, S. and {Apple}, S.~K. and {Arai}, K. and {Araya}, A. and {Araya}, M.~C. and {Arca Sedda}, M. and {Areeda}, J.~S. and {Argianas}, L. and {Aritomi}, N. and {Armato}, F. and {Armstrong}, S. and {Arnaud}, N. and {Arogeti}, M. and {Aronson}, S.~M. and {Arun}, K.~G. and {Ashton}, G. and {Aso}, Y. and {Assiduo}, M. and {Assis de Souza Melo}, S. and {Aston}, S.~M. and {Astone}, P. and {Attadio}, F. and {Aubin}, F. and {AultONeal}, K. and {Avallone}, G. and {Babak}, S. and {Badaracco}, F. and {Badger}, C. and {Bae}, S. and {Bagnasco}, S. and {Bagui}, E. and {Baiotti}, L. and {Bajpai}, R. and {Baka}, T. and {Baker}, T. and {Ball}, M. and {Ballardin}, G. and {Ballmer}, S.~W. and {Banagiri}, S. and {Banerjee}, B. and {Bankar}, D. and {Baptiste}, T.~M. and {Baral}, P. and {Barayoga}, J.~C. and {Barish}, B.~C. and {Barker}, D. and {Barman}, N. and {Barneo}, P. and {Barone}, F. and {Barr}, B. and {Barsotti}, L. and {Barsuglia}, M. and {Barta}, D. and {Bartoletti}, A.~M. and {Barton}, M.~A. and {Bartos}, I. and {Basak}, S. and {Basalaev}, A. and {Bassiri}, R. and {Basti}, A. and {Bates}, D.~E. and {Bawaj}, M. and {Baxi}, P. and {Bayley}, J.~C. and {Baylor}, A.~C. and {Baynard}, II, P.~A. and {Bazzan}, M. and {Bedakihale}, V.~M. and {Beirnaert}, F. and {Bejger}, M. and {Belardinelli}, D. and {Bell}, A.~S. and {Bellie}, D.~S. and {Bellizzi}, L. and {Beltran-Martinez}, D. and {Benoit}, W. and {Bentara}, I. and {Bentley}, J.~D. and {Ben Yaala}, M. and {Bera}, S. and {Bergamin}, F. and {Berger}, B.~K. and {Bernuzzi}, S. and {Beroiz}, M. and {Berry}, C.~P.~L. and {Bersanetti}, D. and {Bertolini}, A. and {Betzwieser}, J. and {Beveridge}, D. and {Bevilacqua}, G. and {Bevins}, N. and {Bhandare}, R. and {Bhatt}, R. and {Bhattacharjee}, D. and {Bhaumik}, S. and {Bhowmick}, S. and {Biancalana}, V. and {Bianchi}, A. and {Bilenko}, I.~A. and {Billingsley}, G. and {Binetti}, A. and {Bini}, S. and {Binu}, C. and {Birnholtz}, O. and {Biscoveanu}, S. and {Bisht}, A. and {Bitossi}, M. and {Bizouard}, M. -A. and {Blaber}, S. and {Blackburn}, J.~K. and {Blagg}, L.~A. and {Blair}, C.~D. and {Blair}, D.~G. and {Bobba}, F. and {Bode}, N. and {Boileau}, G. and {Boldrini}, M. and {Bolingbroke}, G.~N. and {Bolliand}, A. and {Bonavena}, L.~D. and {Bondarescu}, R. and {Bondu}, F. and {Bonilla}, E. and {Bonilla}, M.~S. and {Bonino}, A. and {Bonnand}, R. and {Booker}, P. and {Borchers}, A. and {Borhanian}, S. and {Boschi}, V. and {Bose}, S. and {Bossilkov}, V. and {Boudon}, A. and {Bozzi}, A. and {Bradaschia}, C. and {Brady}, P.~R. and {Branch}, A. and {Branchesi}, M. and {Braun}, I. and {Briant}, T. and {Brillet}, A. and {Brinkmann}, M. and {Brockill}, P. and {Brockmueller}, E. and {Brooks}, A.~F. and {Brown}, B.~C. and {Brown}, D.~D. and {Brozzetti}, M.~L. and {Brunett}, S. and {Bruno}, G. and {Bruntz}, R. and {Bryant}, J.},
        title = "{GWTC-4.0: Population Properties of Merging Compact Binaries}",
      journal = {arXiv e-prints},
         year = 2025,
        month = aug,
          eid = {arXiv:2508.18083},
        pages = {arXiv:2508.18083},
          doi = {10.48550/arXiv.2508.18083},
archivePrefix = {arXiv},
       eprint = {2508.18083},
 primaryClass = {astro-ph.HE},
       adsurl = {https://ui.adsabs.harvard.edu/abs/2025arXiv250818083T}
}

@ARTICLE{Kobayashietal2023,
       author = {{Kobayashi}, Chiaki and {Mandel}, Ilya and {Belczynski}, Krzysztof and {Goriely}, Stephane and {Janka}, Thomas H. and {Just}, Oliver and {Ruiter}, Ashley J. and {Vanbeveren}, Dany and {Kruckow}, Matthias U. and {Briel}, Max M. and {Eldridge}, Jan J. and {Stanway}, Elizabeth},
        title = "{Can Neutron Star Mergers Alone Explain the r-process Enrichment of the Milky Way?}",
      journal = {\apjl},
         year = 2023,
        month = feb,
       volume = {943},
       number = {2},
          eid = {L12},
        pages = {L12},
          doi = {10.3847/2041-8213/acad82},
archivePrefix = {arXiv},
       eprint = {2211.04964},
 primaryClass = {astro-ph.HE},
       adsurl = {https://ui.adsabs.harvard.edu/abs/2023ApJ...943L..12K}
}

@ARTICLE{Dittmannetal2025,
       author = {{Dittmann}, Alexander J. and {Ryan}, Geoffrey and {Combi}, Luciano},
        title = "{Eccentric Binaries Accreting from Thin Disks: Orbital Evolution}",
      journal = {arXiv e-prints},
         year = 2025,
        month = dec,
          eid = {arXiv:2512.11954},
        pages = {arXiv:2512.11954},
archivePrefix = {arXiv},
       eprint = {2512.11954},
 primaryClass = {astro-ph.GA},
       adsurl = {https://ui.adsabs.harvard.edu/abs/2025arXiv251211954D}
}

@ARTICLE{Vetteretal2025,
       author = {{Vetter}, Marco and {R{\"o}pke}, Friedrich K. and {Schneider}, Fabian R.~N. and {Pakmor}, R{\"u}diger and {Ohlmann}, Sebastian and {Mor{\'a}n-Fraile}, Javier and {Lau}, Mike Y.~M. and {Leidi}, Giovanni and {Gagnier}, Damien and {Andrassy}, Robert},
        title = "{Magnetically driven outflows in the 3D common envelope evolution of massive stars}",
      journal = {\aap},
         year = 2025,
        month = jun,
       volume = {698},
          eid = {A133},
        pages = {A133},
          doi = {10.1051/0004-6361/202554685},
archivePrefix = {arXiv},
       eprint = {2504.12213},
 primaryClass = {astro-ph.SR},
       adsurl = {https://ui.adsabs.harvard.edu/abs/2025A&A...698A.133V}
}

@ARTICLE{Dermineetal2013,
       author = {{Dermine}, T. and {Izzard}, R.~G. and {Jorissen}, A. and {Van Winckel}, H.},
        title = "{Eccentricity-pumping in post-AGB stars with circumbinary discs}",
      journal = {\aap},
         year = 2013,
        month = mar,
       volume = {551},
          eid = {A50},
        pages = {A50},
          doi = {10.1051/0004-6361/201219430},
       adsurl = {https://ui.adsabs.harvard.edu/abs/2013A&A...551A..50D}
}

@ARTICLE{Metzger2022,
       author = {{Metzger}, Brian D.},
        title = "{Luminous Fast Blue Optical Transients and Type Ibn/Icn SNe from Wolf-Rayet/Black Hole Mergers}",
      journal = {\apj},
         year = 2022,
        month = jun,
       volume = {932},
       number = {2},
          eid = {84},
        pages = {84},
          doi = {10.3847/1538-4357/ac6d59},
archivePrefix = {arXiv},
       eprint = {2203.04331},
 primaryClass = {astro-ph.HE},
       adsurl = {https://ui.adsabs.harvard.edu/abs/2022ApJ...932...84M}
}

@ARTICLE{CohenSoker2023,
       author = {{Cohen}, Tamar and {Soker}, Noam},
        title = "{Terminating a common envelope jets supernova impostor event with a super-Eddington blue supergiant}",
      journal = {\mnras},
         year = 2023,
        month = jun,
       volume = {522},
       number = {1},
        pages = {885-890},
          doi = {10.1093/mnras/stad1015},
archivePrefix = {arXiv},
       eprint = {2212.08551},
 primaryClass = {astro-ph.HE},
       adsurl = {https://ui.adsabs.harvard.edu/abs/2023MNRAS.522..885C}
}

@ARTICLE{Grichener2025,
       author = {{Grichener}, Aldana},
        title = "{Mergers of compact objects with cores of massive stars: evolutionary pathways, r-process nucleosynthesis and multi-messenger signatures}",
      journal = {\apss},
         year = 2025,
        month = feb,
       volume = {370},
       number = {2},
          eid = {11},
        pages = {11},
          doi = {10.1007/s10509-025-04402-1},
archivePrefix = {arXiv},
       eprint = {2410.18813},
 primaryClass = {astro-ph.HE},
       adsurl = {https://ui.adsabs.harvard.edu/abs/2025Ap&SS.370...11G}
}

@ARTICLE{GrichenerSoker2019,
       author = {{Grichener}, Aldana and {Soker}, Noam},
        title = "{The Common Envelope Jet Supernova (CEJSN) r-process Scenario}",
      journal = {\apj},
         year = 2019,
        month = jun,
       volume = {878},
       number = {1},
          eid = {24},
        pages = {24},
          doi = {10.3847/1538-4357/ab1d5d},
archivePrefix = {arXiv},
       eprint = {1810.03889},
 primaryClass = {astro-ph.SR},
       adsurl = {https://ui.adsabs.harvard.edu/abs/2019ApJ...878...24G}
}

@ARTICLE{Gricheneretal2022,
       author = {{Grichener}, Aldana and {Kobayashi}, Chiaki and {Soker}, Noam},
        title = "{Common Envelope Jet Supernova r-process Yields Can Reproduce [Eu/Fe] Abundance Evolution in the Galaxy}",
      journal = {\apjl},
         year = 2022,
        month = feb,
       volume = {926},
       number = {2},
          eid = {L9},
        pages = {L9},
          doi = {10.3847/2041-8213/ac4f68},
archivePrefix = {arXiv},
       eprint = {2112.08301},
 primaryClass = {astro-ph.HE},
       adsurl = {https://ui.adsabs.harvard.edu/abs/2022ApJ...926L...9G}
}

@ARTICLE{Andrewsetal2015,
       author = {{Andrews}, Jeff J. and {Farr}, W.~M. and {Kalogera}, V. and {Willems}, B.},
        title = "{Evolutionary Channels for the Formation of Double Neutron Stars}",
      journal = {\apj},
         year = 2015,
        month = mar,
       volume = {801},
       number = {1},
          eid = {32},
        pages = {32},
          doi = {10.1088/0004-637X/801/1/32},
archivePrefix = {arXiv},
       eprint = {1410.6797},
 primaryClass = {astro-ph.SR},
       adsurl = {https://ui.adsabs.harvard.edu/abs/2015ApJ...801...32A}
}

@ARTICLE{Manchester2005,
  bibcode={2005AJ....129.1993M},
   author = {{Manchester}, R.~N. and {Hobbs}, G.~B. and {Teoh}, A. and {Hobbs}, M.
	},
    title = "{The Australia Telescope National Facility Pulsar Catalogue}",
  journal = {\aj},
     year = 2005,
    month = apr,
   volume = 129,
    pages = {1993-2006},
      doi = {10.1086/428488},
   adsurl = {http://adsabs.harvard.edu/cgi-bin/nph-bib_query?bibcode=2005AJ....129.1993M&db_key=AST},
  comment = {ATNF catalog paper}
}

@ARTICLE{Goldsteinetal2017,
       author = {{Goldstein}, A. and {Veres}, P. and {Burns}, E. and {Briggs}, M.~S. and {Hamburg}, R. and {Kocevski}, D. and {Wilson-Hodge}, C.~A. and {Preece}, R.~D. and {Poolakkil}, S. and {Roberts}, O.~J. and {Hui}, C.~M. and {Connaughton}, V. and {Racusin}, J. and {von Kienlin}, A. and {Dal Canton}, T. and {Christensen}, N. and {Littenberg}, T. and {Siellez}, K. and {Blackburn}, L. and {Broida}, J. and {Bissaldi}, E. and {Cleveland}, W.~H. and {Gibby}, M.~H. and {Giles}, M.~M. and {Kippen}, R.~M. and {McBreen}, S. and {McEnery}, J. and {Meegan}, C.~A. and {Paciesas}, W.~S. and {Stanbro}, M.},
        title = "{An Ordinary Short Gamma-Ray Burst with Extraordinary Implications: Fermi-GBM Detection of GRB 170817A}",
      journal = {\apjl},
         year = 2017,
        month = oct,
       volume = {848},
       number = {2},
          eid = {L14},
        pages = {L14},
          doi = {10.3847/2041-8213/aa8f41},
archivePrefix = {arXiv},
       eprint = {1710.05446},
 primaryClass = {astro-ph.HE},
       adsurl = {https://ui.adsabs.harvard.edu/abs/2017ApJ...848L..14G}
}

@ARTICLE{MadauDickinson2014,
       author = {{Madau}, Piero and {Dickinson}, Mark},
        title = "{Cosmic Star-Formation History}",
      journal = {\araa},
         year = 2014,
        month = aug,
       volume = {52},
        pages = {415-486},
          doi = {10.1146/annurev-astro-081811-125615},
archivePrefix = {arXiv},
       eprint = {1403.0007},
 primaryClass = {astro-ph.CO},
       adsurl = {https://ui.adsabs.harvard.edu/abs/2014ARA&A..52..415M}
}

@ARTICLE{Fryeretal2012,
       author = {{Fryer}, Chris L. and {Belczynski}, Krzysztof and {Wiktorowicz}, Grzegorz and {Dominik}, Michal and {Kalogera}, Vicky and {Holz}, Daniel E.},
        title = "{Compact Remnant Mass Function: Dependence on the Explosion Mechanism and Metallicity}",
      journal = {\apj},
         year = 2012,
        month = apr,
       volume = {749},
       number = {1},
          eid = {91},
        pages = {91},
          doi = {10.1088/0004-637X/749/1/91},
archivePrefix = {arXiv},
       eprint = {1110.1726},
 primaryClass = {astro-ph.SR},
       adsurl = {https://ui.adsabs.harvard.edu/abs/2012ApJ...749...91F}
}

@ARTICLE{DisbergMandel:2025,
       author = {{Disberg}, Paul and {Mandel}, Ilya},
        title = "{The Kick Velocity Distribution of Isolated Neutron Stars}",
      journal = {\apjl},
         year = 2025,
        month = aug,
       volume = {989},
       number = {1},
          eid = {L8},
        pages = {L8},
          doi = {10.3847/2041-8213/adf286},
archivePrefix = {arXiv},
       eprint = {2505.22102},
 primaryClass = {astro-ph.HE},
       adsurl = {https://ui.adsabs.harvard.edu/abs/2025ApJ...989L...8D}
}

@ARTICLE{Abbottetal2020,
       author = {{Abbott}, B.~P. and {Abbott}, R. and {Abbott}, T.~D. and {Abraham}, S. and {Acernese}, F. and {Ackley}, K. and {Adams}, C. and {Adhikari}, R.~X. and {Adya}, V.~B. and {Affeldt}, C. and {Agathos}, M. and {Agatsuma}, K. and {Aggarwal}, N. and {Aguiar}, O.~D. and {Aiello}, L. and {Ain}, A. and {Ajith}, P. and {Allen}, G. and {Allocca}, A. and {Aloy}, M.~A. and {Altin}, P.~A. and {Amato}, A. and {Anand}, S. and {Ananyeva}, A. and {Anderson}, S.~B. and {Anderson}, W.~G. and {Angelova}, S.~V. and {Antier}, S. and {Appert}, S. and {Arai}, K. and {Araya}, M.~C. and {Areeda}, J.~S. and {Ar{\`e}ne}, M. and {Arnaud}, N. and {Aronson}, S.~M. and {Arun}, K.~G. and {Ascenzi}, S. and {Ashton}, G. and {Aston}, S.~M. and {Astone}, P. and {Aubin}, F. and {Aufmuth}, P. and {AultONeal}, K. and {Austin}, C. and {Avendano}, V. and {Avila-Alvarez}, A. and {Babak}, S. and {Bacon}, P. and {Badaracco}, F. and {Bader}, M.~K.~M. and {Bae}, S. and {Baird}, J. and {Baker}, P.~T. and {Baldaccini}, F. and {Ballardin}, G. and {Ballmer}, S.~W. and {Bals}, A. and {Banagiri}, S. and {Barayoga}, J.~C. and {Barbieri}, C. and {Barclay}, S.~E. and {Barish}, B.~C. and {Barker}, D. and {Barkett}, K. and {Barnum}, S. and {Barone}, F. and {Barr}, B. and {Barsotti}, L. and {Barsuglia}, M. and {Barta}, D. and {Bartlett}, J. and {Bartos}, I. and {Bassiri}, R. and {Basti}, A. and {Bawaj}, M. and {Bayley}, J.~C. and {Baylor}, A.~C. and {Bazzan}, M. and {B{\'e}csy}, B. and {Bejger}, M. and {Belahcene}, I. and {Bell}, A.~S. and {Beniwal}, D. and {Benjamin}, M.~G. and {Berger}, B.~K. and {Bergmann}, G. and {Bernuzzi}, S. and {Berry}, C.~P.~L. and {Bersanetti}, D. and {Bertolini}, A. and {Betzwieser}, J. and {Bhandare}, R. and {Bidler}, J. and {Biggs}, E. and {Bilenko}, I.~A. and {Bilgili}, S.~A. and {Billingsley}, G. and {Birney}, R. and {Birnholtz}, O. and {Biscans}, S. and {Bischi}, M. and {Biscoveanu}, S. and {Bisht}, A. and {Bitossi}, M. and {Bizouard}, M.~A. and {Blackburn}, J.~K. and {Blackman}, J. and {Blair}, C.~D. and {Blair}, D.~G. and {Blair}, R.~M. and {Bloemen}, S. and {Bobba}, F. and {Bode}, N. and {Boer}, M. and {Boetzel}, Y. and {Bogaert}, G. and {Bondu}, F. and {Bonnand}, R. and {Booker}, P. and {Boom}, B.~A. and {Bork}, R. and {Boschi}, V. and {Bose}, S. and {Bossilkov}, V. and {Bosveld}, J. and {Bouffanais}, Y. and {Bozzi}, A. and {Bradaschia}, C. and {Brady}, P.~R. and {Bramley}, A. and {Branchesi}, M. and {Brau}, J.~E. and {Breschi}, M. and {Briant}, T. and {Briggs}, J.~H. and {Brighenti}, F. and {Brillet}, A. and {Brinkmann}, M. and {Brockill}, P. and {Brooks}, A.~F. and {Brooks}, J. and {Brown}, D.~D. and {Brunett}, S. and {Buikema}, A. and {Bulik}, T. and {Bulten}, H.~J. and {Buonanno}, A. and {Buskulic}, D. and {Buy}, C. and {Byer}, R.~L. and {Cabero}, M. and {Cadonati}, L. and {Cagnoli}, G. and {Cahillane}, C. and {Calder{\'o}n Bustillo}, J. and {Callister}, T.~A. and {Calloni}, E. and {Camp}, J.~B. and {Campbell}, W.~A. and {Canepa}, M. and {Cannon}, K.~C. and {Cao}, H. and {Cao}, J. and {Carapella}, G. and {Carbognani}, F. and {Caride}, S. and {Carney}, M.~F. and {Carullo}, G. and {Casanueva Diaz}, J. and {Casentini}, C. and {Caudill}, S. and {Cavagli{\`a}}, M. and {Cavalier}, F. and {Cavalieri}, R. and {Cella}, G. and {Cerd{\'a}-Dur{\'a}n}, P. and {Cesarini}, E. and {Chaibi}, O. and {Chakravarti}, K. and {Chamberlin}, S.~J. and {Chan}, M. and {Chao}, S. and {Charlton}, P. and {Chase}, E.~A. and {Chassande-Mottin}, E. and {Chatterjee}, D. and {Chaturvedi}, M. and {Chatziioannou}, K. and {Cheeseboro}, B.~D. and {Chen}, H.~Y. and {Chen}, X. and {Chen}, Y. and {Cheng}, H.-P. and {Cheong}, C.~K. and {Chia}, H.~Y. and {Chiadini}, F. and {Chincarini}, A. and {Chiummo}, A. and {Cho}, G. and {Cho}, H.~S.},
        title = "{GW190425: Observation of a Compact Binary Coalescence with Total Mass {\ensuremath{\sim}} 3.4 M$_{{\ensuremath{\odot}}}$}",
      journal = {\apjl},
         year = 2020,
        month = mar,
       volume = {892},
       number = {1},
          eid = {L3},
        pages = {L3},
          doi = {10.3847/2041-8213/ab75f5},
archivePrefix = {arXiv},
       eprint = {2001.01761},
 primaryClass = {astro-ph.HE},
       adsurl = {https://ui.adsabs.harvard.edu/abs/2020ApJ...892L...3A}
}

@ARTICLE{Abbotetal2017,
       author = {{Abbott}, B.~P. and {Abbott}, R. and {Abbott}, T.~D. and {Acernese}, F. and {Ackley}, K. and {Adams}, C. and {Adams}, T. and {Addesso}, P. and {Adhikari}, R.~X. and {Adya}, V.~B. and {Affeldt}, C. and {Afrough}, M. and {Agarwal}, B. and {Agathos}, M. and {Agatsuma}, K. and {Aggarwal}, N. and {Aguiar}, O.~D. and {Aiello}, L. and {Ain}, A. and {Ajith}, P. and {Allen}, B. and {Allen}, G. and {Allocca}, A. and {Altin}, P.~A. and {Amato}, A. and {Ananyeva}, A. and {Anderson}, S.~B. and {Anderson}, W.~G. and {Angelova}, S.~V. and {Antier}, S. and {Appert}, S. and {Arai}, K. and {Araya}, M.~C. and {Areeda}, J.~S. and {Arnaud}, N. and {Arun}, K.~G. and {Ascenzi}, S. and {Ashton}, G. and {Ast}, M. and {Aston}, S.~M. and {Astone}, P. and {Atallah}, D.~V. and {Aufmuth}, P. and {Aulbert}, C. and {AultONeal}, K. and {Austin}, C. and {Avila-Alvarez}, A. and {Babak}, S. and {Bacon}, P. and {Bader}, M.~K.~M. and {Bae}, S. and {Bailes}, M. and {Baker}, P.~T. and {Baldaccini}, F. and {Ballardin}, G. and {Ballmer}, S.~W. and {Banagiri}, S. and {Barayoga}, J.~C. and {Barclay}, S.~E. and {Barish}, B.~C. and {Barker}, D. and {Barkett}, K. and {Barone}, F. and {Barr}, B. and {Barsotti}, L. and {Barsuglia}, M. and {Barta}, D. and {Barthelmy}, S.~D. and {Bartlett}, J. and {Bartos}, I. and {Bassiri}, R. and {Basti}, A. and {Batch}, J.~C. and {Bawaj}, M. and {Bayley}, J.~C. and {Bazzan}, M. and {B{\'e}csy}, B. and {Beer}, C. and {Bejger}, M. and {Belahcene}, I. and {Bell}, A.~S. and {Berger}, B.~K. and {Bergmann}, G. and {Bernuzzi}, S. and {Bero}, J.~J. and {Berry}, C.~P.~L. and {Bersanetti}, D. and {Bertolini}, A. and {Betzwieser}, J. and {Bhagwat}, S. and {Bhandare}, R. and {Bilenko}, I.~A. and {Billingsley}, G. and {Billman}, C.~R. and {Birch}, J. and {Birney}, R. and {Birnholtz}, O. and {Biscans}, S. and {Biscoveanu}, S. and {Bisht}, A. and {Bitossi}, M. and {Biwer}, C. and {Bizouard}, M.~A. and {Blackburn}, J.~K. and {Blackman}, J. and {Blair}, C.~D. and {Blair}, D.~G. and {Blair}, R.~M. and {Bloemen}, S. and {Bock}, O. and {Bode}, N. and {Boer}, M. and {Bogaert}, G. and {Bohe}, A. and {Bondu}, F. and {Bonilla}, E. and {Bonnand}, R. and {Boom}, B.~A. and {Bork}, R. and {Boschi}, V. and {Bose}, S. and {Bossie}, K. and {Bouffanais}, Y. and {Bozzi}, A. and {Bradaschia}, C. and {Brady}, P.~R. and {Branchesi}, M. and {Brau}, J.~E. and {Briant}, T. and {Brillet}, A. and {Brinkmann}, M. and {Brisson}, V. and {Brockill}, P. and {Broida}, J.~E. and {Brooks}, A.~F. and {Brown}, D.~A. and {Brown}, D.~D. and {Brunett}, S. and {Buchanan}, C.~C. and {Buikema}, A. and {Bulik}, T. and {Bulten}, H.~J. and {Buonanno}, A. and {Buskulic}, D. and {Buy}, C. and {Byer}, R.~L. and {Cabero}, M. and {Cadonati}, L. and {Cagnoli}, G. and {Cahillane}, C. and {Calder{\'o}n Bustillo}, J. and {Callister}, T.~A. and {Calloni}, E. and {Camp}, J.~B. and {Canepa}, M. and {Canizares}, P. and {Cannon}, K.~C. and {Cao}, H. and {Cao}, J. and {Capano}, C.~D. and {Capocasa}, E. and {Carbognani}, F. and {Caride}, S. and {Carney}, M.~F. and {Carullo}, G. and {Casanueva Diaz}, J. and {Casentini}, C. and {Caudill}, S. and {Cavagli{\`a}}, M. and {Cavalier}, F. and {Cavalieri}, R. and {Cella}, G. and {Cepeda}, C.~B. and {Cerd{\'a}-Dur{\'a}n}, P. and {Cerretani}, G. and {Cesarini}, E. and {Chamberlin}, S.~J. and {Chan}, M. and {Chao}, S. and {Charlton}, P. and {Chase}, E. and {Chassande-Mottin}, E. and {Chatterjee}, D. and {Chatziioannou}, K. and {Cheeseboro}, B.~D. and {Chen}, H.~Y. and {Chen}, X. and {Chen}, Y. and {Cheng}, H.-P. and {Chia}, H. and {Chincarini}, A. and {Chiummo}, A. and {Chmiel}, T. and {Cho}, H.~S. and {Cho}, M. and {Chow}, J.~H. and {Christensen}, N. and {Chu}, Q. and {Chua}, A.~J.~K. and {Chua}, S.},
        title = "{GW170817: Observation of Gravitational Waves from a Binary Neutron Star Inspiral}",
      journal = {\prl},
         year = 2017,
        month = oct,
       volume = {119},
       number = {16},
          eid = {161101},
        pages = {161101},
          doi = {10.1103/PhysRevLett.119.161101},
archivePrefix = {arXiv},
       eprint = {1710.05832},
 primaryClass = {gr-qc},
       adsurl = {https://ui.adsabs.harvard.edu/abs/2017PhRvL.119p1101A}
}

@ARTICLE{Papishetal2015,
       author = {{Papish}, Oded and {Soker}, Noam and {Bukay}, Inbal},
        title = "{Ejecting the envelope of red supergiant stars with jets launched by an inspiralling neutron star}",
      journal = {\mnras},
         year = 2015,
        month = may,
       volume = {449},
       number = {1},
        pages = {288-295},
          doi = {10.1093/mnras/stv345},
       adsurl = {https://ui.adsabs.harvard.edu/abs/2015MNRAS.449..288P}
}

@ARTICLE{JinSoker2024,
       author = {{Jin}, Shilun and {Soker}, Noam},
        title = "{Robust r-process Nucleosynthesis beyond Lanthanides in the Common Envelop Jet Supernovae}",
      journal = {\apj},
         year = 2024,
        month = aug,
       volume = {971},
       number = {2},
          eid = {189},
        pages = {189},
          doi = {10.3847/1538-4357/ad5f8e},
archivePrefix = {arXiv},
       eprint = {2310.08907},
 primaryClass = {astro-ph.HE},
       adsurl = {https://ui.adsabs.harvard.edu/abs/2024ApJ...971..189J}
}

@ARTICLE{Soker2025,
       author = {{Soker}, Noam},
        title = "{Mixing neutron star material into the jets in the common envelope jets supernova r-process scenario}",
      journal = {The Open Journal of Astrophysics},
         year = 2025,
        month = may,
       volume = {8},
          eid = {67},
        pages = {67},
          doi = {10.33232/001c.138777},
archivePrefix = {arXiv},
       eprint = {2502.02411},
 primaryClass = {astro-ph.HE},
       adsurl = {https://ui.adsabs.harvard.edu/abs/2025OJAp....8E..67S}
}

@ARTICLE{Soker2026,
       author = {{Soker}, Noam},
        title = "{The early r-process nucleosynthesis scenarios}",
      journal = {arXiv e-prints},
         year = 2026,
        month = jan,
          eid = {arXiv:2601.15187},
        pages = {arXiv:2601.15187},
          doi = {10.48550/arXiv.2601.15187},
archivePrefix = {arXiv},
       eprint = {2601.15187},
 primaryClass = {astro-ph.HE},
       adsurl = {https://ui.adsabs.harvard.edu/abs/2026arXiv260115187S}
}

@ARTICLE{ThielemannCowan2026,
       author = {{Thielemann}, Friedrich-Karl and {Cowan}, John J.},
        title = "{The r-Process: History, Required Conditions, Astrophysical Sites, and Observations}",
      journal = {arXiv e-prints},
         year = 2026,
        month = jan,
          eid = {arXiv:2601.17246},
        pages = {arXiv:2601.17246},
          doi = {10.48550/arXiv.2601.17246},
archivePrefix = {arXiv},
       eprint = {2601.17246},
 primaryClass = {astro-ph.SR},
       adsurl = {https://ui.adsabs.harvard.edu/abs/2026arXiv260117246T}
}

@ARTICLE{Kobayashietal2020,
       author = {{Kobayashi}, Chiaki and {Karakas}, Amanda I. and {Lugaro}, Maria},
        title = "{The Origin of Elements from Carbon to Uranium}",
      journal = {\apj},
         year = 2020,
        month = sep,
       volume = {900},
       number = {2},
          eid = {179},
        pages = {179},
          doi = {10.3847/1538-4357/abae65},
archivePrefix = {arXiv},
       eprint = {2008.04660},
 primaryClass = {astro-ph.GA},
       adsurl = {https://ui.adsabs.harvard.edu/abs/2020ApJ...900..179K}
}

@ARTICLE{Coteetal2019,
       author = {{C{\^o}t{\'e}}, Benoit and {Eichler}, Marius and {Arcones}, Almudena and {Hansen}, Camilla J. and {Simonetti}, Paolo and {Frebel}, Anna and {Fryer}, Chris L. and {Pignatari}, Marco and {Reichert}, Moritz and {Belczynski}, Krzysztof and {Matteucci}, Francesca},
        title = "{Neutron Star Mergers Might Not Be the Only Source of r-process Elements in the Milky Way}",
      journal = {\apj},
         year = 2019,
        month = apr,
       volume = {875},
       number = {2},
          eid = {106},
        pages = {106},
          doi = {10.3847/1538-4357/ab10db},
archivePrefix = {arXiv},
       eprint = {1809.03525},
 primaryClass = {astro-ph.HE},
       adsurl = {https://ui.adsabs.harvard.edu/abs/2019ApJ...875..106C}
}

@ARTICLE{Garciaetal2021,
       author = {{Garc{\'\i}a}, Federico and {Simaz Bunzel}, Adolfo and {Chaty}, Sylvain and {Porter}, Edward and {Chassande-Mottin}, Eric},
        title = "{Progenitors of low-mass binary black-hole mergers in the isolated binary evolution scenario}",
      journal = {\aap},
         year = 2021,
        month = may,
       volume = {649},
          eid = {A114},
        pages = {A114},
          doi = {10.1051/0004-6361/202038357},
archivePrefix = {arXiv},
       eprint = {2103.03161},
 primaryClass = {astro-ph.HE},
       adsurl = {https://ui.adsabs.harvard.edu/abs/2021A&A...649A.114G}
}

@ARTICLE{Zevinetal2021,
       author = {{Zevin}, Michael and {Bavera}, Simone S. and {Berry}, Christopher P.~L. and {Kalogera}, Vicky and {Fragos}, Tassos and {Marchant}, Pablo and {Rodriguez}, Carl L. and {Antonini}, Fabio and {Holz}, Daniel E. and {Pankow}, Chris},
        title = "{One Channel to Rule Them All? Constraining the Origins of Binary Black Holes Using Multiple Formation Pathways}",
      journal = {\apj},
         year = 2021,
        month = apr,
       volume = {910},
       number = {2},
          eid = {152},
        pages = {152},
          doi = {10.3847/1538-4357/abe40e},
archivePrefix = {arXiv},
       eprint = {2011.10057},
 primaryClass = {astro-ph.HE},
       adsurl = {https://ui.adsabs.harvard.edu/abs/2021ApJ...910..152Z}
}

@ARTICLE{BroekgaardenBerger2021,
       author = {{Broekgaarden}, Floor S. and {Berger}, Edo},
        title = "{Formation of the First Two Black Hole-Neutron Star Mergers (GW200115 and GW200105) from Isolated Binary Evolution}",
      journal = {\apjl},
         year = 2021,
        month = oct,
       volume = {920},
       number = {1},
          eid = {L13},
        pages = {L13},
          doi = {10.3847/2041-8213/ac2832},
archivePrefix = {arXiv},
       eprint = {2108.05763},
 primaryClass = {astro-ph.HE},
       adsurl = {https://ui.adsabs.harvard.edu/abs/2021ApJ...920L..13B}
}

@ARTICLE{Peters1964,
       author = {{Peters}, P.~C.},
        title = "{Gravitational Radiation and the Motion of Two Point Masses}",
      journal = {Physical Review},
         year = 1964,
        month = nov,
       volume = {136},
       number = {4B},
        pages = {1224-1232},
          doi = {10.1103/PhysRev.136.B1224},
       adsurl = {https://ui.adsabs.harvard.edu/abs/1964PhRv..136.1224P}
}

@ARTICLE{Vink2001,
       author = {{Vink}, Jorick S. and {de Koter}, A. and {Lamers}, H.~J.~G.~L.~M.},
        title = "{Mass-loss predictions for O and B stars as a function of metallicity}",
      journal = {\aap},
         year = 2001,
        month = apr,
       volume = {369},
        pages = {574-588},
          doi = {10.1051/0004-6361:20010127},
archivePrefix = {arXiv},
       eprint = {astro-ph/0101509},
 primaryClass = {astro-ph},
       adsurl = {https://ui.adsabs.harvard.edu/abs/2001A&A...369..574V}
}

@ARTICLE{Vink2005,
       author = {{Vink}, Jorick S. and {de Koter}, A.},
        title = "{On the metallicity dependence of Wolf-Rayet winds}",
      journal = {\aap},
         year = 2005,
        month = nov,
       volume = {442},
       number = {2},
        pages = {587-596},
          doi = {10.1051/0004-6361:20052862},
archivePrefix = {arXiv},
       eprint = {astro-ph/0507352},
 primaryClass = {astro-ph},
       adsurl = {https://ui.adsabs.harvard.edu/abs/2005A&A...442..587V}
}

@ARTICLE{Klencki2020,
       author = {{Klencki}, Jakub and {Nelemans}, Gijs and {Istrate}, Alina G. and {Pols}, Onno},
        title = "{Massive donors in interacting binaries: effect of metallicity}",
      journal = {\aap},
         year = 2020,
        month = jun,
       volume = {638},
          eid = {A55},
        pages = {A55},
          doi = {10.1051/0004-6361/202037694},
archivePrefix = {arXiv},
       eprint = {2004.00628},
 primaryClass = {astro-ph.SR},
       adsurl = {https://ui.adsabs.harvard.edu/abs/2020A&A...638A..55K}
}

@ARTICLE{Laplace2020,
       author = {{Laplace}, E. and {G{\"o}tberg}, Y. and {de Mink}, S.~E. and {Justham}, S. and {Farmer}, R.},
        title = "{The expansion of stripped-envelope stars: Consequences for supernovae and gravitational-wave progenitors}",
      journal = {\aap},
         year = 2020,
        month = may,
       volume = {637},
          eid = {A6},
        pages = {A6},
          doi = {10.1051/0004-6361/201937300},
archivePrefix = {arXiv},
       eprint = {2003.01120},
 primaryClass = {astro-ph.SR},
       adsurl = {https://ui.adsabs.harvard.edu/abs/2020A&A...637A...6L}
}

@ARTICLE{vanSon2023,
       author = {{van Son}, L.~A.~C. and {de Mink}, S.~E. and {Chru{\'s}li{\'n}ska}, M. and {Conroy}, C. and {Pakmor}, R. and {Hernquist}, L.},
        title = "{The Locations of Features in the Mass Distribution of Merging Binary Black Holes Are Robust against Uncertainties in the Metallicity-dependent Cosmic Star Formation History}",
      journal = {\apj},
         year = 2023,
        month = may,
       volume = {948},
       number = {2},
          eid = {105},
        pages = {105},
          doi = {10.3847/1538-4357/acbf51},
archivePrefix = {arXiv},
       eprint = {2209.03385},
 primaryClass = {astro-ph.GA},
       adsurl = {https://ui.adsabs.harvard.edu/abs/2023ApJ...948..105V}
}

@ARTICLE{Kruckow2021,
       author = {{Kruckow}, Matthias U. and {Neunteufel}, Patrick G. and {Di Stefano}, Rosanne and {Gao}, Yan and {Kobayashi}, Chiaki},
        title = "{A Catalog of Potential Post-Common Envelope Binaries}",
      journal = {\apj},
         year = 2021,
        month = oct,
       volume = {920},
       number = {2},
          eid = {86},
        pages = {86},
          doi = {10.3847/1538-4357/ac13ac},
archivePrefix = {arXiv},
       eprint = {2107.05221},
 primaryClass = {astro-ph.SR},
       adsurl = {https://ui.adsabs.harvard.edu/abs/2021ApJ...920...86K}
}

@ARTICLE{Beuermann2012,
       author = {{Beuermann}, K. and {Dreizler}, S. and {Hessman}, F.~V. and {Deller}, J.},
        title = "{The quest for companions to post-common envelope binaries. III. A reexamination of HW Virginis}",
      journal = {\aap},
         year = 2012,
        month = jul,
       volume = {543},
          eid = {A138},
        pages = {A138},
          doi = {10.1051/0004-6361/201219391},
archivePrefix = {arXiv},
       eprint = {1206.3080},
 primaryClass = {astro-ph.SR},
       adsurl = {https://ui.adsabs.harvard.edu/abs/2012A&A...543A.138B}
}

@ARTICLE{DittmannRyan2022,
       author = {{Dittmann}, Alexander J. and {Ryan}, Geoffrey},
        title = "{A survey of disc thickness and viscosity in circumbinary accretion: Binary evolution, variability, and disc morphology}",
      journal = {\mnras},
         year = 2022,
        month = jul,
       volume = {513},
       number = {4},
        pages = {6158-6176},
          doi = {10.1093/mnras/stac935},
archivePrefix = {arXiv},
       eprint = {2201.07816},
 primaryClass = {astro-ph.HE},
       adsurl = {https://ui.adsabs.harvard.edu/abs/2022MNRAS.513.6158D}
}

@ARTICLE{ShakuraSunyaev1973,
       author = {{Shakura}, N.~I. and {Sunyaev}, R.~A.},
        title = "{Black holes in binary systems. Observational appearance.}",
      journal = {\aap},
         year = 1973,
        month = jan,
       volume = {24},
        pages = {337-355},
       adsurl = {https://ui.adsabs.harvard.edu/abs/1973A&A....24..337S}
}

@ARTICLE{Lau2022,
       author = {{Lau}, Mike Y.~M. and {Hirai}, Ryosuke and {Price}, Daniel J. and {Mandel}, Ilya},
        title = "{Common envelopes in massive stars II: The distinct roles of hydrogen and helium recombination}",
      journal = {\mnras},
         year = 2022,
        month = nov,
       volume = {516},
       number = {4},
        pages = {4669-4678},
          doi = {10.1093/mnras/stac2490},
archivePrefix = {arXiv},
       eprint = {2206.06411},
 primaryClass = {astro-ph.SR},
       adsurl = {https://ui.adsabs.harvard.edu/abs/2022MNRAS.516.4669L}
}

@ARTICLE{Passy2012,
       author = {{Passy}, Jean-Claude and {De Marco}, Orsola and {Fryer}, Chris L. and {Herwig}, Falk and {Diehl}, Steven and {Oishi}, Jeffrey S. and {Mac Low}, Mordecai-Mark and {Bryan}, Greg L. and {Rockefeller}, Gabriel},
        title = "{Simulating the Common Envelope Phase of a Red Giant Using Smoothed-particle Hydrodynamics and Uniform-grid Codes}",
      journal = {\apj},
         year = 2012,
        month = jan,
       volume = {744},
       number = {1},
          eid = {52},
        pages = {52},
          doi = {10.1088/0004-637X/744/1/52},
archivePrefix = {arXiv},
       eprint = {1107.5072},
 primaryClass = {astro-ph.SR},
       adsurl = {https://ui.adsabs.harvard.edu/abs/2012ApJ...744...52P}
}

@ARTICLE{Tiede2020,
       author = {{Tiede}, Christopher and {Zrake}, Jonathan and {MacFadyen}, Andrew and {Haiman}, Zoltan},
        title = "{Gas-driven Inspiral of Binaries in Thin Accretion Disks}",
      journal = {\apj},
         year = 2020,
        month = sep,
       volume = {900},
       number = {1},
          eid = {43},
        pages = {43},
          doi = {10.3847/1538-4357/aba432},
archivePrefix = {arXiv},
       eprint = {2005.09555},
 primaryClass = {astro-ph.GA},
       adsurl = {https://ui.adsabs.harvard.edu/abs/2020ApJ...900...43T}
}

@ARTICLE{Teasdale2026,
       author = {{Teasdale}, Matthew and {Stamatellos}, Dimitris},
        title = "{The lower mass limit for circumbinary disc fragmentation}",
      journal = {\mnras},
         year = 2026,
        month = jan,
       volume = {545},
       number = {3},
          eid = {staf2129},
        pages = {staf2129},
          doi = {10.1093/mnras/staf2129},
archivePrefix = {arXiv},
       eprint = {2509.20125},
 primaryClass = {astro-ph.SR},
       adsurl = {https://ui.adsabs.harvard.edu/abs/2026MNRAS.545f2129T}
}

@ARTICLE{KratterLodato2016,
       author = {{Kratter}, Kaitlin and {Lodato}, Giuseppe},
        title = "{Gravitational Instabilities in Circumstellar Disks}",
      journal = {\araa},
         year = 2016,
        month = sep,
       volume = {54},
        pages = {271-311},
          doi = {10.1146/annurev-astro-081915-023307},
archivePrefix = {arXiv},
       eprint = {1603.01280},
 primaryClass = {astro-ph.SR},
       adsurl = {https://ui.adsabs.harvard.edu/abs/2016ARA&A..54..271K}
}

@ARTICLE{Siwek2023b,
       author = {{Siwek}, Magdalena and {Weinberger}, Rainer and {Mu{\~n}oz}, Diego J. and {Hernquist}, Lars},
        title = "{Preferential accretion and circumbinary disc precession in eccentric binary systems}",
      journal = {\mnras},
         year = 2023,
        month = feb,
       volume = {518},
       number = {4},
        pages = {5059-5071},
          doi = {10.1093/mnras/stac3263},
archivePrefix = {arXiv},
       eprint = {2203.02514},
 primaryClass = {astro-ph.HE},
       adsurl = {https://ui.adsabs.harvard.edu/abs/2023MNRAS.518.5059S}
}

@ARTICLE{ArtymowiczLubow1994,
       author = {{Artymowicz}, Pawel and {Lubow}, Stephen H.},
        title = "{Dynamics of Binary-Disk Interaction. I. Resonances and Disk Gap Sizes}",
      journal = {\apj},
         year = 1994,
        month = feb,
       volume = {421},
        pages = {651},
          doi = {10.1086/173679},
       adsurl = {https://ui.adsabs.harvard.edu/abs/1994ApJ...421..651A}
}

@ARTICLE{Pringle1981,
       author = {{Pringle}, J.~E.},
        title = "{Accretion discs in astrophysics}",
      journal = {\araa},
         year = 1981,
        month = jan,
       volume = {19},
        pages = {137-162},
          doi = {10.1146/annurev.aa.19.090181.001033},
       adsurl = {https://ui.adsabs.harvard.edu/abs/1981ARA&A..19..137P}
}

@ARTICLE{LaiMunoz2023,
       author = {{Lai}, Dong and {Mu{\~n}oz}, Diego J.},
        title = "{Circumbinary Accretion: From Binary Stars to Massive Binary Black Holes}",
      journal = {\araa},
         year = 2023,
        month = aug,
       volume = {61},
        pages = {517-560},
          doi = {10.1146/annurev-astro-052622-022933},
archivePrefix = {arXiv},
       eprint = {2211.00028},
 primaryClass = {astro-ph.HE},
       adsurl = {https://ui.adsabs.harvard.edu/abs/2023ARA&A..61..517L}
}

@ARTICLE{TutukovYungelson1993,
       author = {{Tutukov}, A.~V. and {Yungelson}, L.~R.},
        title = "{The merger rate of neutron star and black hole binaries.}",
      journal = {\mnras},
         year = 1993,
        month = feb,
       volume = {260},
        pages = {675-678},
          doi = {10.1093/mnras/260.3.675},
       adsurl = {https://ui.adsabs.harvard.edu/abs/1993MNRAS.260..675T}
}

@ARTICLE{DewivandenHeuvel2004,
       author = {{Dewi}, J.~D.~M. and {van den Heuvel}, E.~P.~J.},
        title = "{The formation of the double neutron star pulsar J0737 - 3039}",
      journal = {\mnras},
         year = 2004,
        month = mar,
       volume = {349},
       number = {1},
        pages = {169-172},
          doi = {10.1111/j.1365-2966.2004.07477.x},
archivePrefix = {arXiv},
       eprint = {astro-ph/0312152},
 primaryClass = {astro-ph},
       adsurl = {https://ui.adsabs.harvard.edu/abs/2004MNRAS.349..169D}
}

@ARTICLE{Fragos2019,
       author = {{Fragos}, Tassos and {Andrews}, Jeff J. and {Ramirez-Ruiz}, Enrico and {Meynet}, Georges and {Kalogera}, Vicky and {Taam}, Ronald E. and {Zezas}, Andreas},
        title = "{The Complete Evolution of a Neutron-star Binary through a Common Envelope Phase Using 1D Hydrodynamic Simulations}",
      journal = {\apjl},
         year = 2019,
        month = oct,
       volume = {883},
       number = {2},
          eid = {L45},
        pages = {L45},
          doi = {10.3847/2041-8213/ab40d1},
archivePrefix = {arXiv},
       eprint = {1907.12573},
 primaryClass = {astro-ph.HE},
       adsurl = {https://ui.adsabs.harvard.edu/abs/2019ApJ...883L..45F}
}

@ARTICLE{deMinkBelczynski2015,
       author = {{de Mink}, S.~E. and {Belczynski}, K.},
        title = "{Merger Rates of Double Neutron Stars and Stellar Origin Black Holes: The Impact of Initial Conditions on Binary Evolution Predictions}",
      journal = {\apj},
         year = 2015,
        month = nov,
       volume = {814},
       number = {1},
          eid = {58},
        pages = {58},
          doi = {10.1088/0004-637X/814/1/58},
archivePrefix = {arXiv},
       eprint = {1506.03573},
 primaryClass = {astro-ph.HE},
       adsurl = {https://ui.adsabs.harvard.edu/abs/2015ApJ...814...58D}
}

@ARTICLE{Webbink1984,
       author = {{Webbink}, R.~F.},
        title = "{Double white dwarfs as progenitors of R Coronae Borealis stars and type I supernovae.}",
      journal = {\apj},
         year = 1984,
        month = feb,
       volume = {277},
        pages = {355-360},
          doi = {10.1086/161701},
       adsurl = {https://ui.adsabs.harvard.edu/abs/1984ApJ...277..355W}
}

@ARTICLE{VignaGomez2022,
       author = {{Vigna-G{\'o}mez}, Alejandro and {Wassink}, Michelle and {Klencki}, Jakub and {Istrate}, Alina and {Nelemans}, Gijs and {Mandel}, Ilya},
        title = "{Stellar response after stripping as a model for common-envelope outcomes}",
      journal = {\mnras},
         year = 2022,
        month = apr,
       volume = {511},
       number = {2},
        pages = {2326-2338},
          doi = {10.1093/mnras/stac237},
archivePrefix = {arXiv},
       eprint = {2107.14526},
 primaryClass = {astro-ph.HE},
       adsurl = {https://ui.adsabs.harvard.edu/abs/2022MNRAS.511.2326V}
}

@ARTICLE{Opik1924,
       author = {{{\"O}pik}, E.},
        title = "{Statistical Studies of Double Stars: On the Distribution of Relative Luminosities and Distances of Double Stars in the Harvard Revised Photometry North of Declination -31{\textdegree}}",
      journal = {Publications of the Tartu Astrofizica Observatory},
         year = 1924,
        month = jan,
       volume = {25},
        pages = {1},
       adsurl = {https://ui.adsabs.harvard.edu/abs/1924PTarO..25f...1O}
}

@ARTICLE{Springel2010,
       author = {{Springel}, Volker},
        title = "{E pur si muove: Galilean-invariant cosmological hydrodynamical simulations on a moving mesh}",
      journal = {\mnras},
         year = 2010,
        month = jan,
       volume = {401},
       number = {2},
        pages = {791-851},
          doi = {10.1111/j.1365-2966.2009.15715.x},
archivePrefix = {arXiv},
       eprint = {0901.4107},
 primaryClass = {astro-ph.CO},
       adsurl = {https://ui.adsabs.harvard.edu/abs/2010MNRAS.401..791S}
}

@ARTICLE{Pakmor2016,
       author = {{Pakmor}, R{\"u}diger and {Springel}, Volker and {Bauer}, Andreas and {Mocz}, Philip and {Munoz}, Diego J. and {Ohlmann}, Sebastian T. and {Schaal}, Kevin and {Zhu}, Chenchong},
        title = "{Improving the convergence properties of the moving-mesh code AREPO}",
      journal = {\mnras},
         year = 2016,
        month = jan,
       volume = {455},
       number = {1},
        pages = {1134-1143},
          doi = {10.1093/mnras/stv2380},
archivePrefix = {arXiv},
       eprint = {1503.00562},
 primaryClass = {astro-ph.GA},
       adsurl = {https://ui.adsabs.harvard.edu/abs/2016MNRAS.455.1134P}
}

@ARTICLE{Munoz2013,
       author = {{Mu{\~n}oz}, D.~J. and {Springel}, V. and {Marcus}, R. and {Vogelsberger}, M. and {Hernquist}, L.},
        title = "{Multidimensional, compressible viscous flow on a moving Voronoi mesh}",
      journal = {\mnras},
         year = 2013,
        month = jan,
       volume = {428},
       number = {1},
        pages = {254-279},
          doi = {10.1093/mnras/sts015},
archivePrefix = {arXiv},
       eprint = {1203.1037},
 primaryClass = {astro-ph.IM},
       adsurl = {https://ui.adsabs.harvard.edu/abs/2013MNRAS.428..254M}
}

@ARTICLE{Ho2023,
       author = {{Ho}, Anna Y.~Q. and {Perley}, Daniel A. and {Gal-Yam}, Avishay and {Lunnan}, Ragnhild and {Sollerman}, Jesper and {Schulze}, Steve and {Das}, Kaustav K. and {Dobie}, Dougal and {Yao}, Yuhan and {Fremling}, Christoffer and {Adams}, Scott and {Anand}, Shreya and {Andreoni}, Igor and {Bellm}, Eric C. and {Bruch}, Rachel J. and {Burdge}, Kevin B. and {Castro-Tirado}, Alberto J. and {Dahiwale}, Aishwarya and {De}, Kishalay and {Dekany}, Richard and {Drake}, Andrew J. and {Duev}, Dmitry A. and {Graham}, Matthew J. and {Helou}, George and {Kaplan}, David L. and {Karambelkar}, Viraj and {Kasliwal}, Mansi M. and {Kool}, Erik C. and {Kulkarni}, S.~R. and {Mahabal}, Ashish A. and {Medford}, Michael S. and {Miller}, A.~A. and {Nordin}, Jakob and {Ofek}, Eran and {Petitpas}, Glen and {Riddle}, Reed and {Sharma}, Yashvi and {Smith}, Roger and {Stewart}, Adam J. and {Taggart}, Kirsty and {Tartaglia}, Leonardo and {Tzanidakis}, Anastasios and {Winters}, Jan Martin},
        title = "{A Search for Extragalactic Fast Blue Optical Transients in ZTF and the Rate of AT2018cow-like Transients}",
      journal = {\apj},
         year = 2023,
        month = jun,
       volume = {949},
       number = {2},
          eid = {120},
        pages = {120},
          doi = {10.3847/1538-4357/acc533},
archivePrefix = {arXiv},
       eprint = {2105.08811},
 primaryClass = {astro-ph.HE},
       adsurl = {https://ui.adsabs.harvard.edu/abs/2023ApJ...949..120H}
}

@ARTICLE{Coppejans2020,
       author = {{Coppejans}, D.~L. and {Margutti}, R. and {Terreran}, G. and {Nayana}, A.~J. and {Coughlin}, E.~R. and {Laskar}, T. and {Alexander}, K.~D. and {Bietenholz}, M. and {Caprioli}, D. and {Chandra}, P. and {Drout}, M.~R. and {Frederiks}, D. and {Frohmaier}, C. and {Hurley}, K.~H. and {Kochanek}, C.~S. and {MacLeod}, M. and {Meisner}, A. and {Nugent}, P.~E. and {Ridnaia}, A. and {Sand}, D.~J. and {Svinkin}, D. and {Ward}, C. and {Yang}, S. and {Baldeschi}, A. and {Chilingarian}, I.~V. and {Dong}, Y. and {Esquivia}, C. and {Fong}, W. and {Guidorzi}, C. and {Lundqvist}, P. and {Milisavljevic}, D. and {Paterson}, K. and {Reichart}, D.~E. and {Shappee}, B. and {Stroh}, M.~C. and {Valenti}, S. and {Zauderer}, B.~A. and {Zhang}, B.},
        title = "{A Mildly Relativistic Outflow from the Energetic, Fast-rising Blue Optical Transient CSS161010 in a Dwarf Galaxy}",
      journal = {\apjl},
         year = 2020,
        month = may,
       volume = {895},
       number = {1},
          eid = {L23},
        pages = {L23},
          doi = {10.3847/2041-8213/ab8cc7},
archivePrefix = {arXiv},
       eprint = {2003.10503},
 primaryClass = {astro-ph.HE},
       adsurl = {https://ui.adsabs.harvard.edu/abs/2020ApJ...895L..23C}
}

@ARTICLE{GiacobboMapelli2018,
       author = {{Giacobbo}, Nicola and {Mapelli}, Michela},
        title = "{The progenitors of compact-object binaries: impact of metallicity, common envelope and natal kicks}",
      journal = {\mnras},
         year = 2018,
        month = oct,
       volume = {480},
       number = {2},
        pages = {2011-2030},
          doi = {10.1093/mnras/sty1999},
archivePrefix = {arXiv},
       eprint = {1806.00001},
 primaryClass = {astro-ph.HE},
       adsurl = {https://ui.adsabs.harvard.edu/abs/2018MNRAS.480.2011G}
}

@ARTICLE{Tang2020,
       author = {{Tang}, Petra N. and {Eldridge}, J.~J. and {Stanway}, Elizabeth R. and {Bray}, J.~C.},
        title = "{Dependence of gravitational wave transient rates on cosmic star formation and metallicity evolution history}",
      journal = {\mnras},
         year = 2020,
        month = mar,
       volume = {493},
       number = {1},
        pages = {L6-L10},
          doi = {10.1093/mnrasl/slz183},
archivePrefix = {arXiv},
       eprint = {1912.04474},
 primaryClass = {astro-ph.GA},
       adsurl = {https://ui.adsabs.harvard.edu/abs/2020MNRAS.493L...6T}
}

@ARTICLE{Baibhav2019,
       author = {{Baibhav}, Vishal and {Berti}, Emanuele and {Gerosa}, Davide and {Mapelli}, Michela and {Giacobbo}, Nicola and {Bouffanais}, Yann and {Di Carlo}, Ugo N.},
        title = "{Gravitational-wave detection rates for compact binaries formed in isolation: LIGO/Virgo O3 and beyond}",
      journal = {\prd},
         year = 2019,
        month = sep,
       volume = {100},
       number = {6},
          eid = {064060},
        pages = {064060},
          doi = {10.1103/PhysRevD.100.064060},
archivePrefix = {arXiv},
       eprint = {1906.04197},
 primaryClass = {gr-qc},
       adsurl = {https://ui.adsabs.harvard.edu/abs/2019PhRvD.100f4060B}
}

@ARTICLE{Hobbs2005,
       author = {{Hobbs}, G. and {Lorimer}, D.~R. and {Lyne}, A.~G. and {Kramer}, M.},
        title = "{A statistical study of 233 pulsar proper motions}",
      journal = {\mnras},
         year = 2005,
        month = jul,
       volume = {360},
       number = {3},
        pages = {974-992},
          doi = {10.1111/j.1365-2966.2005.09087.x},
archivePrefix = {arXiv},
       eprint = {astro-ph/0504584},
 primaryClass = {astro-ph},
       adsurl = {https://ui.adsabs.harvard.edu/abs/2005MNRAS.360..974H}
}

@ARTICLE{Tout1996,
       author = {{Tout}, Christopher A. and {Pols}, Onno R. and {Eggleton}, Peter P. and {Han}, Zhanwen},
        title = "{Zero-age main-seqence radii and luminosities as analytic functions of mass and metallicity}",
      journal = {\mnras},
         year = 1996,
        month = jul,
       volume = {281},
       number = {1},
        pages = {257-262},
          doi = {10.1093/mnras/281.1.257},
       adsurl = {https://ui.adsabs.harvard.edu/abs/1996MNRAS.281..257T}
}

@ARTICLE{Tout1997,
       author = {{Tout}, Christopher A. and {Aarseth}, Sverre J. and {Pols}, Onno R. and {Eggleton}, Peter P.},
        title = "{Rapid binary star evolution for N-body simulations and population synthesis}",
      journal = {\mnras},
         year = 1997,
        month = nov,
       volume = {291},
       number = {4},
        pages = {732-748},
          doi = {10.1093/mnras/291.4.732},
       adsurl = {https://ui.adsabs.harvard.edu/abs/1997MNRAS.291..732T}
}

@ARTICLE{Neijssel2019,
       author = {{Neijssel}, Coenraad J. and {Vigna-G{\'o}mez}, Alejandro and {Stevenson}, Simon and {Barrett}, Jim W. and {Gaebel}, Sebastian M. and {Broekgaarden}, Floor S. and {de Mink}, Selma E. and {Sz{\'e}csi}, Dorottya and {Vinciguerra}, Serena and {Mandel}, Ilya},
        title = "{The effect of the metallicity-specific star formation history on double compact object mergers}",
      journal = {\mnras},
         year = 2019,
        month = dec,
       volume = {490},
       number = {3},
        pages = {3740-3759},
          doi = {10.1093/mnras/stz2840},
archivePrefix = {arXiv},
       eprint = {1906.08136},
 primaryClass = {astro-ph.SR},
       adsurl = {https://ui.adsabs.harvard.edu/abs/2019MNRAS.490.3740N}
}

@ARTICLE{Renzo2019,
       author = {{Renzo}, M. and {Zapartas}, E. and {de Mink}, S.~E. and {G{\"o}tberg}, Y. and {Justham}, S. and {Farmer}, R.~J. and {Izzard}, R.~G. and {Toonen}, S. and {Sana}, H.},
        title = "{Massive runaway and walkaway stars. A study of the kinematical imprints of the physical processes governing the evolution and explosion of their binary progenitors}",
      journal = {\aap},
         year = 2019,
        month = apr,
       volume = {624},
          eid = {A66},
        pages = {A66},
          doi = {10.1051/0004-6361/201833297},
archivePrefix = {arXiv},
       eprint = {1804.09164},
 primaryClass = {astro-ph.SR},
       adsurl = {https://ui.adsabs.harvard.edu/abs/2019A&A...624A..66R}
}

@ARTICLE{Mokiem2007,
       author = {{Mokiem}, M.~R. and {de Koter}, A. and {Vink}, J.~S. and {Puls}, J. and {Evans}, C.~J. and {Smartt}, S.~J. and {Crowther}, P.~A. and {Herrero}, A. and {Langer}, N. and {Lennon}, D.~J. and {Najarro}, F. and {Villamariz}, M.~R.},
        title = "{The empirical metallicity dependence of the mass-loss rate of O- and early B-type stars}",
      journal = {\aap},
         year = 2007,
        month = oct,
       volume = {473},
       number = {2},
        pages = {603-614},
          doi = {10.1051/0004-6361:20077545},
archivePrefix = {arXiv},
       eprint = {0708.2042},
 primaryClass = {astro-ph},
       adsurl = {https://ui.adsabs.harvard.edu/abs/2007A&A...473..603M}
}

@ARTICLE{Margutti2019,
       author = {{Margutti}, R. and {Metzger}, B.~D. and {Chornock}, R. and {Vurm}, I. and {Roth}, N. and {Grefenstette}, B.~W. and {Savchenko}, V. and {Cartier}, R. and {Steiner}, J.~F. and {Terreran}, G. and {Margalit}, B. and {Migliori}, G. and {Milisavljevic}, D. and {Alexander}, K.~D. and {Bietenholz}, M. and {Blanchard}, P.~K. and {Bozzo}, E. and {Brethauer}, D. and {Chilingarian}, I.~V. and {Coppejans}, D.~L. and {Ducci}, L. and {Ferrigno}, C. and {Fong}, W. and {G{\"o}tz}, D. and {Guidorzi}, C. and {Hajela}, A. and {Hurley}, K. and {Kuulkers}, E. and {Laurent}, P. and {Mereghetti}, S. and {Nicholl}, M. and {Patnaude}, D. and {Ubertini}, P. and {Banovetz}, J. and {Bartel}, N. and {Berger}, E. and {Coughlin}, E.~R. and {Eftekhari}, T. and {Frederiks}, D.~D. and {Kozlova}, A.~V. and {Laskar}, T. and {Svinkin}, D.~S. and {Drout}, M.~R. and {MacFadyen}, A. and {Paterson}, K.},
        title = "{An Embedded X-Ray Source Shines through the Aspherical AT 2018cow: Revealing the Inner Workings of the Most Luminous Fast-evolving Optical Transients}",
      journal = {\apj},
         year = 2019,
        month = feb,
       volume = {872},
       number = {1},
          eid = {18},
        pages = {18},
          doi = {10.3847/1538-4357/aafa01},
archivePrefix = {arXiv},
       eprint = {1810.10720},
 primaryClass = {astro-ph.HE},
       adsurl = {https://ui.adsabs.harvard.edu/abs/2019ApJ...872...18M}
}

@ARTICLE{Ho2020,
       author = {{Ho}, Anna Y.~Q. and {Perley}, Daniel A. and {Kulkarni}, S.~R. and {Dong}, Dillon Z.~J. and {De}, Kishalay and {Chandra}, Poonam and {Andreoni}, Igor and {Bellm}, Eric C. and {Burdge}, Kevin B. and {Coughlin}, Michael and {Dekany}, Richard and {Feeney}, Michael and {Frederiks}, Dmitry D. and {Fremling}, Christoffer and {Golkhou}, V. Zach and {Graham}, Matthew J. and {Hale}, David and {Helou}, George and {Horesh}, Assaf and {Kasliwal}, Mansi M. and {Laher}, Russ R. and {Masci}, Frank J. and {Miller}, A.~A. and {Porter}, Michael and {Ridnaia}, Anna and {Rusholme}, Ben and {Shupe}, David L. and {Soumagnac}, Maayane T. and {Svinkin}, Dmitry S.},
        title = "{The Koala: A Fast Blue Optical Transient with Luminous Radio Emission from a Starburst Dwarf Galaxy at z = 0.27}",
      journal = {\apj},
         year = 2020,
        month = may,
       volume = {895},
       number = {1},
          eid = {49},
        pages = {49},
          doi = {10.3847/1538-4357/ab8bcf},
archivePrefix = {arXiv},
       eprint = {2003.01222},
 primaryClass = {astro-ph.HE},
       adsurl = {https://ui.adsabs.harvard.edu/abs/2020ApJ...895...49H}
}

@ARTICLE{FoxSmith2019,
       author = {{Fox}, Ori D. and {Smith}, Nathan},
        title = "{Signatures of circumstellar interaction in the unusual transient AT 2018cow}",
      journal = {\mnras},
         year = 2019,
        month = sep,
       volume = {488},
       number = {3},
        pages = {3772-3782},
          doi = {10.1093/mnras/stz1925},
archivePrefix = {arXiv},
       eprint = {1903.01535},
 primaryClass = {astro-ph.HE},
       adsurl = {https://ui.adsabs.harvard.edu/abs/2019MNRAS.488.3772F}
}

@ARTICLE{Nugent2026,
       author = {{Nugent}, Anya E. and {Villar}, V. Ashley and {Metzger}, Brian D. and {Fryer}, Christopher L. and {Burns}, Eric and {Gordon}, Alexa and {Frostig}, Danielle},
        title = "{The Environments of Luminous Fast Blue Optical Transients: Evidence for a Compact Object and Wolf-Rayet Star Merger Origin}",
      journal = {arXiv e-prints},
         year = 2026,
        month = mar,
          eid = {arXiv:2603.23597},
        pages = {arXiv:2603.23597},
          doi = {10.48550/arXiv.2603.23597},
archivePrefix = {arXiv},
       eprint = {2603.23597},
 primaryClass = {astro-ph.HE},
       adsurl = {https://ui.adsabs.harvard.edu/abs/2026arXiv260323597N}
}

@ARTICLE{Wehmeyer2019,
       author = {{Wehmeyer}, B. and {Fr{\"o}hlich}, C. and {C{\^o}t{\'e}}, B. and {Pignatari}, M. and {Thielemann}, F.-K.},
        title = "{Using failed supernovae to constrain the Galactic r-process element production}",
      journal = {\mnras},
         year = 2019,
        month = aug,
       volume = {487},
       number = {2},
        pages = {1745-1753},
          doi = {10.1093/mnras/stz1310},
archivePrefix = {arXiv},
       eprint = {1908.05617},
 primaryClass = {astro-ph.GA},
       adsurl = {https://ui.adsabs.harvard.edu/abs/2019MNRAS.487.1745W}
}

@ARTICLE{Ji2019,
       author = {{Ji}, Alexander P. and {Drout}, Maria R. and {Hansen}, Terese T.},
        title = "{The Lanthanide Fraction Distribution in Metal-poor Stars: A Test of Neutron Star Mergers as the Dominant r-process Site}",
      journal = {\apj},
         year = 2019,
        month = sep,
       volume = {882},
       number = {1},
          eid = {40},
        pages = {40},
          doi = {10.3847/1538-4357/ab3291},
archivePrefix = {arXiv},
       eprint = {1905.01814},
 primaryClass = {astro-ph.HE},
       adsurl = {https://ui.adsabs.harvard.edu/abs/2019ApJ...882...40J}
}

@ARTICLE{HolmbeckAndrews2024,
       author = {{Holmbeck}, Erika M. and {Andrews}, Jeff J.},
        title = "{Total r-process Yields of Milky Way Neutron Star Mergers}",
      journal = {\apj},
         year = 2024,
        month = mar,
       volume = {963},
       number = {2},
          eid = {110},
        pages = {110},
          doi = {10.3847/1538-4357/ad1e52},
archivePrefix = {arXiv},
       eprint = {2310.03847},
 primaryClass = {astro-ph.HE},
       adsurl = {https://ui.adsabs.harvard.edu/abs/2024ApJ...963..110H}
}

@ARTICLE{ChenHsinYu2025,
       author = {{Chen}, Hsin-Yu and {Landry}, Philippe and {Read}, Jocelyn S. and {Siegel}, Daniel M.},
        title = "{Inference of Multichannel r-process Element Enrichment in the Milky Way Using Binary Neutron Star Merger Observations}",
      journal = {\apj},
         year = 2025,
        month = jun,
       volume = {985},
       number = {2},
          eid = {154},
        pages = {154},
          doi = {10.3847/1538-4357/add0af},
archivePrefix = {arXiv},
       eprint = {2402.03696},
 primaryClass = {astro-ph.HE},
       adsurl = {https://ui.adsabs.harvard.edu/abs/2025ApJ...985..154C}
}

@ARTICLE{Saleem2025,
       author = {{Saleem}, Muhammed and {Chen}, Hsin-Yu and {Siegel}, Daniel M. and {Landry}, Philippe and {Read}, Jocelyn S. and {Wang}, Kaile},
        title = "{Mergers Fall Short: Non-merger Channels Required for Galactic Heavy Element Production}",
      journal = {arXiv e-prints},
         year = 2025,
        month = aug,
          eid = {arXiv:2508.06020},
        pages = {arXiv:2508.06020},
          doi = {10.48550/arXiv.2508.06020},
archivePrefix = {arXiv},
       eprint = {2508.06020},
 primaryClass = {astro-ph.HE},
       adsurl = {https://ui.adsabs.harvard.edu/abs/2025arXiv250806020S}
}

@ARTICLE{Dutrey1994,
       author = {{Dutrey}, A. and {Guilloteau}, S. and {Simon}, M.},
        title = "{Images of the GG Tauri rotating ring}",
      journal = {\aap},
         year = 1994,
        month = jun,
       volume = {286},
        pages = {149-159},
       adsurl = {https://ui.adsabs.harvard.edu/abs/1994A&A...286..149D}
}

@ARTICLE{Ragusa2021,
       author = {{Ragusa}, Enrico and {Fasano}, Daniele and {Toci}, Claudia and {Duch{\^e}ne}, Gaspard and {Cuello}, Nicol{\'a}s and {Villenave}, Marion and {van der Plas}, Gerrit and {Lodato}, Giuseppe and {M{\'e}nard}, Fran{\c{c}}ois and {Price}, Daniel J. and {Pinte}, Christophe and {Stapelfeldt}, Karl and {Wolff}, Schuyler},
        title = "{Circumbinary and circumstellar discs around the eccentric binary IRAS 04158+2805 - a testbed for binary-disc interaction}",
      journal = {\mnras},
         year = 2021,
        month = oct,
       volume = {507},
       number = {1},
        pages = {1157-1174},
          doi = {10.1093/mnras/stab2179},
archivePrefix = {arXiv},
       eprint = {2107.13566},
 primaryClass = {astro-ph.SR},
       adsurl = {https://ui.adsabs.harvard.edu/abs/2021MNRAS.507.1157R}
}

@ARTICLE{Akeson2007,
       author = {{Akeson}, R.~L. and {Rice}, W.~K.~M. and {Boden}, A.~F. and {Sargent}, A.~I. and {Carpenter}, J.~M. and {Bryden}, G.},
        title = "{The Circumbinary Disk of HD 98800B: Evidence for Disk Warping}",
      journal = {\apj},
         year = 2007,
        month = dec,
       volume = {670},
       number = {2},
        pages = {1240-1246},
          doi = {10.1086/522579},
archivePrefix = {arXiv},
       eprint = {0708.2390},
 primaryClass = {astro-ph},
       adsurl = {https://ui.adsabs.harvard.edu/abs/2007ApJ...670.1240A}
}

@ARTICLE{GallardoCava2026,
       author = {{Gallardo Cava}, I. and {Alcolea}, J. and {Van Winckel}, H. and {Bujarrabal}, V. and {Santander-Garc{\'\i}a}, M. and {G{\'o}mez-Garrido}, M.},
        title = "{CO observations in rotating circumbinary post-AGB disks}",
      journal = {\aap},
         year = 2026,
        month = jan,
       volume = {705},
          eid = {A29},
        pages = {A29},
          doi = {10.1051/0004-6361/202556445},
archivePrefix = {arXiv},
       eprint = {2511.23348},
 primaryClass = {astro-ph.SR},
       adsurl = {https://ui.adsabs.harvard.edu/abs/2026A&A...705A..29G}
}

@ARTICLE{Hoard2009,
       author = {{Hoard}, D.~W. and {Kafka}, Stella and {Wachter}, Stefanie and {Howell}, Steve B. and {Brinkworth}, Carolyn S. and {Ciardi}, David R. and {Szkody}, Paula and {Belle}, Kunegunda and {Froning}, Cynthia and {van Belle}, Gerard},
        title = "{Observations of V592 Cassiopeiae with the Spitzer Space Telescope{\textemdash}Dust in the Mid-Infrared}",
      journal = {\apj},
         year = 2009,
        month = mar,
       volume = {693},
       number = {1},
        pages = {236-249},
          doi = {10.1088/0004-637X/693/1/236},
archivePrefix = {arXiv},
       eprint = {0811.0363},
 primaryClass = {astro-ph},
       adsurl = {https://ui.adsabs.harvard.edu/abs/2009ApJ...693..236H}
}

@ARTICLE{Liao2021,
       author = {{Liao}, Wei-Ting and {Chen}, Yu-Ching and {Liu}, Xin and {Holgado}, A. Miguel and {Guo}, Hengxiao and {Gruendl}, Robert and {Morganson}, Eric and {Shen}, Yue and {Davis}, Tamara and {Kessler}, Richard and {Martini}, Paul and {McMahon}, Richard G. and {Allam}, Sahar and {Annis}, James and {Avila}, Santiago and {Banerji}, Manda and {Bechtol}, Keith and {Bertin}, Emmanuel and {Brooks}, David and {Buckley-Geer}, Elizabeth and {Carnero Rosell}, Aurelio and {Carrasco Kind}, Matias and {Carretero}, Jorge and {Javier Castander}, Francisco and {Cunha}, Carlos and {D'Andrea}, Chris and {da Costa}, Luiz and {Davis}, Christopher and {De Vicente}, Juan and {Desai}, Shantanu and {Thomas Diehl}, H. and {Doel}, Peter and {Eifler}, Tim and {Evrard}, August and {Flaugher}, Brenna and {Fosalba}, Pablo and {Frieman}, Josh and {Garcia-Bellido}, Juan and {Gaztanaga}, Enrique and {Glazebrook}, Karl and {Gruen}, Daniel and {Gschwend}, Julia and {Gutierrez}, Gaston and {Hartley}, Will and {Hollowood}, Devon L. and {Honscheid}, Klaus and {Hoyle}, Ben and {James}, David and {Krause}, Elisabeth and {Kuehn}, Kyler and {Lima}, Marcos and {Maia}, Marcio and {Marshall}, Jennifer and {Menanteau}, Felipe and {Miquel}, Ramon and {Plazas Malag{\'o}n}, Andr{\'e}s and {Roodman}, Aaron and {Sanchez}, Eusebio and {Scarpine}, Vic and {Schubnell}, Michael and {Serrano}, Santiago and {Smith}, Mathew and {Smith}, R. Chris and {Soares-Santos}, Marcelle and {Sobreira}, Flavia and {Suchyta}, Eric and {Swanson}, Molly and {Tarle}, Gregory and {Vikram}, Vinu and {Walker}, Alistair},
        title = "{Discovery of a candidate binary supermassive black hole in a periodic quasar from circumbinary accretion variability}",
      journal = {\mnras},
         year = 2021,
        month = jan,
       volume = {500},
       number = {3},
        pages = {4025-4041},
          doi = {10.1093/mnras/staa3055},
archivePrefix = {arXiv},
       eprint = {2008.12317},
 primaryClass = {astro-ph.GA},
       adsurl = {https://ui.adsabs.harvard.edu/abs/2021MNRAS.500.4025L}
}

@ARTICLE{Nathaniel2025,
       author = {{Nathaniel}, K. and {Vigna-G{\'o}mez}, A. and {Grichener}, A. and {Farmer}, R. and {Renzo}, M. and {Everson}, R.~W.},
        title = "{Population synthesis of Thorne-{\.Z}ytkow objects: Rejuvenated donors and unexplored progenitors in the common envelope formation channel}",
      journal = {\aap},
         year = 2025,
        month = feb,
       volume = {694},
          eid = {A83},
        pages = {A83},
          doi = {10.1051/0004-6361/202451531},
archivePrefix = {arXiv},
       eprint = {2407.11680},
 primaryClass = {astro-ph.SR},
       adsurl = {https://ui.adsabs.harvard.edu/abs/2025A&A...694A..83N}
}

@ARTICLE{Mandel:2025grb,
       author = {{Mandel}, Ilya and {Salafia}, Om Sharan and {Levan}, Andrew and {Disberg}, Paul},
        title = "{The Maximum Offsets of Binary Neutron Star Mergers from Host Galaxies}",
      journal = {\apjl},
         year = 2026,
        month = mar,
       volume = {999},
       number = {2},
          eid = {L29},
        pages = {L29},
          doi = {10.3847/2041-8213/ae4332},
archivePrefix = {arXiv},
       eprint = {2512.12018},
 primaryClass = {astro-ph.HE},
       adsurl = {https://ui.adsabs.harvard.edu/abs/2026ApJ...999L..29M}
}

@ARTICLE{ArtymowiczLubow1996,
       author = {{Artymowicz}, Pawel and {Lubow}, Stephen H.},
        title = "{Mass Flow through Gaps in Circumbinary Disks}",
      journal = {\apjl},
         year = 1996,
        month = aug,
       volume = {467},
        pages = {L77},
          doi = {10.1086/310200},
       adsurl = {https://ui.adsabs.harvard.edu/abs/1996ApJ...467L..77A}
}

@ARTICLE{Oslowski2011,
       author = {{Os{\l}owski}, S. and {Bulik}, T. and {Gondek-Rosi{\'n}ska}, D. and {Belczy{\'n}ski}, K.},
        title = "{Population synthesis of double neutron stars}",
      journal = {\mnras},
         year = 2011,
        month = may,
       volume = {413},
       number = {1},
        pages = {461-479},
          doi = {10.1111/j.1365-2966.2010.18147.x},
archivePrefix = {arXiv},
       eprint = {0903.3538},
 primaryClass = {astro-ph.GA},
       adsurl = {https://ui.adsabs.harvard.edu/abs/2011MNRAS.413..461O}
}

@ARTICLE{Mandel2021,
       author = {{Mandel}, Ilya and {M{\"u}ller}, Bernhard and {Riley}, Jeff and {de Mink}, Selma E. and {Vigna-G{\'o}mez}, Alejandro and {Chattopadhyay}, Debatri},
        title = "{Binary population synthesis with probabilistic remnant mass and kick prescriptions}",
      journal = {\mnras},
         year = 2021,
        month = jan,
       volume = {500},
       number = {1},
        pages = {1380-1384},
          doi = {10.1093/mnras/staa3390},
archivePrefix = {arXiv},
       eprint = {2007.03890},
 primaryClass = {astro-ph.HE},
       adsurl = {https://ui.adsabs.harvard.edu/abs/2021MNRAS.500.1380M}
}

@ARTICLE{Chattaraj2026,
       author = {{Chattaraj}, Abhishek and {Andrews}, Jeff J. and {Bavera}, Simone S. and {Briel}, Max and {Chattopadhyay}, Debatri and {Fragos}, Tassos and {Gossage}, Seth and {Kalogera}, Vicky and {Kovlakas}, Konstantinos and {Kruckow}, Matthias U. and {Liotine}, Camille and {Rocha}, Kyle A. and {Srivastava}, Philipp M. and {Sun}, Meng and {Teng}, Elizabeth and {Xing}, Zepei and {Zapartas}, Emmanouil},
        title = "{Forming Double Neutron Stars Using Detailed Binary Evolution Models with POSYDON: Comparison to the Galactic Systems}",
      journal = {\apj},
         year = 2026,
        month = jan,
       volume = {997},
       number = {1},
          eid = {52},
        pages = {52},
          doi = {10.3847/1538-4357/ae1f93},
archivePrefix = {arXiv},
       eprint = {2508.00186},
 primaryClass = {astro-ph.SR},
       adsurl = {https://ui.adsabs.harvard.edu/abs/2026ApJ...997...52C}
}

@ARTICLE{GhodlaEldridge2023,
       author = {{Ghodla}, Sohan and {Eldridge}, J.~J.},
        title = "{Sustained super-Eddington accretion around neutron stars \& black holes}",
      journal = {\mnras},
         year = 2023,
        month = aug,
       volume = {523},
       number = {2},
        pages = {1711-1717},
          doi = {10.1093/mnras/stad1558},
archivePrefix = {arXiv},
       eprint = {2305.09919},
 primaryClass = {astro-ph.HE},
       adsurl = {https://ui.adsabs.harvard.edu/abs/2023MNRAS.523.1711G}
}

@ARTICLE{RickerTaam2012,
       author = {{Ricker}, Paul M. and {Taam}, Ronald E.},
        title = "{An AMR Study of the Common-envelope Phase of Binary Evolution}",
      journal = {\apj},
         year = 2012,
        month = feb,
       volume = {746},
       number = {1},
          eid = {74},
        pages = {74},
          doi = {10.1088/0004-637X/746/1/74},
archivePrefix = {arXiv},
       eprint = {1107.3889},
 primaryClass = {astro-ph.SR},
       adsurl = {https://ui.adsabs.harvard.edu/abs/2012ApJ...746...74R}
}

@ARTICLE{Sand2020,
       author = {{Sand}, Christian and {Ohlmann}, Sebastian T. and {Schneider}, Fabian R.~N. and {Pakmor}, R{\"u}diger and {R{\"o}pke}, Friedrich K.},
        title = "{Common-envelope evolution with an asymptotic giant branch star}",
      journal = {\aap},
         year = 2020,
        month = dec,
       volume = {644},
          eid = {A60},
        pages = {A60},
          doi = {10.1051/0004-6361/202038992},
archivePrefix = {arXiv},
       eprint = {2007.11000},
 primaryClass = {astro-ph.SR},
       adsurl = {https://ui.adsabs.harvard.edu/abs/2020A&A...644A..60S}
}

@ARTICLE{Gonzales2022,
       author = {{Gonz{\'a}lez-Bol{\'\i}var}, Miguel and {De Marco}, Orsola and {Lau}, Mike Y.~M. and {Hirai}, Ryosuke and {Price}, Daniel J.},
        title = "{Common envelope binary interaction simulations between a thermally pulsating AGB star and a low mass companion}",
      journal = {\mnras},
         year = 2022,
        month = dec,
       volume = {517},
       number = {3},
        pages = {3181-3199},
          doi = {10.1093/mnras/stac2301},
archivePrefix = {arXiv},
       eprint = {2205.09749},
 primaryClass = {astro-ph.SR},
       adsurl = {https://ui.adsabs.harvard.edu/abs/2022MNRAS.517.3181G}
}

@ARTICLE{LIGO2026,
       author = {{The LIGO Scientific Collaboration} and {the Virgo Collaboration} and {the KAGRA Collaboration} and {Abac}, A.~G. and {Abe}, A. and {Abouelfettouh}, I. and {Acernese}, F. and {Ackley}, K. and {Adam}, A. and {Adhicary}, S. and {Adhikari}, D. and {Adhikari}, R.~X. and {Adkins}, V.~K. and {Afroz}, S. and {Agapito}, A. and {Agarwal}, D. and {Agathos}, M. and {Aggarwal}, N. and {Aggarwal}, S. and {Aguiar}, O.~D. and {Ahrend}, I.-L. and {Aiello}, L. and {Ain}, A. and {Ajith}, P. and {Akutsu}, T. and {Albers}, L. and {Ali}, W. and {Al-Kershi}, S. and {Allene}, C. and {Allocca}, A. and {Al-Shammari}, S. and {Alvarez}, J.~A. and {Alvarez-Lopez}, S. and {Amar}, W. and {Amarasinghe}, O. and {Amato}, A. and {Amicucci}, F. and {Amra}, C. and {Anand}, A.~B. and {Anand}, C. and {Ananyeva}, A. and {Anderson}, S.~B. and {Anderson}, W.~G. and {Andia}, M. and {Ando}, M. and {Andrade-Oliveira}, F. and {Andr{\'e}s-Carcasona}, M. and {Andrey}, J.~L. and {Andri{\'c}}, T. and {Anglin}, J. and {Anna}, J. and {Antelis}, J.~M. and {Antier}, S. and {Antonini}, F. and {Aoki}, T. and {Aoumi}, M. and {Appavuravther}, E.~Z. and {Appelt}, E.~A. and {Appert}, S. and {Apple}, S.~K. and {Arai}, K. and {Araya}, A. and {Araya}, M.~C. and {Arca Sedda}, M. and {Arciprete}, F. and {Areeda}, J.~S. and {Aritomi}, N. and {Armato}, F. and {Armstrong}, S. and {Arnaud}, N. and {Arogeti}, M. and {Aronson}, S.~M. and {Ashton}, G. and {Aso}, Y. and {Asprea}, L. and {Assiduo}, M. and {Assis de Souza Melo}, S. and {Aston}, S.~M. and {Astone}, P. and {Aswathi}, P.~S. and {Attadio}, F. and {Aubin}, F. and {AultONeal}, K. and {Avallone}, G. and {Avdeev}, N. and {Avila}, E.~A. and {Babak}, S. and {Badger}, C. and {Bae}, S. and {Bagnasco}, S. and {Baimukhametova}, S. and {Baiotti}, L. and {Baka}, T. and {Baker}, K.~A. and {Baker}, T. and {Balbi}, G. and {Baldi}, G. and {Baldicchi}, N. and {Ball}, M. and {Ballardin}, G. and {Ballelli}, M. and {Ballmer}, S.~W. and {Banagiri}, S. and {Banerjee}, B. and {Bankar}, D. and {Baptiste}, T.~M. and {Baral}, P. and {Baratti}, M. and {Barayoga}, J.~C. and {Baric}, K. and {Barish}, B.~C. and {Barker}, D. and {Barman}, N. and {Barone}, F. and {Barr}, B. and {Barrios}, M. and {Barsotti}, L. and {Barsuglia}, M. and {Barta}, D. and {Barton}, M.~A. and {Bartos}, I. and {Basalaev}, A. and {Bassiri}, R. and {Basti}, A. and {Bawaj}, M. and {Bayley}, J.~C. and {Baylor}, A.~C. and {Baynard}, II, P.~A. and {Bazzan}, M. and {Bedakihale}, V.~M. and {Beirnaert}, F. and {Bejger}, M. and {Bell}, A.~S. and {Bellani}, C. and {Bellie}, D.~S. and {Beltran-Martinez}, D. and {Benedetti}, E. and {Benoit}, W. and {Bentara}, I. and {Ben Yaala}, M. and {Bera}, S. and {Bergamin}, F. and {Berger}, B.~K. and {Beroiz}, M. and {Berry}, C.~P.~L. and {Berry}, I. and {Bersanetti}, D. and {Bertheas}, T. and {Bertolini}, A. and {Betzwieser}, J. and {Beveridge}, D. and {Bevins}, N. and {Bezerra-Sobrinho}, J. and {Bhandare}, R. and {Bhatt}, R. and {Bhattacharjee}, A. and {Bhattacharjee}, D. and {Bhattacharyya}, S. and {Bhaumik}, S. and {Biancalana}, V. and {Bianchi}, F. and {Bilenko}, I.~A. and {Bilicki}, M. and {Billingsley}, G. and {Binetti}, A. and {Bini}, S. and {Biot}, S. and {Birnholtz}, O. and {Biscoveanu}, S. and {Bisht}, A. and {Bitossi}, M. and {Bizouard}, M.-A. and {Blaber}, S. and {Blackburn}, J.~K. and {Blagg}, L.~A. and {Blair}, C.~D. and {Blair}, D.~G. and {Bloch}, M. and {Bode}, N. and {Boettner}, N. and {Bogdan}, P. and {Boileau}, G. and {Boldrini}, M. and {Bolingbroke}, G.~N. and {Bonavena}, L.~D. and {Bonhomme}, V.~A. and {Bonilla}, E. and {Bonilla}, M.~S. and {Bonino}, A. and {Bonnand}, R. and {Borchers}, A. and {Borghi}, N. and {Boschi}, V. and {Bose}, S. and {Bossilkov}, V. and {Bothra}, Y. and {Boudon}, A. and {Boybeyi}, T.~D. and {Boyle}, M. and {Bozzi}, A.},
        title = "{GWTC-5.0: Population Properties of Merging Compact Binaries}",
      journal = {arXiv e-prints},
         year = 2026,
        month = may,
          eid = {arXiv:2605.27226},
        pages = {arXiv:2605.27226},
          doi = {10.48550/arXiv.2605.27226},
archivePrefix = {arXiv},
       eprint = {2605.27226},
 primaryClass = {astro-ph.HE},
       adsurl = {https://ui.adsabs.harvard.edu/abs/2026arXiv260527226T}
}

@ARTICLE{GlanzPerets2021,
       author = {{Glanz}, Hila and {Perets}, Hagai B.},
        title = "{Common envelope evolution of eccentric binaries}",
      journal = {\mnras},
         year = 2021,
        month = oct,
       volume = {507},
       number = {2},
        pages = {2659-2670},
          doi = {10.1093/mnras/stab2291},
archivePrefix = {arXiv},
       eprint = {2105.02227},
 primaryClass = {astro-ph.SR},
       adsurl = {https://ui.adsabs.harvard.edu/abs/2021MNRAS.507.2659G}
}

@ARTICLE{Oh2016,
       author = {{Oh}, Seungkyung and {Kroupa}, Pavel},
        title = "{Dynamical ejections of massive stars from young star clusters under diverse initial conditions}",
      journal = {\aap},
         year = 2016,
        month = may,
       volume = {590},
          eid = {A107},
        pages = {A107},
          doi = {10.1051/0004-6361/201628233},
archivePrefix = {arXiv},
       eprint = {1604.00006},
 primaryClass = {astro-ph.GA},
       adsurl = {https://ui.adsabs.harvard.edu/abs/2016A&A...590A.107O}
}

@ARTICLE{Chevance2020,
       author = {{Chevance}, M{\'e}lanie and {Kruijssen}, J.~M. Diederik and {Hygate}, Alexander P.~S. and {Schruba}, Andreas and {Longmore}, Steven N. and {Groves}, Brent and {Henshaw}, Jonathan D. and {Herrera}, Cinthya N. and {Hughes}, Annie and {Jeffreson}, Sarah M.~R. and {Lang}, Philipp and {Leroy}, Adam K. and {Meidt}, Sharon E. and {Pety}, J{\'e}r{\^o}me and {Razza}, Alessandro and {Rosolowsky}, Erik and {Schinnerer}, Eva and {Bigiel}, Frank and {Blanc}, Guillermo A. and {Emsellem}, Eric and {Faesi}, Christopher M. and {Glover}, Simon C.~O. and {Haydon}, Daniel T. and {Ho}, I.-Ting and {Kreckel}, Kathryn and {Lee}, Janice C. and {Liu}, Daizhong and {Querejeta}, Miguel and {Saito}, Toshiki and {Sun}, Jiayi and {Usero}, Antonio and {Utomo}, Dyas},
        title = "{The lifecycle of molecular clouds in nearby star-forming disc galaxies}",
      journal = {\mnras},
         year = 2020,
        month = apr,
       volume = {493},
       number = {2},
        pages = {2872-2909},
          doi = {10.1093/mnras/stz3525},
archivePrefix = {arXiv},
       eprint = {1911.03479},
 primaryClass = {astro-ph.GA},
       adsurl = {https://ui.adsabs.harvard.edu/abs/2020MNRAS.493.2872C}
}

@ARTICLE{Haydon2020,
       author = {{Haydon}, Daniel T. and {Kruijssen}, J.~M. Diederik and {Chevance}, M{\'e}lanie and {Hygate}, Alexander P.~S. and {Krumholz}, Mark R. and {Schruba}, Andreas and {Longmore}, Steven N.},
        title = "{An uncertainty principle for star formation - III. The characteristic emission time-scales of star formation rate tracers}",
      journal = {\mnras},
         year = 2020,
        month = oct,
       volume = {498},
       number = {1},
        pages = {235-257},
          doi = {10.1093/mnras/staa2430},
archivePrefix = {arXiv},
       eprint = {1810.10897},
 primaryClass = {astro-ph.GA},
       adsurl = {https://ui.adsabs.harvard.edu/abs/2020MNRAS.498..235H}
}

@ARTICLE{Bobrick2017,
       author = {{Bobrick}, Alexey and {Davies}, Melvyn B. and {Church}, Ross P.},
        title = "{Mass transfer in white dwarf-neutron star binaries}",
      journal = {\mnras},
         year = 2017,
        month = may,
       volume = {467},
       number = {3},
        pages = {3556-3575},
          doi = {10.1093/mnras/stx312},
archivePrefix = {arXiv},
       eprint = {1702.02377},
 primaryClass = {astro-ph.HE},
       adsurl = {https://ui.adsabs.harvard.edu/abs/2017MNRAS.467.3556B}
}

\clearpage
\appendix
\restartappendixnumbering

\section{Galactic merging double neutron star observations}
\label{app:galactic_merging_dns}

Galactic DNSs that are expected to merge within a Hubble time provide an additional observational comparison to the chirp-mass distributions predicted by our population models. In Table~\ref{tab:galactic_merging_dns}, we list the properties of the eight observed merging Galactic DNSs shown in Fig.~\ref{fig:ChirpMass}. Their chirp masses are calculated from the measured component masses, while their remaining GW inspiral times are estimated  from the present-day orbital parameters according to the \cite{Peters1964} equations. 

\begin{table}[h!]
\centering
\caption{Observed and derived properties of the merging Galactic DNS systems shown in Fig.~\ref{fig:ChirpMass}. $P_{\rm m}$ and $e_{\rm m}$ are the present-day measured orbital period and eccentricity, respectively. $M_{\rm p}$ and $M_{\rm c}$ are the masses of the observed pulsar and its companion, and $\mathcal{M}$ is the chirp mass of the binary. $\tau_{\rm GW}$ is the remaining gravitational-wave inspiral time. The period and eccentricity are taken from \citealt{Grichener2026} (and references therein), while the component masses come from the references listed in
the final column.
}
\label{tab:galactic_merging_dns}
\begin{tabular}{lccccccc}
\hline
Name &
$P_{\rm m}$ &
$e_{\rm m}$ &
$M_{\rm p}$ &
$M_{\rm c}$ &
$\mathcal{M}$ &
$\tau_{\rm GW}$ &
Ref \\
&
[days] & &
[$M_{\odot}$] &
[$M_{\odot}$] &
[$M_{\odot}$] &
[Gyr] & \\
\hline
J1946+2052
& 0.078 & 0.064
& 1.2842 & 1.2477
& 1.102 & 0.044 & (1) \\

J1757--1854
& 0.183 & 0.606
& 1.3384 & 1.3945
& 1.189 & 0.075 & (2) \\

J0737--3039A/B
& 0.102 & 0.088
& 1.338185 & 1.248868
& 1.125 & 0.085 & (3) \\

B1913+16
& 0.323 & 0.617
& 1.438 & 1.390
& 1.231 & 0.301 & (4) \\

J1913+1102
& 0.206 & 0.090
& 1.599 & 1.290
& 1.249 & 0.467 & (5) \\

J0509+3801
& 0.380 & 0.586
& 1.36 & 1.45
& 1.222 & 0.575 & (6) \\

J1756--2251
& 0.320 & 0.181
& 1.341 & 1.230
& 1.118 & 1.66 & (7) \\

B1534+12
& 0.421 & 0.274
& 1.3330 & 1.3455
& 1.166 & 2.74 & (8) \\
\hline
\end{tabular}

\begin{flushleft}
\footnotesize
References:
(1) \citealt{Meng2025};
(2) \citealt{Singha2026};
(3) \citealt{Kramer2021};
(4) \citealt{Weisberg2016};
(5) \citealt{Miao2026};
(6) \citealt{Lynch2018};
(7) \citealt{Ferdman2014};
(8) \citealt{Fonseca2014}.
\end{flushleft}
\end{table}

\end{document}